\documentclass[pre,twocolumn,english,superscriptaddress,floatfix,longbibliography]{revtex4-2}
\usepackage{graphicx,amsmath,amssymb,natbib,color,soul,easyReview,comment,lineno,appendix}

\usepackage{mhchem}
\usepackage[absolute,overlay]{textpos} % Add the textpos package for positioning
\usepackage{graphicx}% Include figure files
\usepackage{dcolumn}% Align table columns on decimal point
\usepackage{bm}% bold math
\usepackage[utf8]{inputenc}
\usepackage[T1]{fontenc}
\usepackage{mathptmx}
\usepackage{etoolbox}
\usepackage{overpic}
\usepackage[hidelinks]{hyperref} % or colorlinks=true,linkcolor=blue,...
\usepackage{color,xcolor}
\definecolor{red}{rgb}{0.8, 0.0, 0.0}
\definecolor{purp}{rgb}{0.4, 0.2, 0.8}

\begin{document}

\title{Recent Advances in Metallic Glasses}

%Now it is alphabetical order
\author{Silvia Bonfanti}
\affiliation{NOMATEN Centre of Excellence, National Center for Nuclear Research, ul. A. So\l{}tana 7, 05-400 Swierk/Otwock, Poland.}
\affiliation{Center for Complexity and Biosystems, Department of Physics "Aldo Pontremoli", University of Milan, Via Celoria 16, 20133 Milano, Italy.}
\author{Ralf Busch}
\affiliation{Saarland University, Chair of Metallic Materials, Campus C6.3, 66123 Saarbrücken, Germany}

\author{Jesper Byggm\"astar}
\affiliation{Department of Physics, P.O. Box 43, FI-00014 University of Helsinki, Finland.}

\author{Jeppe C. Dyre}
\affiliation{\textit{Glass and Time}, IMFUFA, Department of Science and Environment, Roskilde University, DK-4000 Roskilde, Denmark}

\author{Jürgen Eckert}
\affiliation{Department of Materials Science, Montanuniversität Leoben, Jahnstraße 12, 8700 Leoben, Austria}
\affiliation{Erich Schmid Institute of Materials Science, Austrian Academy of Sciences, Jahnstraße 12, 8700 Leoben, Austria}

\author{Spencer Fajardo}
\affiliation{Department of Materials Science and Engineering, Johns Hopkins University, Baltimore, MD 21218, USA}

\author{Michael L. Falk}
\affiliation{Department of Materials Science and Engineering, Johns Hopkins University, Baltimore, MD 21218, USA}
\affiliation{Department of Mechanical Engineering, Johns Hopkins University, Baltimore, MD 21218, USA}
\affiliation{Department of Physics and Astronomy, Johns Hopkins University, Baltimore, MD 21218, USA}
\affiliation{Hopkins Extreme Materials Institute, Johns Hopkins University, Baltimore, MD 21218, USA}

\author{Isabella Gallino}
\affiliation{Technical University of Berlin, Chair of Metallic Materials, Ernst-Reuter Platz 1, 10587 Berlin, Germany}

\author{Jamie J. Kruzic}
\affiliation{School Mechanical and Manufacturing Engineering, University of New South Wales (UNSW Sydney), Sydney NSW 2052, Australia}

\author{Jiayin Lu}
\affiliation{Department of Mathematics, University of California, Los Angeles, CA 90095, USA}
\affiliation{Department of Mathematics, University of Wisconsin--Madison, WI 53706, USA}

\author{Giulio Monaco}
\affiliation{Dipartimento di Fisica e Astronomia “Galileo Galilei”, Università degli Studi di Padova, Via F. Marzolo, 8, Padova, 35131, PD, Italy}

\author{Misaki Ozawa}
\affiliation{Université Grenoble Alpes, CNRS, LIPhy, 38000 Grenoble, France}

\author{Anshul D. S. Parmar}
\affiliation{NOMATEN Centre of Excellence, National Center for Nuclear Research, ul. A. So\l{}tana 7, 05-400 Swierk/Otwock, Poland.}

\author{Chris H. Rycroft}
\affiliation{Department of Mathematics, University of Wisconsin--Madison, WI 53706, USA}
\affiliation{Mathematics Group, Lawrence Berkeley Laboratory, Berkeley, CA 94720, USA}

\author{Srikanth Sastry}
\affiliation{Jawaharlal Nehru Center for Advanced Scientific Research, Bengaluru, India}

\date{\today}

\begin{abstract}
    This paper reviews recent advances in the field of metallic glasses, focusing on the development of novel experimental techniques and \textit{in silico} models. 
    We discuss progress in experimental characterization, additive manufacturing, multiscale modeling approaches, and the growing role of machine learning in understanding and designing these complex materials.     
    On the experimental side, we highlight measurements of thermophysical properties of supercooled liquids via fast chip calorimetry and enhancements in mechanical properties through rejuvenation treatments. This work underscores the crucial role of short-range order and medium-range order in controlling metallic glass mechanical properties. Recent progress in structural probes allows \textit{in situ} observations of deformation mechanisms, positioning the field well to further advance our understanding of mechanical properties.
    Additive manufacturing of metallic glasses is discussed as one encouraging new manufacturing route for metallic glasses. We examine laser powder-bed fusion process physics and the central trade-off between amorphicity and densification, including heat affected zone devitrification and defects formation, together with emerging mitigation strategies and applications.
    On the theoretical and simulation side, we review advances in nanoscale, mesoscale, and continuum modeling of metallic glasses that have led to promising approaches by which multiscale schemes can incorporate data sourced from atomic-scale simulations. These efforts have helped to elucidate the connection between the glass structure and mechanical and rheological responses. 
    We also cover the development of machine learning interatomic potentials for metallic glasses, along with machine learning driven prediction of glass forming ability and inverse design methods. 
    Finally, challenges and directions for future research are presented and discussed.     
\end{abstract}

\maketitle

\tableofcontents
\section{Introduction}
Metallic glasses (MGs) are a unique class of materials characterized by a disordered atomic structure, formed when a molten alloy is rapidly cooled to prevent crystallization~\cite{greer1995metallic, KruzicReview2016,gan17,wan25b}.
% Properties, applications, impact
Basically all liquids form glasses when cooled sufficiently fast  
\cite{ang95,deb01,dyr06,berthier2011theoretical}, but compared to other types of glasses and crystalline metals MGs feature exceptional properties~\cite{greer2007bulk} including, for example, high strength and hardness~\cite{inoue2000stabilization}, high elastic strain limits, corrosion and radiation resistance~\cite{wang2004bulk}, low energy loss, soft magnetic behavior~\cite{tiberto2007magnetic}, and excellent processability~\cite{schroers2010processing}.
These unique characteristics are driving a growing interest for a wide range of promising applications~\cite{suryanarayana2017bulk,gao2022recent,greer2023metallic,sohrabi2024manufacturing}.
For example, Zr-based MGs have been used in consumer electronics such as smartphone and smartwatch casings, where their superior strength, scratch resistance, and moldability provide advantages compared to conventional metals~\cite{inoue2024development}. Fe-based MGs are used in transformer cores to reduce magnetic losses and improve the energy efficiency of power grids~\cite{hasegawa2008impacts}. In aerospace and biomedical sectors, the light weight, remarkable mechanical features, and excellent corrosion resistance of MGs make them good candidates for structural components and biocompatible medical implants~\cite{li2016recent}.
Despite these promising applications and extensive research in both industry and academia, large-scale use of MGs remains limited. This is related to persistent challenges in experiments, manufacturing, and theoretical understanding.
Below, we briefly summarize the main challenges of metallic glasses, in Sections~A-D, including glass forming ability, thermomechanical history, brittleness and ductility, and local structural effects. Next, in Section~E, we discuss current multiscale characterization tools aiming at illuminating these open questions, on length scales ranging from the atomic to the macroscopic.

\begin{figure*}[htp]
    \centering
    \begin{minipage}[c]{0.45\textwidth}
        \centering
        \begin{overpic}[width=\linewidth]{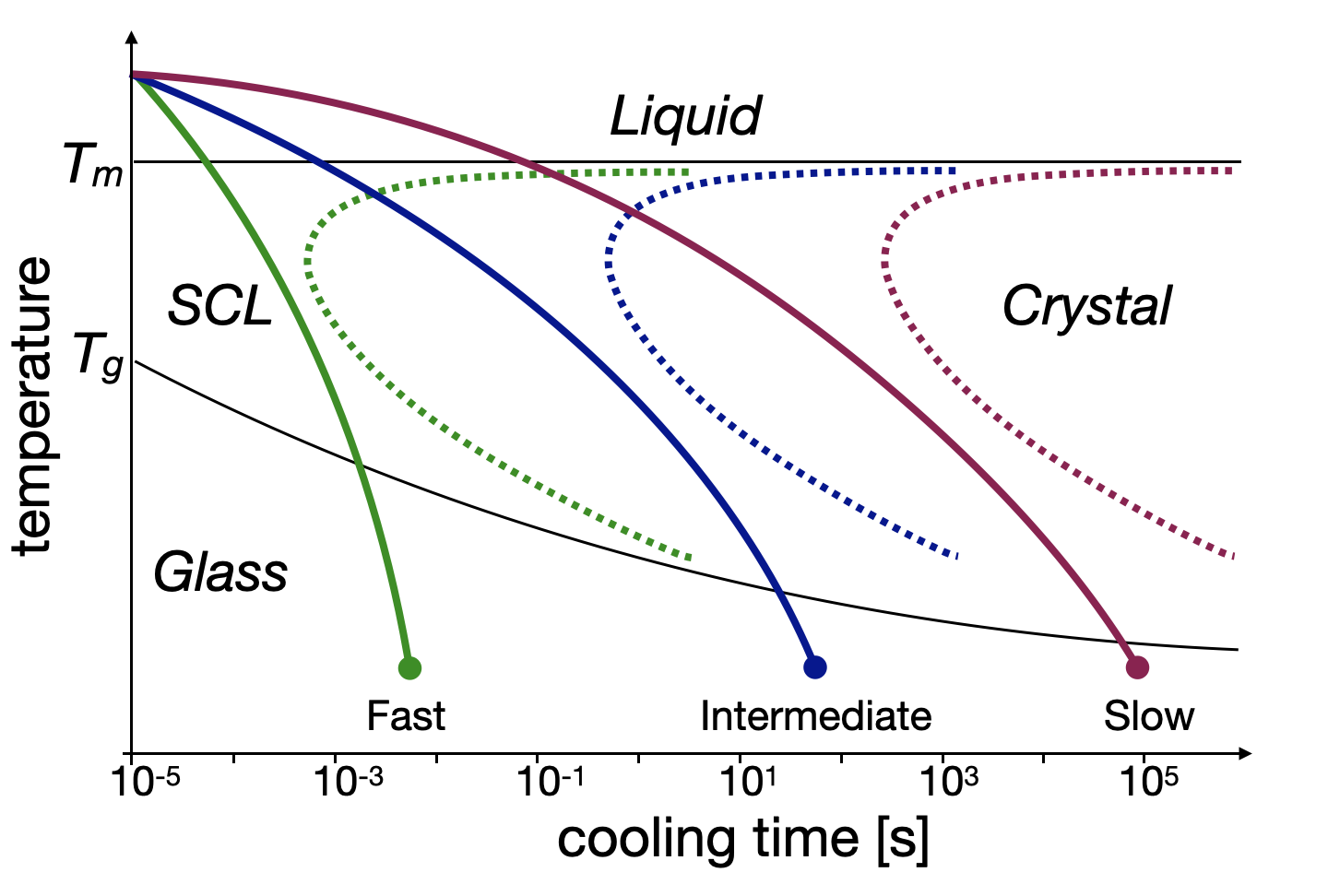}
            \put(-5,70){\textbf{a)}}  % (x%, y%) from bottom-left
        \end{overpic}
    \end{minipage}
    \hspace{1cm}
    \begin{minipage}[c]{0.43\textwidth}
        \centering
        \begin{overpic}[width=\linewidth]{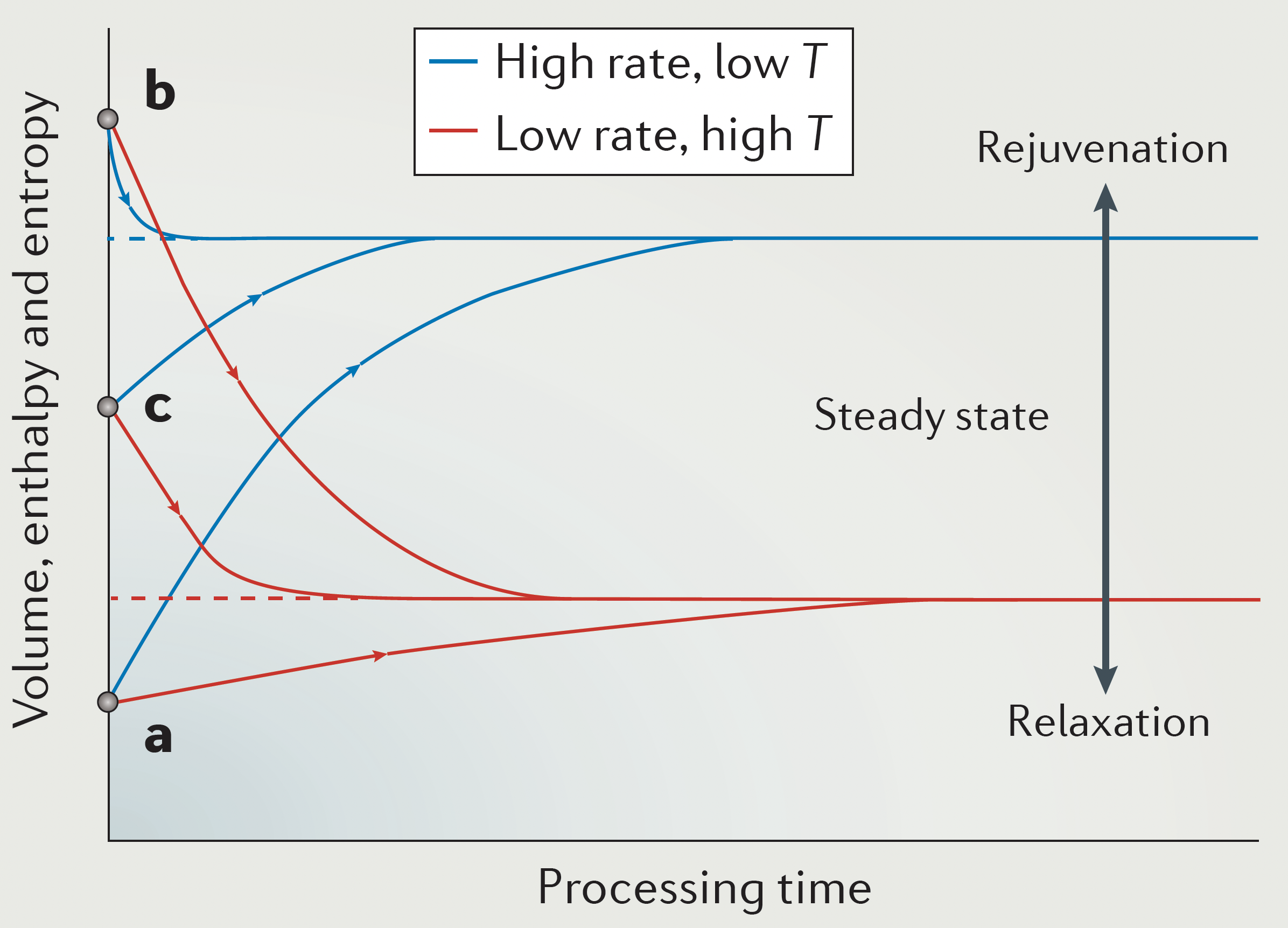}
        \put(-5,70){\textbf{b)}}  % (x%, y%) from bottom-left
        \end{overpic}
    \end{minipage}    
    \vspace{0.5cm} % vertical space between rows   
    \begin{minipage}[c]{0.43\textwidth}
        \centering
        \begin{overpic}[width=\linewidth]{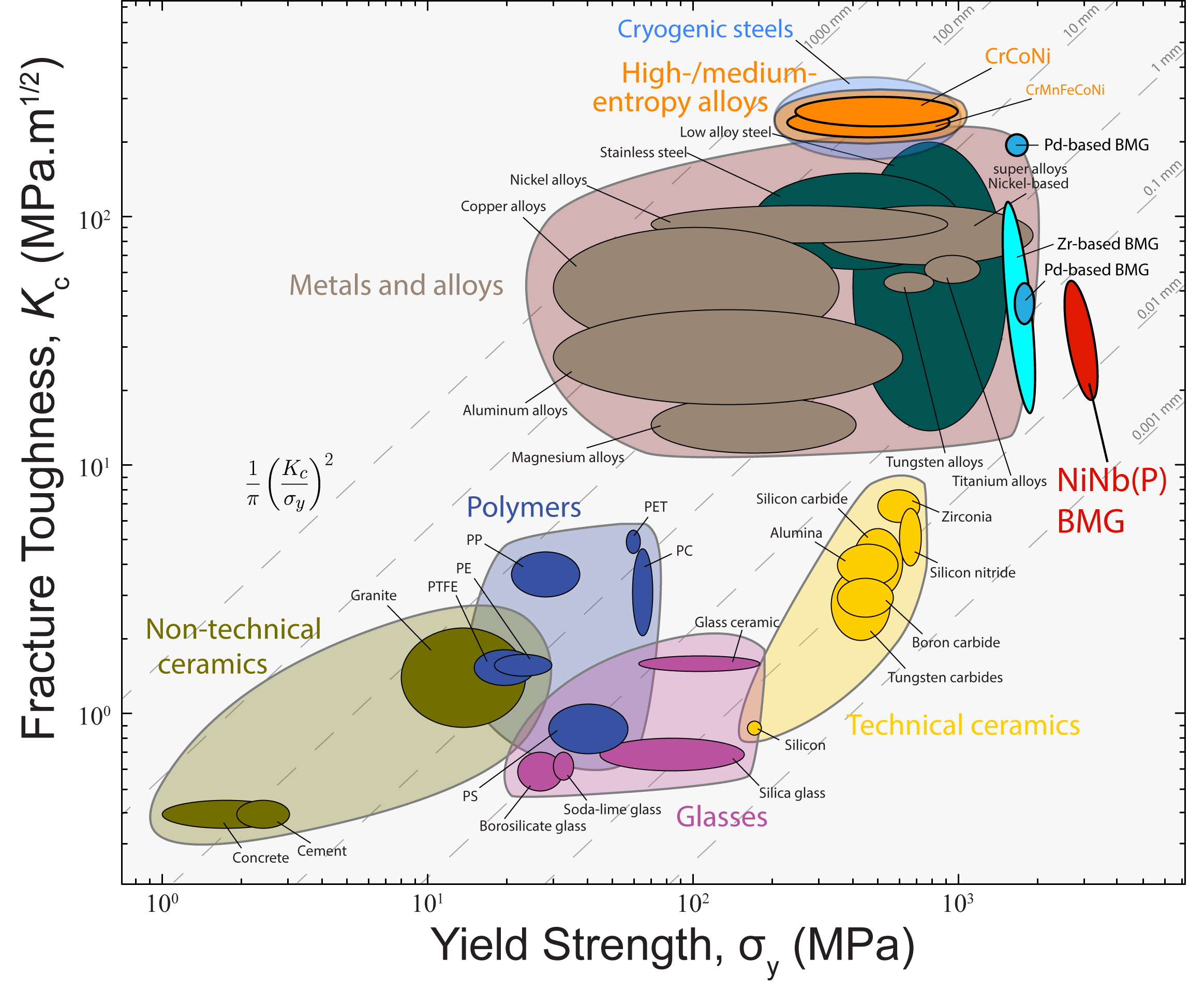}
        \put(-5,75){\textbf{c)}}  % (x%, y%) from bottom-left
        \end{overpic}
    \end{minipage}
    \hspace{1cm}
    \begin{minipage}[c]{0.45\textwidth}
        \centering
        \begin{overpic}[width=\linewidth]{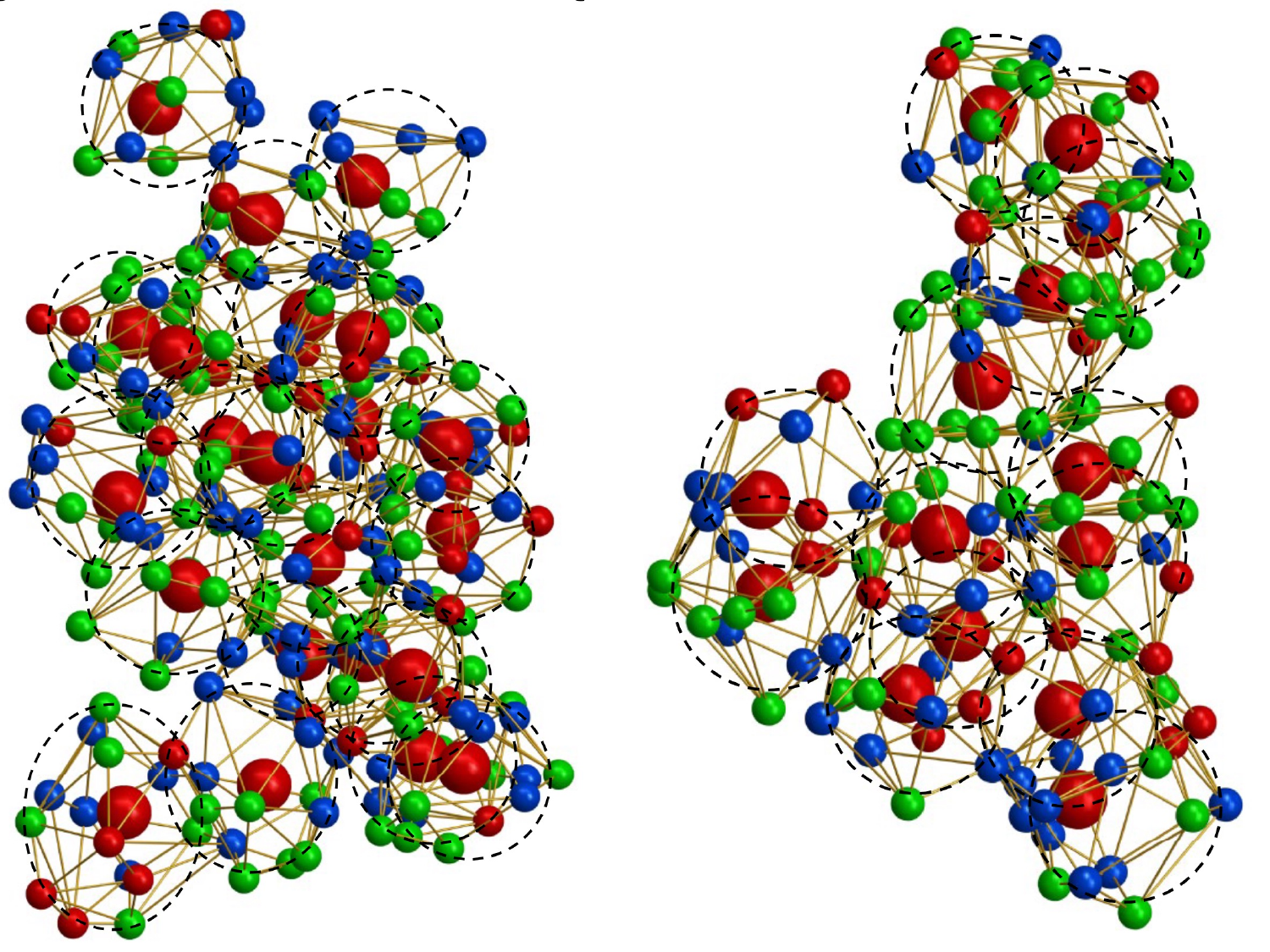}
        \put(-5,70){\textbf{d)}}  % (x%, y%) from bottom-left
        \end{overpic}
    \end{minipage} 
	\caption{Challenges of metallic glasses.
    {a)} \textit{Glass Forming Ability}: Schematic time-temperature-transformation diagram showing the critical cooling rate to bypass crystallization. Fast, intermediate, and slow cooling regimes correspond to cooling rates~$R_c\sim10^{5}$--$10^{6}$~K/s (1960s), $R_c\sim10^{2}$~K/s (1970s--1980s), and $R_c\sim1$~K/s (1990s), respectively. SCL denotes the supercooled liquid region. The underlying conceptual framework is discussed in Ref.~\cite{gallino_busch_book}. 
    {b)} \textit{Thermomechanical History}: Schematic illustrating various possibilities for thermomechanically relaxing or rejuvenating metallic glasses. Points \textbf{a}, \textbf{b}, and \textbf{c} represent possible initial energy states for a glass that may evolve to higher or lower energy after exposure to mechanical deformation and/or elevated temperatures. Figure reproduced from Ref.~\cite{sun2016thermomechanical} with permission from Springer-Nature. 
    {c)} \textit{Brittleness and Ductility}: An Ashby plot of yield strength versus fracture toughness showing the wide range of fracture toughness values that have been achieved for metallic glasses with relatively good fracture toughness based on Pd, Zr, or Ni. Figure reproduced from Ref.~\cite{Li_JALCOM2025} with open access CC BY license. %\CHR{Maybe nothing can be done about this, but some of the labels on this figure are extremely small and difficult to read.}
    {d)} \textit{Local structure and its influence on properties}:  3D atomic packing of face centered cubic like (at left) and hexagonal close packed like (at right) medium range order in a metallic glass sample revealed by atomic electron tomography. Solute centers are shown as large red spheres with solvent atoms shown in blue and green.  Figures reproduced from Ref.~\cite{Yang_Nature2021} with permission from Springer-Nature.}        	
    \label{fig:TTT_rate}
\end{figure*}

\subsection{Glass Forming Ability}
%The ability of an alloy to form a metallic glass, referred to as glass forming ability (GFA), depends on a combination of compositional and processing factors. Materials with a high atomic packing fraction and no obvious long-range ordering tend to be better metallic glass formers \cite{Busch2007}. 
%Enhancement of glass forming ability is generally achieved by mixing different elements. For example, the addition of elements with different atomic sizes or valence states can disrupt the crystalline structures and promote glass formation. At the same time, the cooling rates and the thermomechanical treatment are crucial to avoid crystal nucleation~\cite{sun2016thermomechanical}. 
A major difficulty in MG research is understanding their glass forming ability (GFA), which describes how easily a metallic alloy can form an amorphous structure during cooling by suppressing crystallization~\cite{kelton2010nucleation}. Monatomic metals are generally extremely poor at forming glasses, but utilizing picosecond-pulsed laser ablation in a liquid medium it has recently been demonstrated that glasses may be produced from even the poorest monatomic glass formers, the fcc metals \cite{ton24}. The GFA of alloys -- our focus henceforth -- depends significantly on the composition and elemental concentration, which determine the critical cooling rate, i.e., the minimum rate to prevent crystal formation~\cite{schroers2010processing}. This, in turn, determines the maximum size, or critical casting thickness, of a purely amorphous sample~\cite{johnson1999bulk}.

Alloys with low GFA, such as some Fe-based compositions, can only be produced as very thin splats or films with a size less than 10~$\mu$m thick, while those with moderate GFA form ribbons or foils typically 20–100~$\mu$m thick~\cite{lu2002new}. 
Advances in alloy processing conditions, such as copper mold casting, have enabled alloys such as Zr-Cu-Ni-Al or Pd-based systems to be processed as bulk metallic glasses (BMGs), defined as MGs with a minimum casting thickness greater than 1~mm~\cite{peker1993highly}. Currently, the largest BMGs can reach dimensions of a few centimeters~\cite{suryanarayana2017bulk}.
Bulk formation is challenging because, in larger samples, heat cannot diffuse out quickly enough to maintain the critical cooling rate throughout the material, leading to crystallization in the slower cooling interior.
The design of BMGs that can be cast into larger sizes while remaining amorphous is a key goal of the field. Improving our understanding of GFA is crucial to this end.  

Overall, the lack of a clear theoretical framework and predictive models for GFA remains a central challenge in metallic glass research~\cite{lu2002new}. Consequently, researchers assess GFA using various metrics, but there is no universal model to predict which alloys will exhibit high GFA~\cite{schroers2010processing}.
Current criteria for evaluating GFA can be classified into five main categories:
(i) Empirical rules, that have been proposed during the years based on observed trends in different compositions, such as the \textit{confusion principle} (glass formation becomes easier with increasing number of elements~\cite{greer1993confusion}), and Inoue's three criteria for high GFA in bulk MGs: 1) multicomponent alloy systems (typically three or more elements), 2) significant atomic size differences among the main constituent elements (greater than 12\%), and 3) negative heats of mixing among major constituent elements~\cite{inoue2000stabilization,wang2007roles}.
Other features include proximity to deep eutectic compositions (which are specific alloy ratios where the liquidus temperature reaches a minimum) and presence of metalloids in specific proportions~\cite{johnson2016quantifying,gallino_busch_book}. 
(ii) Physical properties, including critical cooling rate, viscosity, atomic mobility, and diffusion behavior~\cite{schroers2010processing,busch2000}.
(iii) Structural and topological parameters, considering factors such as atomic size differences, atomic packing efficiency, and topological frustration~\cite{miracle2004structural,sheng2006atomic}.
(iv) Transformation temperatures, based on characteristic temperatures such as the glass transition temperature ($T_g$), crystallization onset temperature ($T_x$), and liquidus temperature ($T_l$)~\cite{inoue2000stabilization,lu2002new}. 
For example, Turnbull's criterion identifies the reduced glass transition temperature $T_{rg}= T_g/T_l)$. 
According to it, good glass formers typically have a high $T_{rg}$, i.e. a relatively high $T_g$ compared to $T_l$~\cite{turnbull1969under}.
%\SB{subsequent work is done to refine this.}
(v) Thermodynamic modeling, involving parameters such as the heat of mixing, free energy differences between liquid and crystal phases, and phase diagrams considerations~\cite{takeuchi2000calculations,suryanarayana2017bulk}.
%The GFA depends on many factors, including atomic size differences, mixing enthalpy, and cooling rate \cite{Busch2000}. Materials with a high atomic packing fraction and no obvious long-range ordering tend to be better glass formers \cite{Busch2007}. However, a clear theoretical understanding and a predictive rules are still missing.
Among these factors, cooling rate plays a central practical role. Figure~\ref{fig:TTT_rate}a) illustrates the evolution of critical cooling rates needed for MG formation using a time-temperature-transformation (TTT) diagram. Early MGs in the 1960s (green curve) required extremely high cooling rates (\(10^6\)~K/s), limiting samples to about 50~\(\mu\)m thickness~\cite{chen1968evidence}. 
Mid-1970s alloys (blue curve) needed a reduced critical rate of approximately 100~K/s, enabling thicknesses of 100–500~$\mu$m. By the 1990s, BMGs (purple curve) could be produced at around 1~K/s, enabling castings with thicknesses greater than 1~mm. 
The gray region indicates crystallization under slower cooling, leading to the formation of crystalline mixtures. The supercooled liquid (SCL) region corresponds to a metastable state between the equilibrium liquid and the glassy state, characterized by rapid temperature changes over short time intervals~\cite{loffler2003bulk}. 

Recently, there has been a growing interest in using machine learning models trained on experimental and simulation data to predict the GFA of alloy compositions~\cite{forrest2022machine}. This paper also discusses this research trend below.

\subsection{Thermal and mechanical history}
Metallic glasses are non-equilibrium materials, and their properties strongly depend both on how they are prepared and their subsequent thermal and mechanical history. Factors such as the cooling rate along with subsequent annealing and/or applied stress can significantly alter the thermodynamic, kinetic, and mechanical properties of metallic glasses. Figure~\ref{fig:TTT_rate}b) provides a schematic representation taken from Ref.~\cite{sun2016thermomechanical} illustrating how different initial processing can give rise to glasses with different initial energy states, labeled \textbf{a}, \textbf{b}, and \textbf{c}. Then, subsequent exposure to combinations of stress and temperature may cause the energy state to evolve over time into higher (rejuvenated) or lower (relaxed) energy states. However, while these concepts are readily accepted, our ability to precisely control the structure and properties of metallic glasses via various processing methods remains an unsolved challenge. Moreover, another important issue is physical aging, where even at ambient temperature the structure and properties of a metallic glass may change over time. This occurs, in particular,  for metallic glasses with glass transition temperatures close to room temperature, whereby rearrangements at the atomic level can reduce free volume (the excess space in the disordered atomic packing that allows atoms to move), relieve residual stresses, and affect mechanical properties such as hardness, ductility, and toughness. For practical applications, material performance must remain stable over time, so understanding and controlling aging is critical for low $T_g$ metallic glasses.

%\SB{Metallic glasses are non-equilibrium materials, meaning their properties strongly depend on preparation conditions and subsequent thermal or mechanical treatments. Thermal and mechanical history — such as cooling rate, annealing, and applied stress — can significantly alter their thermodynamic, kinetic, and mechanical behavior.
%Another critical phenomenon is physical aging: over time, even under constant external conditions, metallic glasses undergo slow atomic rearrangements as the structure relaxes toward a lower-energy configuration. This process leads to increased density, hardness, and elastic modulus, but at the same time reduces ductility and promotes embrittlement. Interestingly, recent studies show that aging does not occur continuously but through discrete, avalanche-like atomic events, evidenced by step-like changes in resistivity and bursts in creep experiments. Moreover, aging is not governed solely by the reduction of free volume, but also involves the relaxation of internal stresses and changes in local atomic order.
%Although metallic glasses remain amorphous during aging, prolonged relaxation can bring the material closer to crystallization, especially near the glass transition temperature. Mechanical deformation, such as shear or compression, can partially reverse aging by reintroducing disorder into the structure, a process known as mechanical rejuvenation.
%Controlling aging and its associated effects is thus essential for ensuring the long-term mechanical stability and reliability of metallic glasses in practical applications.}

\subsection{Brittleness and Ductility}
Many of the excellent mechanical properties of metallic glasses, such as their near-theoretical strengths and high elastic limits, are inherent traits related to their amorphous structures. However, high strength is not a valuable property for engineering applications if it is accompanied by excessive brittleness, a common problem for ceramics and crystals. In terms of fracture toughness, metallic glasses can demonstrate a wide range of behavior. Some metallic glasses are reported to be among the toughest metallic materials \cite{Demetriou_NatMater2011}, while others are nearly as brittle as ceramic glasses, especially those based on Fe, Mg, or rare earth elements and those that have been heavily relaxed by annealing \cite{KruzicReview2016, SUN2015211}. Furthermore, even metallic glasses that can demonstrate relatively good fracture toughness, such as those based on Pd, Zr, or Ni, may exhibit a wide range of toughness values that can be difficult to control or predict. Figure~\ref{fig:TTT_rate}c) shows an Ashby plot of yield strength versus fracture toughness that illustrates this issue. Although the metallic glasses shown in Fig.~\ref{fig:TTT_rate}c) always exhibit excellent yield strength as a result of their amorphous structures, a wide range of fracture toughness values have been measured, even for the same composition. The free volume and energy state are known to be important factors in affecting fracture toughness of metallic glasses \cite{LAUNEY2008500, Li_Acta2019}. 
%In fact, molecular dynamics simulations of a simple model system have recently shown that, depending on their preparation protocol, glasses can yield in qualitatively different ways \cite{ozawa2018random}: well annealed (low energy) glasses often display a brittle response, with the emergence of sharp shear bands, while rejuvenated (high energy) glasses tend to display a ductile response. Simulations employing the cyclic shear protocol reveal very similar results \cite{BhaumikPNAS2021,YehPRL2020}}. 
Besides, structural heterogeneity at different length-scales will also play a role, but its impact as well as the roles or other structural features such as short and medium range order are less understood \cite{Nomoto_MaterTod2021}. Elucidating the detailed processing-structure-property relationships that control brittleness and ductility in metallic glasses is a critical area for further research to enable metallic glasses to achieve their full potential as structural materials.

\subsection{Local structure and its influence on properties}
A major barrier to developing improved structure-property relationships for metallic glasses is the detailed characterization of atomic structures \cite{wei19}. Unlike crystalline metals, which exhibit long-range atomic order and contain well-defined crystalline defects such as dislocations and grain boundaries that can be directly observed using various experimental techniques, metallic glasses lack such ordered structures. Instead, they possess a disordered atomic arrangement similar to that of a frozen liquid, with atoms packed in an irregular, non-periodic manner, making the identification of defects and local structures more challenging. However, they also exhibit short- and medium-range order (SRO and MRO) that affect their properties. Both simulations, such as molecular dynamics (MD), and experimental techniques, such as synchrotron or neutron scattering, transmission electron microscopy (TEM), etc., are used to study the atomic structure of metallic glasses, but all have significant limitations. In MD simulations, proper atomic potentials do not currently exist to simulate the 4+ component metallic glasses compositions with interesting mechanical properties, and extending the findings for binary systems to more complex metallic glass compositions is not straightforward \cite{ZHOU2024101311}. On the experimental side, synchrotron or neutron scattering can get averaged statistical data for SRO and MRO sites, but struggle to characterize the details of the different local atomic arrangements that contribute to the average scattering signal \cite{Billinge_Science_2007}. As shown in Fig.~\ref{fig:TTT_rate}d), TEM-based methods such as atomic electron tomography show a remarkable ability in mapping the local atomic arrangements. Yet, the broad applicability of this technique has been limited to a few published results \cite{Yang_Nature2021,Yuan_NaterMat2022}. Overall, it is necessary to continue to develop advanced simulation and experimental methods for revealing the local atomic structures of metallic glasses to better understand how they influence the mechanical and thermophysical properties, and better theoretical descriptions of structure are also needed~\cite{wei19,wan25}. 
%If we can better link structure to properties, we will be able to produce metallic glasses with improved and repeatable properties that may open up new engineering applications. 

%Recent approaches to designing metallic glasses have seen significant advancements, particularly through the application of modeling, machine learning (ML) and deep learning techniques. These methods have enabled more efficient and effective exploration and design of new metallic glass (MG) compositions.
%Recent years have witnessed rapid developments in the design and understanding of MGs, driven by novel computational, experimental, and data-driven approaches. 
%For example, machine learning (ML) has enabled the prediction of GFA and alloy properties with remarkable accuracy. Advanced simulations and modeling techniques have facilitated the exploration of atomic-scale mechanisms governing MG behavior, while additive manufacturing (AM) technologies now offer the ability to fabricate complex geometries, including hierarchical metamaterials, with tailored amorphous properties.

\subsection{Multiscale characterization}
\begin{figure*}[ht]
	\centering
 	\includegraphics[width=1\textwidth]{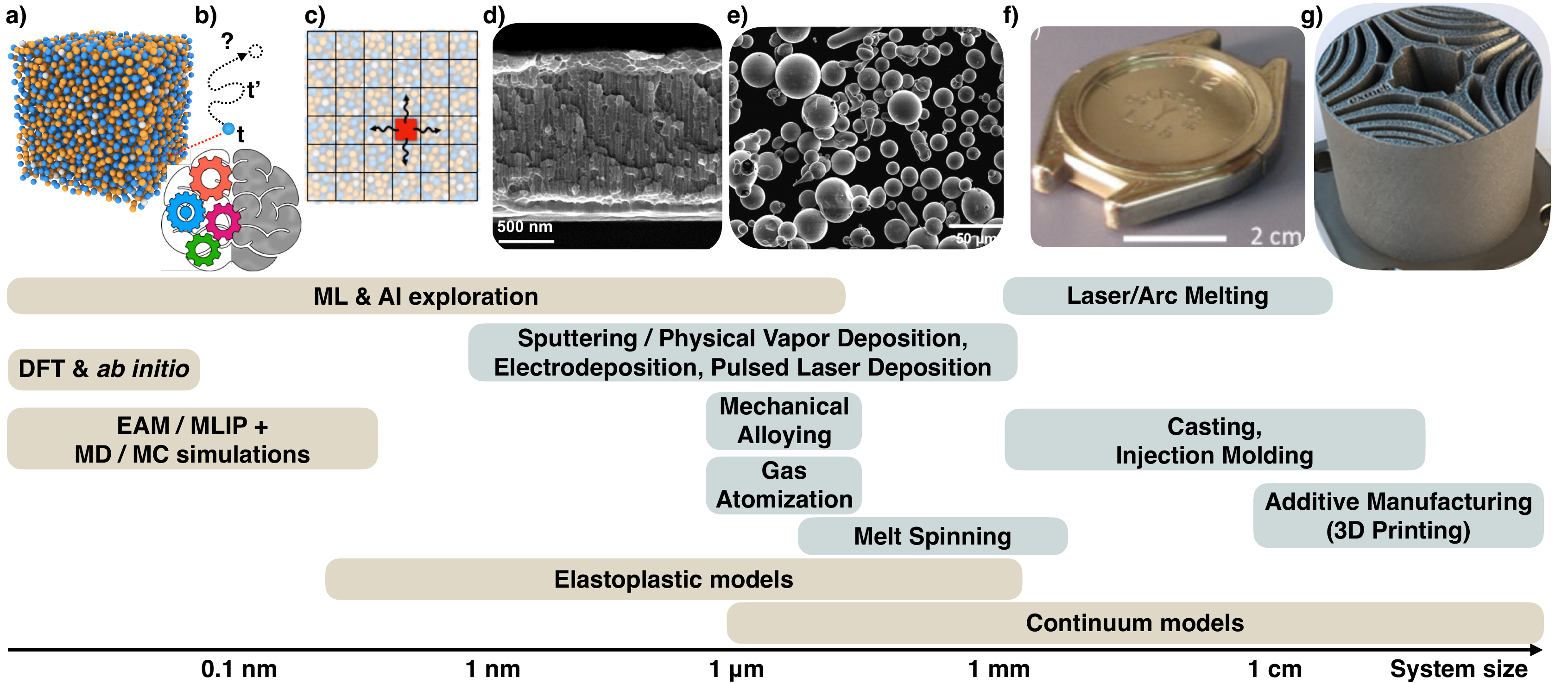}    
	\caption{Multiscale characterization of metallic glasses. Overview of metallic glasses (MGs) from atomic to macroscopic dimensions. \textit{From left to right}: 
    a)~Atomic scale representation of a ZrCuAl MGs~\cite{makinen2024bayesian} from molecular dynamics simulations. 
    b)~Schematic of machine learning predictions for MG trajectories, where $t$ and $t'$ represent two different times. 
    c) Schematic representation of elastoplastic models, in which each lattice site represents a coarse-grained region containing a group of particles.
    d)~Cross-section electron microscopy image of a MG thin film produced with vapor deposition, showing nanostructured morphology, from Ref.~\cite{brognara2022effect}. 
    e)~Scanning electron micrographs of micron-sized gas atomized MG powders, from Ref.~\cite{Bosong_Li2024AM}. 
    f)~Example of bulk MG samples from Ref.~\cite{schroers2010processing}. 
    g)~Soft-magnetic additive manufactured MG (via laser podwer bed fusion) with internal complex geometries~\cite{THORSSON2022110483} (metamaterials). \textit{At the bottom}: Popular modeling approaches (beige) and manufacturing methods (light blue), are arranged according to the length scales at which they operate. Figures reproduced with permissions.} 
%\MO{Atomic electron tomograph shown in Fig. 1d is not mentioned in the pannel and text?} \SB{I am fixing this. Suction casting?}    
	\label{fig:lengthscale}
\end{figure*}

The behavior of metallic glasses depends on multiple length-scales, from the local atomic structure ($\sim 0.1$ nm) to the bulk scale relevant for real-world applications ($\sim 1$ cm), as illustrated in Fig.~\ref{fig:lengthscale}. Different experimental methods, numerical models, and machine learning techniques are applied at these various length scales.

{\it Atomic scale} ($\sim 0.1$ nm):
At this scale, the detailed shape of the interatomic potentials is crucial, as it determines local structural and compositional orders. Yet, this information is not accessible in experiments. Hence, \textit{ab initio} quantum mechanical calculations such as density functional theory (DFT) have been performed~\cite{sheng2006atomic}. Although DFT provides highly accurate results, it is computationally extremely expensive~\cite{perdew1996generalized}. To reduce costs, classical effective potentials fitted from \textit{ab initio} calculation are often used. A notable example is the embedded atom model (EAM)~\cite{daw1984embedded}. Such models allow to study thermodynamic, kinetic, and mechanical behaviors of metallic glasses using molecular dynamics (MD) simulations. An example of configuration for the ZrCuAl MGs~\cite{makinen2024bayesian} from MD is reported in Fig.~\ref{fig:lengthscale}a), where different colors represent different atomic species. 
Even simpler models like the Lennard-Jones potential can further reduce computational costs at the expense of detailed microscopic interactions~\cite{allen2017computer}. Machine learning (ML) is particularly effective at this scale for developing accurate interatomic potentials derived from quantum mechanical data,  resulting in a new class of interaction potentials that are termed Machine Learning Interatomic Potentials (MLIP)~\cite{behler2011atom}. Besides, ML techniques can predict or forecast particle rearrangements that influence the kinetic and mechanical properties of metallic glasses based on static configurations~\cite{Richard_PhysRevMaterials.4.113609,jung2025roadmap}.
A sketch of ML prediction framework for particle trajectories is reported in Fig.~\ref{fig:lengthscale}b).

{\it Mesoscopic scale} ($\sim 1$ nm - 1 $\mu$m):
At this scale, local plasticity and elasticity are key factors in determining the mechanical behavior of metallic glasses~\cite{schuh2007mechanical}. Here, coarse-grained and lattice-based models, commonly known as elastoplastic models, are highly effective~\cite{bulatov1994stochastic,falk1998dynamics,homer2009mesoscale,ding2014soft,budrikis2017universal,nic18,wang2018spatial,wang2024atomistically}. 
In these models, each lattice site represents a mesoscopic block containing a group of particles. This coarse-graining approach significantly reduces computational costs while retaining essential ingredients like elasticity and plasticity. As a result, it enables the simulation of much larger length scales than molecular dynamics simulations can reach. A schematic representation of elastoplastic models working principle is reported in Fig.~\ref{fig:lengthscale}c).
Experimentally, the mesoscopic regime corresponds mainly to MG thin films with thicknesses ranging from a few nanometers to tens of microns~\cite{chu2012thin,yiu2020thin}. Several vapor-phase and electrochemical techniques have been developed to fabricate MG at these sizes. 
One of the most widely used methods is \textit{physical vapor deposition} (PVD), specifically magnetron sputtering~\cite{liu2012deposition,chuang2013mechanical,ghidelli2015extrinsic,etiemble2017innovative,ketov2017formation}. In this process, a MG target is bombarded with energetic ions (typically argon plasma) that cause the ejection of atoms or atomic clusters and subsequent deposition onto a substrate where the MG thin film grows. Sputtering is a versatile tool for exploring the amorphous phase space allowing good control over film thickness, uniformity, and multi-element composition. Moreover, due to the inherently high quench rates and low surface mobility during deposition, sputtering often produces fully amorphous films even for compositions that would crystallize under slower cooling conditions. An electron microscopy image of a MG thin film produced via vapor deposition technique is shown in Fig.~\ref{fig:lengthscale}d).
Another important technique is \textit{pulsed laser deposition} (PLD), that uses high-energy laser pulses to vaporize a material from a target, forming a cloud of plasma that is deposited as an amorphous film~\cite{ghidelli2021novel,bignoli2025extending}. It offers precise stoichiometry and is suitable for complex alloys. However, it has limited thickness uniformity, low deposition rates, and poor scalability. Energetic particles can also damage the growing film. 
%from here
The \textit{electrodeposition} method allows for fabrication of MG films and nanostructures via electrochemical reduction from of  metal ions from an electrolyte solution to a substrate~\cite{guo2017ni,saji2018electrodeposition,li2019recent,li2021effects}. The amorphization is achieved by incorporating glass-forming elements (e.g., phosphorus or boron) via bath additives, together with tuning parameters such as current density to suppress crystallization. The chemistry of the bath determines the ratio of elements deposited~\cite{guo2017ni}.
It is scalable and compatible with patterned substrates, but has limited control over composition and is restricted to a narrow range of alloy systems~\cite{li2025thermodynamic}. 
Thin films fabricated by these described approaches are crucial experimental platforms for probing the mesoscopic deformation behavior of metallic glasses. Such films are used for nanoindentation, in situ Scanning Electron Microscopy (SEM) or Transmission Electron Microscopy (TEM), and mechanical tests experiments to study velocity-dependent plasticity and shear bands.  
Importantly, the results for structural relaxation or heterogeneity found in films may not fully represent those of bulk metallic glasses formed by conventional casting~\cite{schnabel2016electronic,yiu2020thin,bi2018multiscale}. 

\textit{Sub-macroscopic scale} (1~$\mu$m - 1~mm): At this scale, MGs exhibit size-dependen mechanical behavior~\cite{jang2010transition}, influenced by surface quality, internal heterogeneities, and deformation instabilities~\cite{guo2007tensile}, such as shear band formation and crack initiation~\cite{greer2013shear}.
Experimentally, this scale corresponds to MG ribbons, powders and micropillars~\cite{sohrabi2024manufacturing}.
Ribbons are manufactured by \textit{melt spinning} technique, where molten alloy is ejected onto a fast spinning wheel for rapid cooling, producing amorphous thin strips about 20–100~$\mu$m thick~\cite{inoue2000stabilization}. 
Metallic glass powders (Fig.\ref{fig:lengthscale}, panel e)) are made through \textit{gas atomization}, where molten alloy is sprayed into gas to form micron-sized spherical particles (10–200~$\mu$m), or \textit{mechanical alloying}, in which rough metal chunks are ground and mixed in a high-energy mill ~\cite{suryanarayana2001mechanical}. 
Micropillars, tiny columns (1–100~$\mu$m), are fabricated from bulk MGs or films using focused ion beam (FIB) milling for compression tests~\cite{yang2009effects}.
%These are tested with micro-compression tools, often in SEM for real time views of shear bands and plasticity~\cite{bei2007effects}. 
This scale bridge the gap between nano-scale films (where surface effects dominate) and bulk MGs (where internal heterogeneities prevail).

\textit{Macroscopic scale} (1~mm – 1~cm and above): 
At this scale, MGs are considered as bulk metallic glasses (BMGs) and exhibit mechanical properties dominated by their internal structure, such as large-scale heterogeneities and residual stresses, with fracture patterns~\cite{trexler2010mechanical}. Experimentally, this scale includes bulk MG samples produced through methods such as arc melting, conventional casting, and injection molding, alongside additive manufacturing (AM).
\textit{Arc melting} involves melting high purity metal species in an electric arc under a controlled atmosphere. The liquid alloy, once mixed, is rapidly poured into a copper mold, where it cools fast (rates up to 10$^2$–10$^3$~K/s) to form an amorphous rod plate or rod (see Fig.~\ref{fig:lengthscale}f)) %\SB{check method}
, typically 1~mm to several cm thick~\cite{inoue2000stabilization}.
\textit{Conventional Casting} is a basic method to manufacture BMGs pouring the molten alloy into a mold (e.g., copper or steel), similar to traditional metal casting, but with particular attention to rapid cooling to maintain the amorphous structure. (1~mm to cm scale), ideal for alloys with good glass-forming ability~\cite{greer2013shear}.
To obtain complex MG geometries and precision shapes (as in the case of consumer electronics or medical devices), the technique used is \textit{injection molding}, whose fundamental difference from previous methods is that the liquid is processed in its supercooling window, not at very high temperatures  (consistency of e.g., honey). Subsequently, the supercooled liquid is injected at high pressure into a detailed mold that rapidly cools it to solidify it into an amorphous part. 
%This technique requires alloys with a wide supercooled region
\textit{Additive manufacturing} (AM), or three-dimensional (3D) printing,
has emerged as a transformative tool for manufacturing MGs~\cite{sohrabi2021additive,zhang20213d}. By enabling localized rapid heating and cooling, AM has opened pathways to bypass the high cooling rate usually required for metallic glass formation. 
Unlike conventional casting, which is constrained by the need for rapid cooling rates (>10\textsuperscript{3} K/s) to prevent crystallization, AM processes achieve cooling rates of 10\textsuperscript{4}--10\textsuperscript{7} K/s, enabling the fabrication of large amorphous components \cite{sohrabi2021additive}. 
This capability is particularly evident in processes like \textit{powder bed fusion}, where controlled thermal gradients allow the preservation of the amorphous structure. This technology also allows for the fabrication of intricate geometries, such as lattices or metamaterials, that were previously unattainable using conventional manufacturing methods. Injection molding, in fact, fills a mold all at once (subtractive/forming approach), while AM builds the part gradually (additive approach). Advances in this field include the production of fully amorphous or crystalline-amorphous composites, enabling the creation of parts with customized microstructures and enhanced properties~\cite{madge2021laser}, reconciling scalability with high cooling rates. However, this progress is not without limitations. The ability to achieve and maintain an amorphous structure during 3D printing is highly dependent on the alloy system and its inherent glass-forming ability~\cite{ouyang2021understanding}. Additionally, achieving amorphous structures with minimal defects requires precise control of numerous parameters~\cite{jia2021scanning}, that are often interdependent, creating a complex parameter space that must be optimized to balance thermal gradients, avoid defects such as porosity or thermal cracking, and maintain the amorphous structure. 
For the modeling side, continuum modeling of metallic glasses aims to describe their mechanical behavior at scales much larger than the atomic level by treating the material as a continuous medium. These models capture key features such as shear band formation, plastic flow, and viscoplastic deformation by using constitutive laws that effectively incorporate the disordered nature of the atomic structure. 
%This field is still in its infancy.
%In addition, mechanical performance is always constrained by the brittleness associated with shear banding during deformation, with little influence from the manufacturing process. To bridge these gaps a multidisciplinary effort that combines insights from simulations, experiments, and emerging technologies like generative ML models, is needed. 

The goal of this review paper is to summarize recent developments in metallic glass research, with a particular focus on novel experimental techniques, additive manufacturing, computational modeling, and machine learning applications, addressing the challenges outlined above.
The manuscript is organized as follows: Experimental breakthroughs in thermophysical and mechanical property characterization in Section~\ref{sec:recent_experiment}, Additive manufacturing (AM) processes in Section~\ref{sec:AM},  Advancements in modeling techniques, from atomistic (Section~\ref{sec:nano_modeling}) to multi-scale simulations (Section~\ref{sec:multiscale_modeling}), and Integration of ML for predicting GFA and designing novel compositions in Section~\ref{sec:ML_GFA}. By highlighting these advances and their interconnections, we aim to present a comprehensive perspective on the future directions and challenges in Section~\ref{sec:future}.

\section{Recent Experimental Advances} 
\label{sec:recent_experiment}
\subsection{Thermophysical properties of the supercooled liquid phase}
Modern multi-component bulk metallic glasses (BMG) are much more stable with respect to crystallization when heated into the supercooled liquid than the early metallic glasses. Some BMGs can be heated 50~K up to 120~K above the glass transition temperature $T_{g}$ with 20~K/min before crystallization sets in. That enabled experimental access of the deeply supercooled liquid state (SCL), allowing thermophysical properties of the supercooled liquid to be studied and connected to the glass forming ability. Accurate temperature- and time-dependent changes in enthalpy, viscosity and structure were made possible to monitor by advanced methods such as chip calorimetry, high-brilliance synchrotron X-ray scattering and viscosity measurements via electrostatic levitation under microgravity conditions \cite{gallino_busch_book}. 
%In the following, we review some important advances.

\begin{figure*}[ht]
	\centering
        \includegraphics[width=1\textwidth]{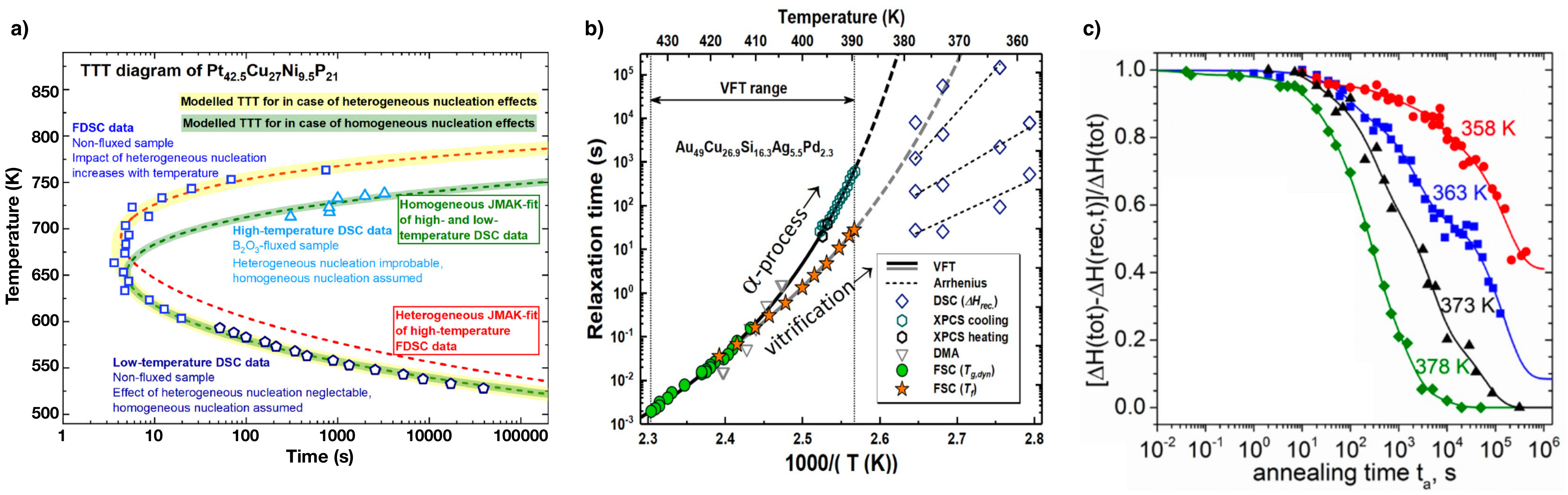}
	\caption{Examples of thermophysical properties measured in bulk metallic glasses.     
    a) Detection of a time temperature transformation (TTT) diagram around the nose for crystallization is enabled by chip-calorimetry (FDSC, open squares) complementing standard calorimetry experiments (DSC, pentagons and triangles) which only cover longer times. The combinations of these calorimetric techniques opens the possibility of disentangling homogeneous and heterogeneous nucleation effects using appropriate models (yellow and green dashed lines). 
    Figure reproduced from Ref.~\cite{frey2022} with open access CC BY license.
    b) Activation plot of a Au-based BMG showing $\alpha$-relaxation time data obtained by X-ray photon correlation spectroscopy (XPCS) and dynamic mechanical analyzer (DMA) and vitrification kinetics obtained using various calorimetric approaches (DSC, FSC). The $\alpha$-relaxation and the vitrification kinetics show a clear decoupling which becomes stronger at lower temperatures. This decoupling is also dependent on the cooling rate (open diamonds).
    The dashed lines are fits to the data using the Vogel-Fulcher-Tammann expression, and the dotted lines are fits using the Arrhenius ansatz. Figure reproduced with permissions from Ref.~\cite{monnier2020}.
    c) Normalized enthalpy relaxation during physical aging at different temperatures, showing hierarchical decays. Figure reproduced with permissions from Ref.~\cite{gallino2018}.
    %Thermophysical properties measurements of bulk metallic glasses. (a) TTT-diagram, enthalpy change during physical aging showing hierarchical enthalpy decays and intermittent microscopic dynamics of BMGs.   
    %\cite{frey2022,gallino2018,monnier2020,gallinoentropy,hechler2018}
    %\CHR{Licensing info?}
    }
    \label{new_figure3}
\end{figure*}

%\textit{Glass forming ability --- } 
The understanding of the factors that influence the glass forming ability (GFA) of metallic glasses is of great technological importance in itself. However, as explained above, it remains one of the complex topics that are difficult to assess both experimentally and computationally, and none of the simple criteria proposed over the years has demonstrated universal applicability. The location of the nose in the time temperature transformation (TTT) diagram, see panel a) of Fig.~\ref{new_figure3} \cite{neuber2020,frey2022}, defines the critical cooling rate of a melt necessary to bypass crystallization and to transform it into a glass. The GFA reflects the ability of the supercooled liquid to bypass the ``nose'' for primary crystallization during the solidification process and, thus, experimental investigations rely on the measurement of TTT-diagrams \cite{buschJOM2017}. With the emergence of fast chip-calorimetry, one is able to access shorter and shorter time scales near the nose and to fully detect isothermal TTT-diagrams. The nucleation and growth theory can be applied to describe the data and, depending on the BMG-forming system, one observes that the glass formation can be promoted by a low driving force for crystallization, slow crystallization kinetics, a high interfacial energy between the liquid and the crystalline phase, or any combination of these factors \cite{gross2017}. Calculation of phase diagram (CALPHAD) methods are expected to become valuable tools for significantly improving the understanding of the GFA of bulk metallic glasses, as they can provide the crystallization driving force of the first crystalline phase to appear in the supercooled liquid, and this information can be used to fit the experimentally determined TTT diagram \cite{Palumbo2008, Ma2025, Rahimi2025}.

%\textit{Vitrification kinetics and observation of multiple aging decays ---} 
Recently, chip calorimetry studies of the glass transition during cooling have challenged the generally accepted description of the mechanism for vitrification \cite{monnier2020,dilisio2023}. In the deep SCL state, the vitrification does not appear to be triggered exclusively by the main structural ($\alpha$) relaxation process. It is observed to occur with a milder temperature dependence than the $\alpha$-relaxation, and it is accompanied by a more pronounced decoupling between vitrification kinetics and atomic mobility the slower the system is cooled, as shown in panel (b) of Fig.~\ref{new_figure3}. This is directly connected to the activation energy spectrum for relaxation modes, implying multiple mechanisms for atomic diffusion. As a consequence, the limiting fictive temperature, $T_f$, i.e., the temperature at which a glass formed after cooling at a given rate would be at equilibrium, is found to be lower than that expected by accounting only for the $\alpha$-relaxation. Observation of this so-called ``$T_g$-depression'' was also observed in other glass forming systems \cite{cangialosi2012}. In BMGs, this is of high importance because it is believed to be directly connected to the heterogeneity of cooperative atomic rearrangements; the slower mechanisms for atomic mobility are responsible for delaying vitrification to lower temperatures, even if those mechanisms do not contribute to the $\alpha$-relaxation process \cite{monnier2020}. 
%The implications of this heterogeneity are that faster mechanisms for atomic mobility are responsible for maintaining the supercooled liquid system in (metastable) equilibrium, which delays vitrification to lower temperatures, even if those mechanisms do not contribute to the $\alpha$-relaxation process \cite{monnier2020}. 
Recently, it has also been observed that the apparent decoupling of the timescales for vitrification kinetics from the time scales for the $\alpha$-relaxation process is more pronounced for small sample sizes \cite{dilisio2023}. Moreover, during isothermal aging experiments at temperatures much below the glass transition temperature, $T_g$, multiple enthalpy relaxation decays towards the supercooled liquid can be observed, see panel c) of Fig.~\ref{new_figure3} \cite{gallino2018}, which again reflects the multi-component nature of BMGs as seen above for vitrification. Multiple relaxation decays imply multiple mechanisms for atomic diffusion \cite{yu14,gallinoJOM2017,yu17,gao25}. Especially at low temperatures, where there are fewer active degrees of freedom: some of the slower relaxation processes of the supercooled system may stay frozen while others, controlled by smaller atoms, are active \cite{gallinoJOM2017}. 

\begin{figure*}[ht]
	\centering
        \includegraphics[width=1\textwidth]{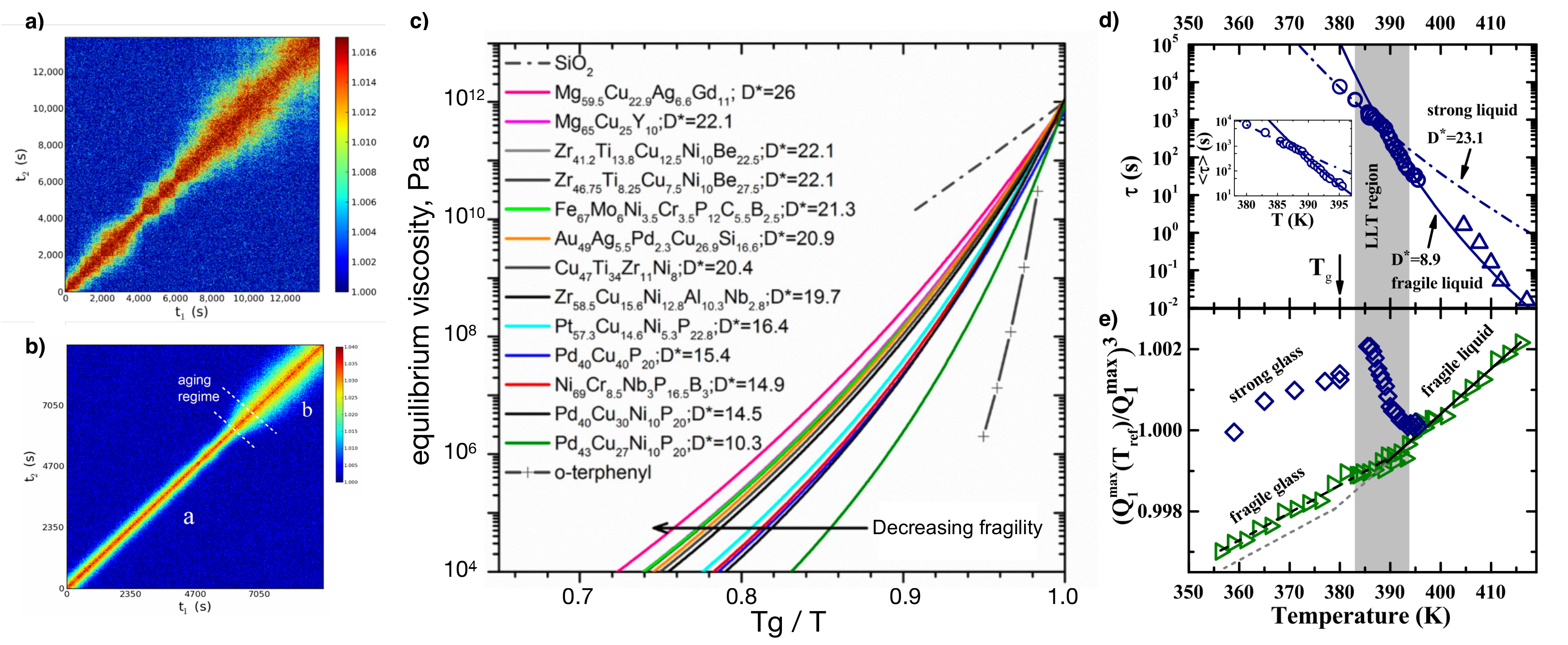}
	\caption{
    a) Two-times autocorrelation functions measured by XPCS during low temperature aging of a BMG showing highly heterogeneous intermittent aging dynamics indicative of a complex energy landscape. Figure reproduced with permission from Ref.~\cite{evenson2015}.
    b) Two-times autocorrelation function showing heterogeneous aging behavior consisting  of periods of stationary dynamics (labeled a and b) interconnected with fast-motion events. Figure reproduced with permission from Re.~\cite{gallino2018}.
    c) Fragility plot of viscosity versus the $T_g$-scaled inverse temperature for 14 metallic glass-formers, in comparison to SiO$_2$ and o-terphenyl. The solid lines are the fits of equilibrium viscosity data to the VFT-equation. Figure reproduced under CC-BY licence from Ref.~\cite{gallino2017fragility}.
    d) Temperature dependence of the relaxation time $\tau$ measured by XPCS (circles) and DMA (triangles) in the Au$_{49}$Cu$_{26.9}$Si$_{16.3}$Ag$_{5.5}$Pd$_{2.3}$ metallic glass former. The change in the slope of the experimental data is associated to a liquid-liquid transition (highlighted in grey) between two liquids of different fragility. The inset is a zoom of the transition range. 
    e) The same LLT is measured by high-intensity X-ray diffraction applying the same thermal protocol used for the XPCS analysis in panel d) (blue diamonds). The observable reported on the y-axis as a function of the temperature, $[Q_p(T_{ref})/Q_p(T)]^3$, is the inverse of the cube of the position of the first sharp diffraction peak of the static structure factor normalized to that measured at $T_{ref}$=395.5~K, and is a proxy of the volume. Note that this observable depends on the cooling rate: during continuous cooling with 1.5~K/min (green triangles), the LLT is no longer visible. The grey dashed line is the standard behavior in absence of the LLT as shown by the green triangles. 
    Figures reproduced from Refs.~\cite{hechler2018} with permissions. 
    }
	\label{new_figure4}
\end{figure*}

%\textit{Microscopic dynamics ---} 
More information on the microscopic dynamics at different length-scales has been gained by X-ray photon correlation spectroscopy (XPCS), a synchrotron-based technique used extensively in the last decade to probe metallic glass-forming liquids \cite{ruta2012,evenson2015,gallino2018,hechler2018,neuber2022,cornet2023}. This technique has revealed that the dynamics are intermittent and highly heterogeneous during low-temperature aging experiments, contrary to the common assumption of a steady slowing down of the dynamics usually observed in macroscopic studies \cite{ruta2012,evenson2015}. This is for example shown in panel a) of Fig.~\ref{new_figure4}, where a two-times plot is reported. Each point at coordinates ($t_1$, $t_2$) in this plot corresponds to the product of the scattered intensities measured at those times, and the color-bar shows the range of values obtained. In this plot, the thickness of the higher-intensity region along the main diagonal corresponds to the relaxation time, which then shows strong fluctuations during aging. Physical aging seems also to be triggered by cooperative atomic rearrangements, driven by the relaxation of internal stresses \cite{evenson2015}. During long-term annealing at low temperature the glass configuration gets trapped in deep local energy minima where the atomic dynamics are observed in XPCS to be stationary and persistent (constant relaxation time) and physical aging may occur intermittently, see panel b) of Fig.~\ref{new_figure4} \cite{gallino2018}. Understanding the physical aging mechanism of glasses is not only of general scientific interest \cite{nar71,scherer,hec15,mck17,rut17,monnier2020,rie22,boh24} but also of great practical importance as BMG performances depend on their annealing state. Aging, in fact, can cause severe embrittlement and reduce the resistance of metallic glasses to fracture and fatigue \cite{LAUNEY2008500}. Studies of this issue are currently being extended to a broader range of glass states, including those characterized by faster relaxations. With the advent of modern detectors and more and more brilliant synchrotrons, the temporal resolution in time-resolved diffraction studies has improved tremendously, and it is now possible to study the microscopic dynamics of glasses and supercooled liquids using XPCS over several orders of magnitude timescales and in combination with other techniques, e.g., fast calorimetry \cite{Sun2025}, or under extreme conditions like high pressure \cite{cornet2023}. 

%\textit{Connection between the kinetic fragility, thermodynamics and structural changes in the supercooled liquid ---}
The possibility to probe the atomic dynamics across the glass-transition has also provided a tool to study the connection between the kinetic fragility, thermodynamics and structural changes. According to the picture developed by Austen Angell for glass-formers, BMG-forming liquids display kinetics that are intermediate between network liquids that exhibit Arrhenius kinetics, such as SiO\textsubscript{2}, and molecular liquids that exhibit highly super-Arrhenius kinetics, such as o-terphenyl, see panel c) of Fig.~\ref{new_figure4}. Intuitively, this can be understood by noting that the strength of metallic bonding, and in particular its temperature dependence, lies between that of strong covalent bonds and weaker interactions such as hydrogen or van der Waals bonds. The slowdown of the liquid kinetics in BMG-formers is connected to a viscosity rise of many orders of magnitude in a small temperature range. In the vicinity of the glass transition temperature, this is related to an increase in the activation energy for viscous flow, while interestingly the structural features change only little. Recent synchrotron X-ray results clearly show that the origin of the viscous slowdown in deeply supercooled BMG-forming liquids is linked to structural heterogeneities at the nanometer length scale, where medium range order gets established. For strong liquids, the medium range order is more persistent when heated above $T_g$. In the Adam-Gibbs scenario \cite{ada65,dyr09} this is connected with a slower increase of the configurational entropy and a high activation energy for cooperative rearrangements. A similar scale of 1~nm is found for the cooperative length scale at the glass transition by chip calorimetry using a step-response analysis \cite{monnier2020}.

%\textit{Crossovers during the liquid-liquid (LLT) transition ---}
The study of the combined structural and dynamical properties of metallic liquids has also revealed new examples of liquid-liquid transitions (LLTs). In fact, glass-forming liquids with intermediate fragility may undergo a fragile-to-strong LLT transition. Among these systems are water, Si, Ge, and bulk metallic glasses. Viscosity measurements by levitating oscillating droplets in an electrostatic levitator located either on the ground or in a reduced-gravity aircraft during parabolic flights suggest that these findings are universal to BMG forming systems: in most metallic glass formers a kinetic crossover from fragile-to-strong behavior occurs in the supercooled liquid. Diffraction and XPCS measurements have proven that LLTs in metallic glass-formers are connected to both a structural and a dynamic crossover \cite{hechler2018}, see panel d) and e) of Fig.~\ref{new_figure4}. In particular, panel d) shows a change of slope in the temperature dependence of the $\alpha$-relaxation time of the Au$_{49}$Cu$_{26.9}$Si$_{16.3}$Ag$_{5.5}$Pd$_{2.3}$ metallic glass former in the supercooled liquid phase. This change of slope, highlighted in the insert, can be associated to a LLT between two liquids with different fragility. This transition has also a structural signature measured by high-intensity X-ray diffraction, as shown in panel e) of Fig.~\ref{new_figure4}. In particular, the temperature dependence of the inverse of the cube of the position of the first sharp diffraction peak of the static structure factor normalized to that measured at $T_{ref}$=395.5~K, $[Q_p(T_{ref})/Q_p(T)]^3$, is reported there. This quantity is a proxy for the sample volume, and shows evidence of the LLT when the sample is cooled applying the same thermal protocol used for the XPCS analysis shown in panel d) (blue diamonds). It is interesting that if continuous cooling with 1.5~K/min is used instead (green triangles), the LLT is no longer visible. The existence of different liquid phases is very interesting for BMGs as it offers the possibility to both better understand and possibly even control the structures appearing in the glass by appropriate choice of the temperature at which the liquid is quenched into the glass and of the quenching rate used in the process.

\subsection{Mechanical properties}
 The advent of modern multicomponent BMGs in the 1990s enabled the characterization of a wide range of mechanical properties due to the relatively large samples that could be produced, and a wealth of mechanical property data has now been collected for BMGs. This has revealed that mechanical properties related to plastic deformation, e.g., hardness, compression/bending ductility, fracture toughness, fatigue, etc., are highly sensitive to the processing methods and thermomechanical history of the BMG \cite{KruzicReview2016}. It is now generally accepted that BMGs with low fictive temperature, which may be created either by slow cooling from the melt or by relaxation via sub-$T_g$ annealing, become embrittled \cite{LAUNEY2008500,Rycroft_PRL2012}. Conversely, BMGs that have been fast cooled into higher fictive temperatures or rejuvenated into higher energy states are softer, with enhanced ductility and toughness \cite{sun2016thermomechanical,Ketkaew_NatCom2018}. This has led to extensive recent experimental advancements in improving ductility and fracture toughness either by precisely controlling fictive temperature (e.g., by thermoplastic forming \cite{Ketkaew_NatCom2018}) or by inducing structural rejuvenation by various methods such as radiation \cite{Xie_JNuclear2024}, elastostatic compression \cite{Li_MSEA2022,Costa_Acta2024}, cryogenic thermal cycling \cite{Ketov_Nature2015, Li_Acta2019, Ketkaew_Acta2020}, high-pressure torsion \cite{Bian_MSEA2019,Ebner_Acta2018}, cold rolling \cite{Stolpe_Acta2014,Xia_JALCOM2017,Li_MSEA2020}, and mechanical imprinting \cite{Li_MSEA2015,Scudino_SciRep2018}. 
 
A key remaining experimental challenge is to develop a detailed description of the related processing-structure-property relationships for BMGs. Further complicating this challenge is that many processing and rejuvenation methods induce highly heterogeneous glassy structures where the energy state and local mechanical properties change over dimensions that can be many micrometres, or even hundreds of micrometres, in length scale \cite{Li_MSEA2022, Li_Acta2019,Li_MSEA2020, Scudino_SciRep2018, Li_MSEA2015, Best_APL2019}. Such heterogeneities are thought to further enhance ductility and toughness by promoting shear band proliferations \cite{Li_Acta2019,Scudino_SciRep2018, Li_MSEA2015, Liu_Science2007}, but make it more difficult to create a three-dimensional picture of the glassy nanostructural and microstructural features controling the mechanical properties.   

The development of processing-structure-property relationships for crystalline metals is based on: 1) our understanding of ordered lattice structures, which are measurable by diffraction methods, and 2) the direct observations of crystalline defects visible by various microscopy and tomography methods. While long-range order is absent in BMGs, the presence of short and medium-range order (SRO and MRO) can be observed using various diffraction methods such as synchrotron X-ray diffraction, neutron diffraction, and nanobeam electron diffraction. In particular, a powerful tool used in conjunction with synchrotron studies is the electromagnetic \cite{Mohr2023}  or electrostatic levitation technique \cite{Kordel2011}, by which liquid droplets can be studied {\it in-situ} with respect to their structure during undercooling all the way from the equilibrium liquid down into the glassy state \cite{Stolpe2016}. Overall, these studies show that, while crystalline defects do not exist in BMGs, locally soft spots (e.g., Eshelby-like inclusions) and regions surrounding clusters of close-packed atoms are responsible for initiating inelastic atomic displacements in metallic glasses under strain~\cite{Sopu_PRL2017, Kang_AdvMater2023, Cubuk_Science2017, ding2014soft, Kondori_EML2016}. Furthermore, by utilizing experimental diffraction techniques, an understanding of how MRO within the glass structure strongly influences the local plastic deformation response has recently been obtained \cite{Nomoto_MaterTod2021, Nomoto_PRM2022, Davani_JAP2020, Liu_JMST2023}. 

\begin{figure*}[ht]
	\centering
 	\includegraphics[width=1.0\textwidth]{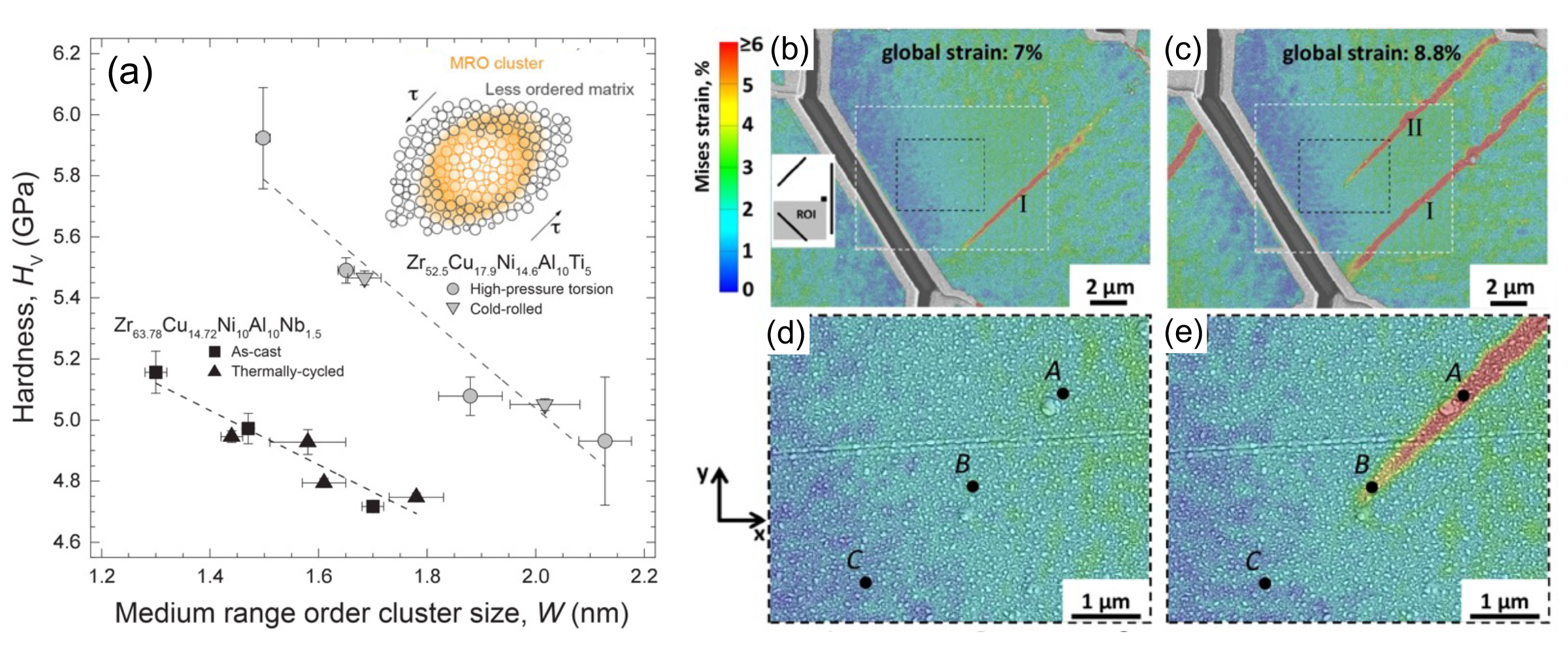}  
	\caption{
    a) Linear correlation between medium range order (MRO) cluster size and local hardness measured for two different BMG compositions. The correlation between hardness and MRO was maintained even after the hardness was altered by cryogenic thermal cycling, high pressure torsion, or cold rolling and a similar linear correlation was found between hardness and MRO volume fraction. Inset shows a schematic of an MRO cluster acting as a nucleus for a shear transformation zone. Figure reprinted from Ref.~\cite{Nomoto_MaterTod2021} with permission from Elsevier. 
    b)-e) Direct observations of shear band birth, propagation, and arrest visualized by the von Mises strains. From b) to c) the progressive extension of shear band I is observed with increasing strain along with the nucleation and arrest of shear band II while d) and e) give a higher magnification view of the regions within the black rectangles. Figure reproduced from Ref.~\cite{Glushko_NatComm2024} with open access CC BY license.}
	\label{figureMechProp}
\end{figure*}

In particular, nanobeam electron diffraction and fluctuation electron microscopy studies have revealed a linear relationship between decreasing local hardness with increasing MRO cluster size (Fig. \ref{figureMechProp}a) and volume fraction at the nanoscale for various Zr-based and Ni-based BMGs \cite{Nomoto_MaterTod2021, Li_JALCOM2025, Li_MSEA2022}. While the scaling of the relationship depends on the exact BMG composition, the measured linear relationship between hardness and MRO is identical, even after using various rejuvenation treatments to alter the overall distribution of hardness heterogeneities within the BMG microstructure \cite{Nomoto_MaterTod2021, Li_MSEA2022, Nomoto_PRM2022}. In other words, locations with identical measured hardness in various rejuvenated and unrejuvenated samples had the same MRO cluster size and volume fraction regardless of the sample processing history. Furthermore, a model of ductile phase softening has been developed based on the concept that high symmetry face centered cubic like (FCC-like) MRO clusters act as soft spots within a harder matrix where icosahedral structure is thought to dominate \cite{Nomoto_MaterTod2021}. This represents a significant step forward in using experimental measurements to explain how prossessing induced changes to the glassy structure control the mechanical properties of BMGs. 

Many recent experimental developments to look into the structure of BMGs are based on the use of very bright X-ray and electron sources, which poses the serious problem of the effects of these beams on the materials under investigation. For example, it is known that oxide and chalcogenide glasses are modified, in terms of both atomic structure and dynamics, by intense X-ray beams \cite{Ruta2017,Martinelli2023,Baglioni2024}. In particular, the X-ray beam induces atomic displacements that lead to a rejuvenation of the glass \cite{Baglioni2024}. BMGs are generally considered to be more forgiving in this regard, but it has recently been reported in electron correlation microscopy experiments at 300 kV acceleration voltage an inverse relation between electron dose rate and the characteristic time of the measured density fluctuations in a Pd$_{40}$Ni$_{40}$P$_{20}$ metallic glass \cite{Vaerst2023}. This result clearly implies that, similarly to the X-ray case, intense electron beams do induce measurable atomic displacements in BMGs as well. This seems related to the rejuvenation process that also metals undergo under heavy irradiation \cite{Xie_JNuclear2024}. Recent instrumentation developments coupling flash calorimetry with the X-ray analogue of electron correlation microscopy, i.e., X-ray photon correlation spectroscopy \cite{Martinelli2024}, and with X-ray diffraction \cite{Sun2025} are now able to provide calorimetric information during the X-ray experiments, e.g., monitoring glass-rejuvenation under X-ray exposure in real time, and thus will contribute to better understand this issue.   

Another ongoing experimental challenge is that, unlike for crystalline materials, there is only a limited experimental ability to create a three-dimensional picture of atomic positions in BMGs. This has hindered the ability to develop theories or computational models of the deformation process outside the binary systems (e.g., Zr-Cu) that can be accurately modeled from first principles. Furthermore, most experimental studies have measured MRO using single scalar quantities (size, volume fraction) that do not distinguish between different types of MRO that may affect mechanical properties differently, e.g., FCC-like versus icosahedral-like order. Recently, an experimental breakthrough in this regard has been achieved using atomic electron tomography to reveal a three-dimensional picture of the atomic arrangements inside individual SRO and MRO clusters in partially amorphous nanoparticles and thin films \cite{Yang_Nature2021, Yuan_NaterMat2022}. Such results have been able to resolve the detailed sizes and distribution of four different crystal-like MRO types (FCC, body centered cubic (BCC), simple cubic (SC), and hexagonal close packed (HCP)) in eight-component partially amorphous nanoparticles (see Fig.~\ref{fig:TTT_rate}d)) with no observed icosahedra \cite{Yang_Nature2021}. Further studies using this technique on monatomic amorphous Ta and Pd have revealed that their MRO comprises mainly pentagonal bipyramid networks with a lesser amount of icosahedra \cite{Yuan_NaterMat2022}. Combined, these results suggest high symmetry crystal-like MRO may be more prevalent in multicomponent metallic glasses compared to monoatomic ones, which would help explain their favorable mechanical properties.

%\textit{Short range order, medium range order, and cluster connection ---} The thermodynamic and kinetic properties of supercooled liquids and glasses are rooted in the atomic structure of the alloys, which depends on the alloy composition as well as on its degree of relaxation or rejuvenation. Since there is no long range order, the structural differences are found in the topological and chemical short range order, in the medium range order as well as in the degree of connectivity of the short range ordered clusters. The structural differences critically determine the mechanical properties of the alloys as will be also discussed in the next section. The enormous advances in high energy X-ray diffraction (HEXRD) over the past decade using synchrotron radiation, e.g., at DESY in Germany or at the ESRF in France, lead to the very accurate determination of structural differences in glass-forming alloy liquids and glasses as a function of temperature. 
The role of short range order (SRO), medium range order (MRO) and cluster connection emerges clearly in recent studies of the interesting family of (Pd,Pt)-Cu-Ni-P alloys \cite{Gross2019}. This family includes Pd\textsubscript{42.5}Cu\textsubscript{30}Ni\textsubscript{7.5}P\textsubscript{20}, which is the best glass former of all BMG with a critical casting thickness of 80~mm \cite{nishiyama2012world}.
By variation of the Pd to Pt ratio in the (Pd,Pt)\textsubscript{42.5}Cu\textsubscript{27}Ni\textsubscript{9.5}P\textsubscript{21} alloy, changes in SRO and MRO can be tracked. The Pd-rich alloys are dominated by icosahedral SRO whereas the Pt- rich alloy shows pronounced MRO dominated by trigonal prisms. In addition, the Pt-rich alloy shows a much more pronounced amount of three atom connections between adjacent clusters that are quantitatively associated with a decreased strain rate sensitivity during nano-indentation and with a brittle behavior of the Pd-rich alloys \cite{NeuberSciRep2022}. 
The change from ductile to brittle fracture is also reflected in the decrease of residual enthalpy and free volume as the liquid freezes in at the glass transition with increasing Pd content reducing the amount of shear transformation zones (STZs) in the alloy~\cite{neuber2021}, where STZ is a small cluster of atoms in an amorphous solid that undergoes cooperative rearrangement under stress, serving as the basic unit of plastic deformation~\cite{ARGON197947}.
Moreover, aging (structural relaxation) in (Pd,Pt)-based alloys leads to an increase in three-atom connections, which is accompanied by embrittlement and a reduction in volume. A similar link between structural changes and aging-induced embrittlement has also been reported in the well-known Zr-based alloy Vit105~\cite{Ruschel2024}.  Samples with a progressively lower fictive temperature exhibit a lower enthalpic state, coupled with a reduced amount of free volume, which is responsible for a continuous embrittlement. High-Energy X-ray Diffraction (HEXRD) experiments again reveal a correlation between the increase in rigid three-atom cluster connections with the reduction in the fracture strain, as a measure of ductility, indicating a strong correlation with the thermal history. While the atomic connections seem to have a crucial contribution to the ductility, changes of the short- and medium-range order seem to be equally important in this case as well. 

Overall, the collective body of evidence using various techniques suggests that crystal-like, e.g., FCC, HCP, BCC, or SC, clusters and polyhedral clusters, e.g., icosahedral and pentagonal bipyramids, should be expected to co-exist in various ratios for different metallic glasses \cite{Nomoto_MaterTod2021, Yang_Nature2021, Yuan_NaterMat2022, Hwang_PRL2012, Im_PRM2021}. Unfortunately, so far, atomic electron tomography has been limited to the analysis of thin films and nanoparticles and thus the distribution of MRO types in bulk samples of well-studied BMG compositions has not yet been measured. Thus, while great progress has been made in the ability to experimentally assess MRO and its relationship to plastic deformation, there is still a tremendous amount of research that needs to be done in this area to fully understand the structure-property relationships that control the mechanical behavior of BMGs.  

Another recent methodological breakthrough in the nanostructural characterization of BMGs has been achieved via atom probe tomography, by which researchers are now able to observe nanoscale solute clustering in multicomponent BMGs \cite{Nomoto2025APT}. This approach allows for the characterization and quantification of the three-dimensional distribution, chemical composition, and volume fraction of nanoscale solute-rich clusters in samples extracted from any BMG, i.e., unlike atomic electron tomography this approach is not limited to thin film and nanoparticle samples. In addition, correlations have been found between the size and volume fraction of the solute-rich clusters and the local hardness in two Zr-based BMGs, with both BMGs becoming softer with more solute-rich clustering \cite{Nomoto2025APT}. However, the drawback of atom probe tomography is that the topological SRO or MRO of the solute-rich clusters, if any, cannot be directly measured. Thus, while the trends in mechanical behavior with MRO clusters \cite{Nomoto_MaterTod2021, Li_MSEA2022, Nomoto_PRM2022} and solute-rich clustering \cite{Nomoto2025APT} appear similar, there is still considerable work to be done to establish the full three-dimensional atomic structure of metallic glasses and the detailed relationships between topological ordering, solute clustering, free volume, and mechanical behavior.

Considering free volume, shear transformation zones (STZs), and shear bands to be the deformation carriers in BMGs, there is still no ability to observe them “in action” like crystalline deformation carriers, e.g., dislocations, twins, and stress-induced martensitic transformations, \textit{in situ} in a transmission electron microscope (TEM). In this regard, recent advances have used digital image correlation and {\it in situ} loading experiments in a high-resolution scanning electron microscope (SEM) to measure local strains during initiation, propagation, or arrest of shear bands (Figs. \ref{figureMechProp}b)-e) in a metallic glass thin film \cite{Glushko_NatComm2024}. Such experimental advances have enabled the development of a continuum mechanical formulation to accurately describe shear band behavior in metallic glasses. The combined experiments and theory suggest that shear bands generally propagate in a progressive manner, and that observations of structural changes and heating are the consequence of dissipated energy from the plastic deformation process \cite{Glushko_NatComm2024}. 
%Such experiments open the door for extensive future in situ deformation studies of plastic deformation in metallic glasses using both SEM and TEM based methods.  

While novel approaches to directly observe the effects of STZs, such as their displacement and stress fields, are still under development, an indirect route to probe STZs in bulk metallic glasses (BMGs) has been found in the study of the so-called $\beta$-relaxation. This is also known as Johari-Goldstein relaxation~\cite{Johari1970}, which branches off the $\alpha$-relaxation in the deeply-supercooled liquid phase, and is the only relaxation process active in the glass. Therefore, it is probably not surprising that it plays an important role in a number of relevant mechanical properties, as, for instance, the plastic response of the material \cite{Yu2013}. In BMGs, it has been mainly characterized via dynamical mechanical analysis \cite{Yu2010} and, more recently, differential scanning calorimetry \cite{Hu2009,Yang2020}. The observation that the potential energy barrier involved in STZs matches the activation energy of the $\beta$-relaxation \cite{Yu2010} establishes a strong connection between shear transformation zones and the quasi-local regions in the glass matrix where the $\beta$-relaxation is active. This clearly suggests the possibility of studying STZs via the analysis of the $\beta$-relaxation in BMGs. In fact, STZs and $\beta$-relaxations seem to reflect the mechanical properties of BMGs in a very similar way \cite{Yu2013}. While these studies cannot of course provide a tool to look at the STZs in action, they establish an interesting connection between two fundamental aspects of the physics of glasses, i.e., deformation mechanisms and relaxation dynamics \cite{Yu2013}.

\section{Additive manufacturing}\label{sec:AM}
\subsection{Motivation and process overview}
\begin{figure*}[htbp!]
	\centering
 	\includegraphics[width=\textwidth]{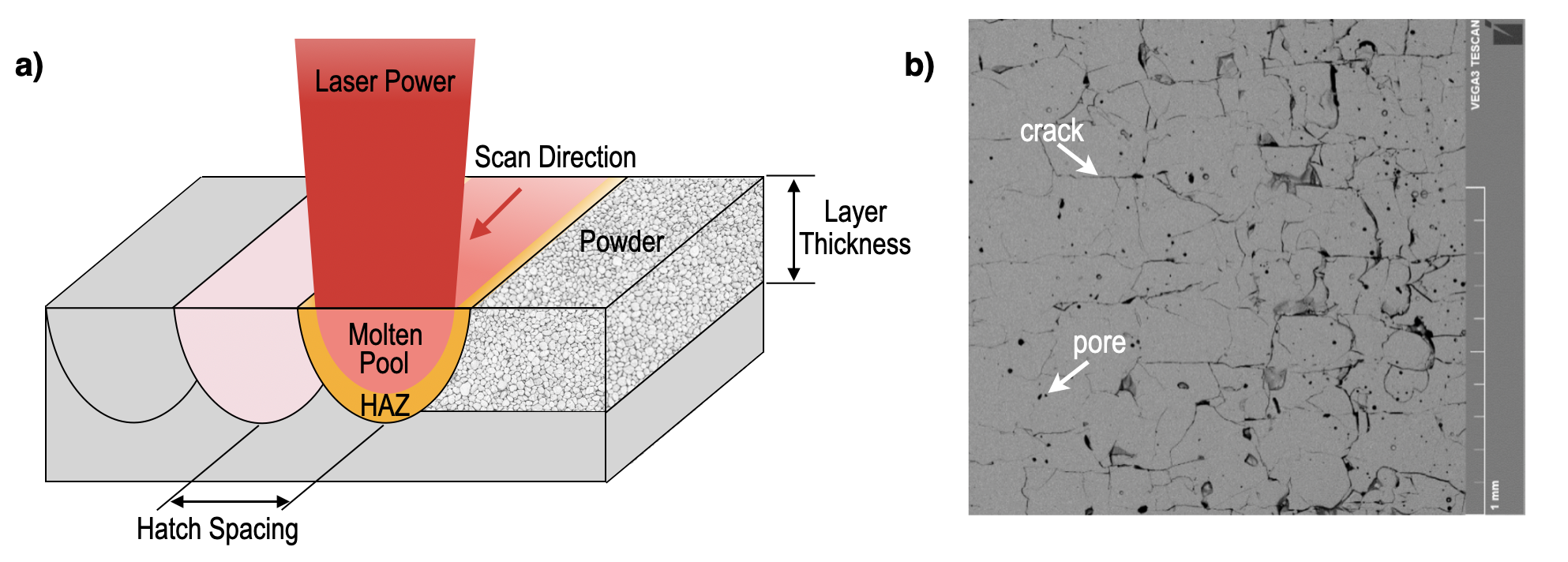}  
	\caption{Laser-based additive manufacturing (3D~printing) of metallic glasses: working principle and microstructure. 
    a)~Schematic diagram of the laser powder bed fusion (LPBF) process. The laser power is directed onto the powder bed, forming a molten pool with an underlying heat-affected zone (HAZ), where thermal effects from reheating might induce microstructural response, such as devitrification. An adjacent prior melt track is shown, with the hatch spacing indicated between the tracks.
    b)~SLM Fe-MG microstructure showing amorphous regions, pores, and cracks. Figure adapted from~\cite{malachowska2023selective}, CC BY 4.0.}
	\label{schematic_3DP}
\end{figure*}

Industrial applications of metallic glasses have traditionally been limited by issues such as poor machinability and small critical casting dimensions required to achieve an amorphous structure \cite{KruzicReview2016, sohrabi2024manufacturing}. However, over the last decade, additive manufacturing (AM) processes have overcome these issues, and are able to produce amorphous metallic structures with few constraints on dimensions or geometric complexity \cite{XP_Li2017Review, sohrabi2021additive, Lashgari2022AM_Review}. Importantly, additive manufacturing has been successfully applied for numerous alloys with both high and low GFA, including systems based on zirconium \cite{XP_Li2016AM, Bordeenithikasem2018AM}, titanium \cite{Deng:2018eb, Schoenrath2024AM}, copper \cite{FREY2023AM}, nickel \cite{Soares2024, Frey2025}  
and iron \cite{Pauly2013AM, THORSSON2022110483, Mahbooba:2018kj}. 
Powder Bed Fusion (PBF) is the most popular method for 3D~printing metallic glasses. It is a family of AM processes that exploit an energy source to selectively melt fine metallic powder of the target composition, layer by layer, to produce three-dimensional structures. Depending on the type of source and fusion mechanisms, PBF can be distinguished into laser-based PBF or electron beam-based PBF. Laser-based PBF is the dominant method for additively manufacturing MGs, as it allows precise control of laser parameters to achieve high cooling rates (typically $10^3$–$10^7$~K/s), essential for maintaining the amorphous structure and supporting various MG compositions.
It comprises subgroups such as laser powder bed fusion (LPBF), which is essentially synonymous with selective laser melting (SLM) in metallic contexts, focusing on complete fusion for high-density results.
In this review, we focus on LPBF processes~\cite{joshi2023metal}, given their prevalence in MG manufacturing, highlighting challenges, parameter optimization, mechanical properties, and applications in the following sub-sections.
Figure~\ref{schematic_3DP}a) schematically illustrates the LPBF process.

%The LPBF process explained - Fig 6a
The LPBF process occurs in a controlled, inert atmosphere (e.g. argon or vacuum) to minimize oxidation and contamination. It begins with the deposit of a thin layer of powder (20–100~$\mu$m, e.g., Zr-based MG particles have a typical size of 15–63~$\mu$m~\cite{Bosong_Li2024AM}) on a building plate. Then, a laser scans and fully melts selected regions according to a predefined pattern, forming molten pools that solidify rapidly. Afterwards, the plate lowers to allow a new layer of powder to be added until the process is complete. 
%the challenges, HAZ, lack of fusion, porosity and cracks
Despite the advantages of 3D~printing to manufacture MGs, several processing challenges remain, primarily due to the thermal cycles inherent in the layer-by-layer fusion protocol. 
A key issue that might arise is \textit{crystallization}~\cite{liu2020crystallization,zrodowski2023control} that compromises the mechanical properties~\cite{madge2021laser}. 
It occurs mainly in the so-called heat-affected zone (HAZ), the region near the molten pool subjected to high temperatures without fully melting, due to repeated reheating from successive layers. This reheating exposes the metastable amorphous structure to temperatures above the glass transition but below melting, allowing cumulative time for nucleation and growth of crystals over multiple thermal cycles, as governed by time-temperature-transformation kinetics.
In addition, \textit{cracks} often form, caused by rapid thermal gradients and contraction during solidification, particularly in low-toughness MGs, and can be influenced by microstructural heterogeneity~\cite{sohrabi2021additive,liu2023pores}.
Another issue is \textit{porosity}, which mainly arises from gas entrapment in the molten pools or incomplete powder fusion, and results in voids that act as stress concentrators and reduce density~\cite{liu2023pores,luo2023mechanical}. 
Similarly, \textit{lack of fusion} defects can occur when there is insufficient molten material to fill the spaces between particles or layers. This is mainly due to a suboptimal energy supplied by the laser to locally melt the powder, resulting in weak bonding between layers, and anisotropic properties~\cite{zhang2022research}. 
Figure~\ref{schematic_3DP}b) shows the microstructure of a 3D~printed Fe-based MG~\cite{malachowska2023selective} with amorphous regions, pores, and cracks.

% the solution (?): parameter optimization
%For MGs, LPBF enables dense (>99%), high-strength parts but requires parameter tuning to avoid crystallization or defects like porosity.
Due to the metastable nature of metallic glasses, several strategies are needed to overcome the challenges of AM. We classify these into \textit{systemic approaches}, which modify or improve the AM process through external tools or innovative strategies, without altering the fundamental parameters of the machine, and \textit{process-centric approaches}, which focus on optimizing the intrinsic LPBF parameters themselves. 

\textit{Systemic approaches} encompass broader improvements, including advanced scanning techniques, remelting methods, \textit{in-situ} monitoring, and alloy design.
Optimized scanning techniques mitigate thermal stresses~\cite{pauly2017processing}. \textit{Chessboard scanning}, for instance, separates layers into alternating squares for even heat distribution and reduced stresses, while \textit{random scanning} randomizes paths to homogenize gradients and minimize anisotropy.
Compared to \textit{concentric scanning}, which uses circular paths and risks pore accumulation and high gradients due to overlaps, chessboard and random approaches are more effective for MGs, reducing cracks in Fe-based alloys by up to 70\% through stress relief~\cite{zou2020selective}. 
Remelting methods such as \textit{dual-laser remelting} involve an initial melting followed by a second passage to refine microstructures and reduce defects such as porosity or crystallization~\cite{zhang2022research}. \textit{In-situ} monitoring improves control through real-time thermal imaging and feedback systems that adjust parameters to maintain cooling rates above critical thresholds for amorphicity, using sensors such as near-infrared cameras to detect anomalies in the molten pool and predict defects~\cite{moshiri2023performance}. Finally, alloy design involves selecting suitable material compositions, adding elements that can improve the GFA and therefore printability, preventing detrification. Novel routes to discover MG compositions with high GFA are discussed in Section~\ref{sec:ML_GFA}.

\textit{Process-centric approaches} are discussed in detail in the next paragraph. Their aim is to optimize machine parameters for better control of the molten pool and minimization of defects (e.g., porosity and crystallization)~\cite{Bosong_Li2024AM, Hadibeik2024AM}.

\begin{figure*}[htbp!]
	\centering
 	\includegraphics[width=1.0\textwidth]{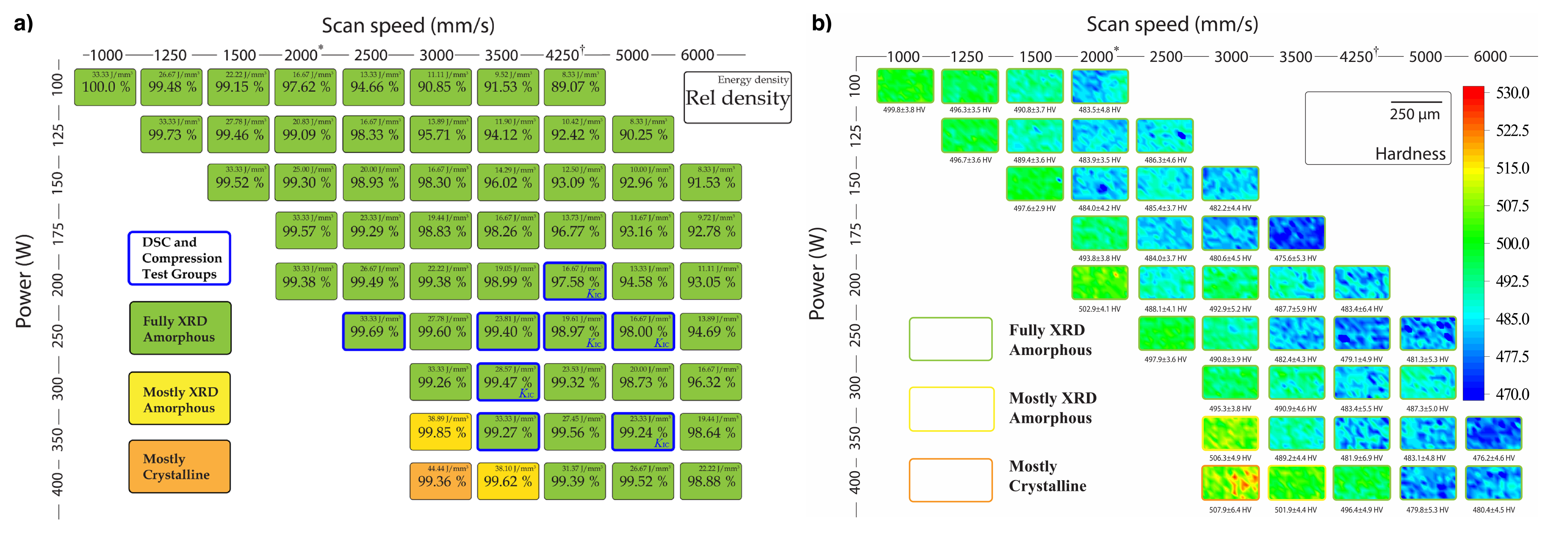}  
	\caption{
    a)~The classic tradeoff of amorphicity versus density is illustrated for a Zr$_{59.3}$Cu$_{28.8}$Nb$_{1.5}$Al$_{10.4}$ MG fabricated by laser powder bed fusion. Green cells indicate fully amorphous samples for various laser powers and scan speeds, with the relative density and laser energy values given in each cell. 
    %b)The variation in relative density can be observed directly in the micrographs with lack of fusion becoming prevalent at low values of laser energy.  Figure reproduced from \cite{Bosong_Li2024AM} with open access CC BY license.
    b)~Microhardness maps (0.6~mm$\times$~0.6~mm) for fully dense LPBF-fabricated Zr$_{59.3}$Cu$_{28.8}$Nb$_{1.5}$Al$_{10.4}$ MG samples, along with the corresponding laser power and scan speed. The average hardness values~(HV) are indicated below each hardness map, while the color outline indicates the measured XRD amorphicity. Figures reproduced from Ref.~\cite{Bosong_Li2024AM} with open access CC BY license.}
	\label{figureAM}
\end{figure*}

\subsection{Processing parameters and trade-offs}
%remove the par in te before text
Achieving a fully amorphous structure in LPBF-fabricated MGs requires careful optimization of processing parameters, including laser power~$P$, scanning speed~$v$, hatch spacing~$h$, and layer thickness~$t$, which determine the laser energy density $E = \frac{P}{v \cdot h \cdot t}$~\cite{prashanth2017energy}. 
Typical ranges for MG processing include laser power of 50--400~W, scanning speeds of 500--6000 mm/s, hatch spacing of 50--150 $\mu$m, and layer thickness of 20--50$\mu$m \cite{Bosong_Li2024AM, Hadibeik2024AM}. 

%Part written by Jay: The dense-amorphous tradeoff
A trade-off arises between the amorphous phase content and the density of the produced material. High laser energy densities in the LPBF processes increase material density, but lead simultaneously to rapid devitrification in the heat-affected zone (HAZ), which reduces the amorphous phase content. 
As demonstrated by Ouyang \textit{et al.}~ \cite{ouyang2021understanding}, crystallization primarily occurs as a result of repeated heating within the HAZ, and the printability of certain alloys is determined by their resistance to devitrification under rapid heating and cooling cycles. In contrast, low laser energy densities more easily achieve an amorphous structure but also increase porosity, which is detrimental to the mechanical properties.  Overcoming this ``dense-amorphous'' trade-off is a well documented challenge in the metallic glass additive manufacturing literature as illustrated in Fig.~\ref{figureAM}. The green cells in Fig. \ref{figureAM}a) indicate that a fully amorphous structure, as measured by laboratory X-ray diffractometry, can be achieved over a wide range of laser powers and scanning speeds. However, as the laser energy density decreases in the upper right corner, although the samples are still fully amorphous, there is no longer enough laser energy to fully fuse the particles, leading to the large amounts of lack of fusion defects. %seen in Fig. \ref{figureAM}b. 
Nonetheless, for good glass formers, a wide range of LPBF process parameters can be used to give a range of mechanical properties, as will be discussed in the next sub-section.  

%oxygen content
Environmental factors, such as oxygen content in the powder and the build chamber, can also affect the final structure and properties of additive manufactured BMG samples. Elevated oxygen levels ($>$500~ppm) in the final printed samples can increase brittleness and promote crystallization in Zr-based BMGs, narrowing the dense-amorphous processing window \cite{Wegner2021, Bordeenithikasem2018AM, Bosong_Li2024AM, Best2020AM}. However, it has yet to be determined how important the role of oxygen will be for BMGs with less affinity for oxygen uptake, such as those based on nickel \cite{Soares2024, Frey2025} and iron \cite{Pauly2013AM, THORSSON2022110483, Mahbooba:2018kj}.

Powder-related factors also influence AM of metallic glasses. Finer particles can improve packing density, layer uniformity, and feature resolution, but they risk oxidation and, in reactive systems, ignition. Spherical particles with a narrow size distribution are desired to achieve uniform spreading of powder.  Temperature parameters are also important. A moderate preheating of the chamber and/or build plate (typically tens to a few hundred $^{\circ}$C) can reduce thermal gradients and residual stress. 
%
%Overall, high-GFA materials (e.g., Zr$_{59.3}$Cu$_{28.8}$Nb$_{1.5}$Al$_{10.4}$) are able to avoid crystallization over broader ranges of process parameters than poor glass formers, which gives more options for balancing other factors such as density, residual stresses, and mechanical properties. 
%In general, there are a wide range of processing parameters that can be altered in the LPBF process. While the dense-amorphous trade-off will always be a concern for BMGs, for relatively good glass formers both simulations and experiments have shown that a wide range of LPBF parameters can produce fully dense and amorphous samples \cite{YANG_2022_ADDMAN, Bosong_Li2024AM, Hadibeik2024AM}. 
In general, there is a wide range of processing parameters that can be altered in the LPBF process. While the dense-amorphous trade-off will always be a concern for BMGs, high-GFA materials are able to avoid crystallization while achieving full density over broader ranges of process parameters than poor glass formers~\cite{YANG_2022_ADDMAN, Bosong_Li2024AM, Hadibeik2024AM}. This gives more options for balancing other factors such as residual stresses and mechanical properties. 

\subsection{Mechanical properties}
LPBF fabricated BMGs often have lower reported strength, ductility, and/or fracture toughness values compared to BMG castings of the same composition \cite{pauly2017processing, Bordeenithikasem2018AM, Deng:2018eb, Best2020AM, Best2020MSEA, Wegner2021, Bosong_Li2024AM}. Reduced ductility and fracture toughness can be a significant problem since these are usually limiting properties for using BMGs in engineering applications \cite{Demetriou_NatMater2011, Ritchie_NM_2011, KruzicReview2016}. Fracture toughness in particular, is used as an engineering design parameter, and precracked fracture toughness values have been reported in the range of  24-38~MPa$\sqrt{m}$ for LPBF produced Zr-based BMGs \cite{Best2020AM, Best2020MSEA, Bosong_Li2024AM}. While these values are competitive with some crystalline engineering alloys (e.g., tool steels, high strength aluminum alloys, etc.), they are much lower than what has been reported for many cast BMG samples which can range well over 100~MPa$\sqrt{m}$ \cite{Demetriou_NatMater2011, Best2020MSEA, Best2020AM, Ketkaew_Acta2020, Li_Acta2019, He_Acta_2012, Wen_Chen_Acta_2016}.
%Moreover, by altering the laser heat input via the laser power and scanning speed, a large range of structural states and relaxation enthalpies can be achieved at nominally identical density and amorphicity levels \cite{Bosong_Li2024AM, Hadibeik2024AM}. 
%

The systematic evolution of the mechanical properties of produced LPBF samples with respect to machine parameters such as laser power and scan speed is addressed in~\cite{Bosong_Li2024AM}. Figure~\ref{figureAM}b) shows representative Vickers hardness maps and average hardness values. Higher laser heat input tends to relax the layers underneath the melt pool, increasing the strength and hardness while decreasing the relaxation enthalpy, ductility, and fracture toughness. As the laser energy density increases towards the lower left corner, the samples become progressively harder due to structural relaxation of the heat-affected zones until eventually the heat input becomes so great that crystallization begins to occur. As expected, the softest samples exhibit the best compression ductility and fracture toughness values up to 6\% and 38~MPa$\sqrt{m}$, respectively \cite{Bosong_Li2024AM}. Furthermore, softer samples show larger FCC-like medium range order clusters within the amorphous structure \cite{Bosong_Li2024AM}, which is in agreement with studies of conventionally processed BMG samples \cite{Nomoto_MaterTod2021, Nomoto_PRM2022}.

As mentioned above, enhanced oxygen content introduced from the powders is thought to contribute to embrittling LPBF fabricated BMGs relative to cast materials \cite{Bordeenithikasem2018AM, Deng:2018eb, Best2020AM, Best2020MSEA, Wegner2021, Bosong_Li2024AM}. However, lowering the oxygen level by four times in the BMG with composition Zr$_{59.3}$Cu$_{28.8}$Nb$_{1.5}$Al$_{10.4}$ was only able to increase the fracture toughness from 25 to 38~MPa$\sqrt{m}$~\cite{Bosong_Li2024AM}, which suggests that other factors are important in affecting the toughness of LPBF fabricated BMGs. Furthermore, the use of powder in the LPBF process means that minimum oxygen levels will generally be higher than for cast samples. This motivates the development of other methods for enhancing the mechanical properties of LPBF fabricated BMGs.

%this text is from amazemet - it has 0 references i cannot relate to anything 
%Despite the potential for adverse effects of devitrification on the mechanical properties of metallic glasses, there are also examples of amorphous/crystalline composites with enhanced mechanical properties produced via LPBF, using both ex-situ and in-situ composite fabrication methods. In one example, a novel approach has been developed to create fully dense composite materials with a precisely designed distribution of amorphous and crystalline phases. The approach achieves control over both localized crystallization in the HAZ and global overheating, allowing for precise microstructure manipulation and material property tuning specific to different regions of the manufactured objects. This opens up new design possibilities for the LPBF technology. \SB{----------REFERENCES MISSING-----------------} It has been demonstrated that precise tuning of dual-laser remelting in conduction and keyhole modes, combined with optimized scanning line separation, achieves exceptionally high heating and cooling rates while maintaining high amorphous phase content and material density. This advanced scanning strategy decouples microstructure design from powder densification concerns by minimizing thermal accumulation effects between adjacent melting tracks.

Opportunities to improve mechanical properties arise from designing and manufacturing multiphase MG matrix \textit{composite} systems using mixed powders~\cite{Chang_2025_JALCOM} and/or by \textit{in-situ} crystallization.
As MGs are brittle amorphous alloys, the idea is to intentionally introduce ductile toughening sites dispersed within the glassy matrix. This is achieved by selectively controlling the melting during LPBF: by optimizing parameters like laser power and scan speed, high-melting-temperature reinforcements (e.g., particles or precipitates) remain unmelted. 
For instance, in Ta-reinforced Zr-based MGs, when LPBF parameters are optimized to keep Ta unmelted during fabrication, a ductile-phase-reinforced composite with enhanced compressive ductility and toughness is achieved~\cite{Zhang_2019_JALCOM}.
This toughening is particularly useful for brittle Fe-based BMGs, where adding Cu or CuNi particles enables crack-free LPBF samples with improved strength and ductility~\cite{Li_2018_MatDes}.
%
%A similar effect can be achieved with mixed powders, which involve physically blending separate high-melting-temperature powders (e.g., Nb) with MG powders prior to LPBF, allowing the Nb particles to remain unmelted for ductile reinforcement~\cite{Chang_2025_JALCOM}. 

%
Alternatively, LPBF can promote \textit{in-situ} crystallization during fabrication. For carefully designed BMG compositions that crystallize into ductile phases, increasing laser energy density can encourage crystallization in heat-affected zones, yielding mechanical enhancements, as shown with Cu${50}$Zr${50}$ powders~\cite{Zhang_2021_SM}.
More complex microstructures can also emerge from mixed elemental powders that fully melt into a homogeneous liquid but phase-separate in the supercooled state, generating ductile phases in both melt pools and heat-affected zones, as demonstrated in the Ti-Zr-Cu system~\cite{Gao_2019_SM}.
Finally, laser rescanning strategies can be employed to crystallize BMG samples at selected locations during a second laser scan to give intricately patterned composite structures~\cite{zrodowski2023control}.
Although this rescanning technique has so far produced only brittle crystals, combining it with strategies for ductile-phase crystallization could enable BMG matrix composites with exceptional mechanical performance.

\subsection{Emerging laser powder bed fusion applications}
In this subsection, we describe some applications of LPBF additive manufactured MGs. In the following, we categorize them into different types and discuss future perspectives.

\begin{figure}[htbp!]
	\centering
 	\includegraphics[width=0.5
\textwidth]{{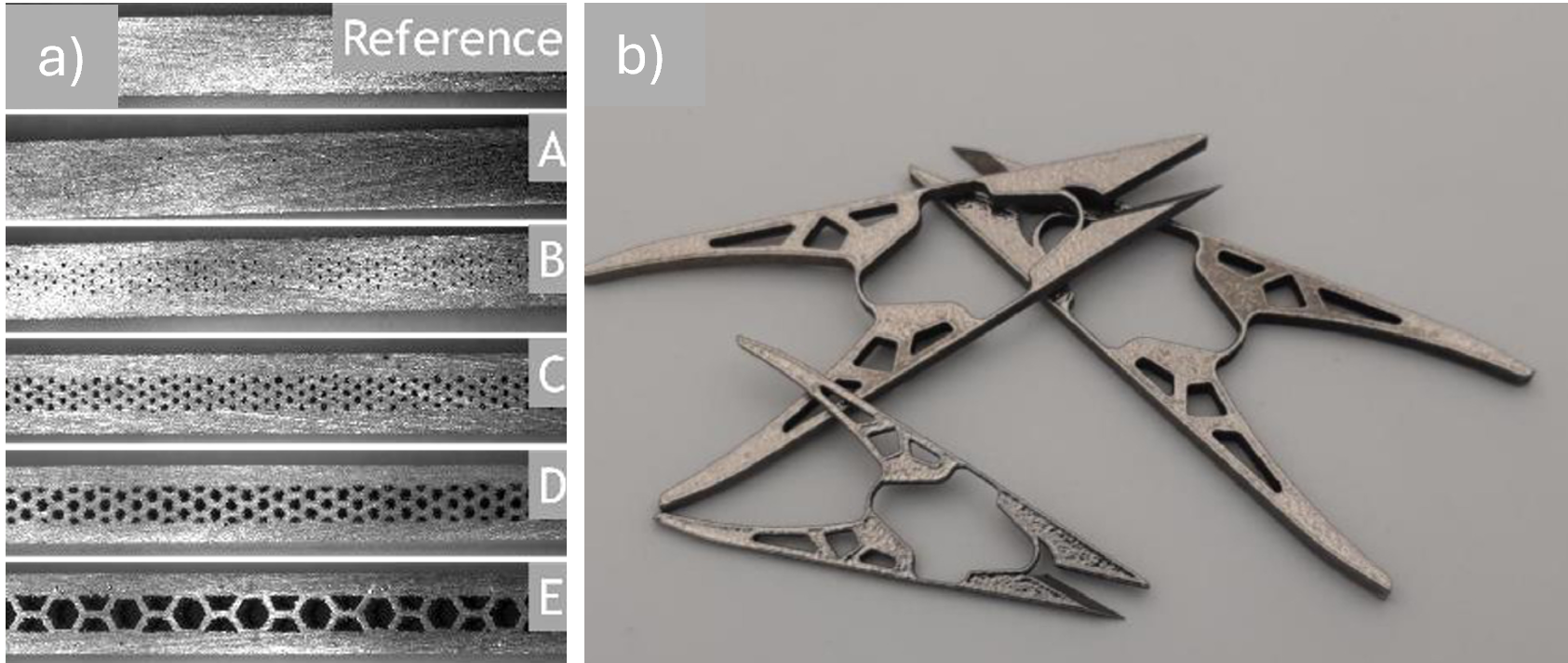}}  
	\caption{First examples of intricate fully amorphous structures fabricated by laser powder bed fusion with Zr$_{59.3}$Cu$_{28.8}$Nb$_{1.5}$Al$_{10.4}$ BMG. 
    a) Honeycomb structures. Figure reproduced from Ref.~\cite{wegner2019} with permissions. 
    b) Compliant forceps. Figure reproduced from Ref.~\cite{Wegner2021a} under CC BY-NC-ND license.
   %\CHR{Could the caption describe the size of these structures? They do not have scale bars.}
   }
	\label{figureAM2}
\end{figure}

\textit{Mechanical applications --- } 
The combination of high strength and high elastic deformation of MGs \cite{wan12} offers potential for applications in cellular structures and compliant mechanisms. 
\textit{Cellular structures}, such as lattices or mechanical metamaterials~\cite{bonfanti2024computational}, are lightweight structures possessing a complex internal geometry that endows them with outstanding mechanical properties. The fabrication of cellular structures made of AM-MGs results in superior mechanical performances that are ideal for aerospace, energy absorption, and medical fields.
A seminal work by Wegner \textit{et al.}~\cite{wegner2019}, demonstrates that Zr-based MGs manufactured in honeycomb structures (see Fig.~\ref{figureAM2}a)), shows improved quasi-plastic behavior under compression, and strength outperforming the crystalline counterparts. The honeycombs show elastic elongations up to 2\% and improved energy absorption, which is attributed to cell buckling.
Cellular MG structures are also promising candidates for compliant mechanisms.
\textit{Compliant mechanisms} rely on the elastic flexibility of a material and geometry to achieve complex movements~\cite{greer2023metallic}.
Traditional materials have limited motion due to low elastic limits, but AM metallic glasses unlock enhanced performance~\cite{homer2014new}. In a recent study~\cite{Wegner2021a} a Zr-based BMG was printed in a compliant mechanism inspired by forceps (see Fig.~\ref{figureAM2}b)). This designed structure is able to achieve a rotational elastic motion increase of about 300\% compared to Ti$_6$Al$_4$V, with yield strengths approaching 2~GPa and elastic limits of~2\%, overcoming casting limitations. In a subsequent work~\cite{frey2023laser}, Cu-Ti-Zr-Ni BMGs were successfully manufactured, achieving bending strengths up to 2.5~GPa, and surpassing many Zr-based AM MGs. Fracture toughness was the key to minimize cracks, with implications for robust compliant mechanisms. \\

\textit{Soft magnetic applications ---} 
Soft magnetic materials have a significant role in improving the energy efficiency of energy conversion devices~\cite{silveyra2018soft}. This is because they help reduce energy losses when the magnetic field changes. Much of this loss comes from hysteresis, which depends on how easily the magnetic domains of the material can switch. To address this, efforts are made to design materials with low magneto-crystalline anisotropy energy, in order to facilitate domain switching and reduce losses, improving overall efficiency.
Beyond conventional steels, amorphous alloys offer distinct advantages~\cite{ozden2021laser}.
Additive manufacturing of MGs, especially Fe-based ones, has transformed soft magnetic applications by enabling the fabrication of complex geometries with low coercivity (thanks to the intrinsic absence of magneto-crystalline anisotropy due to the amorphous structure)~\cite{azuma2020recent,lofstrand2024stress} and high magnetization~\cite{makino2008fesibp}. This results in a reduction of core losses in devices such as transformers, sensors, and electric motors~\cite{greer2023metallic}.
Recent progress includes the production of complex soft magnetic parts for motors, such as stators and rotors~\cite{THORSSON2022110483, Sadanand2024, Rodriguez2024}. Specifically, SLM technique has been used to produce a large-scale amorphous rotor from an Fe-Si-Cr-B-C powder alloy~\cite{THORSSON2022110483}, achieving record dimensions (see Fig.~\ref{fig:lengthscale}g)). 
The complex internal geometry is designed to efficiently channel magnetic flux. Characterization using synchrotron XRD, calorimetry, SEM, TEM, and magnetic testing reveals a predominantly amorphous structure with partial crystallization in the molten pools, leading to anisotropic magnetization. The work highlights the potential of SLM to overcome the limitations of melting for BMGs in electric motor applications, underscoring sustainability thanks to the abundance of elements and the reduction of waste. In another study~\cite{Sadanand2024} the LPBF processability of a Fe-Si-B-Nb-Cu alloy using a pulsed laser, and varying laser power and scan speed, is explored. Samples achieved good density and dimensional accuracy, but exhibited crystalline/amorphous composites. The crystalline regions are made of fine grains with random orientations. The magnetic properties show similar saturation magnetization to amorphous ribbons but higher coercivity, inversely correlated with amorphous fraction and grain size. Nanograined crystalline regions with random orientations and lack-of-fusion pores/cracks along crystalline/amorphous interfaces were found in another work~\cite{Rodriguez2024}. The saturation magnetization and coercivity were found to be inversely related to the amorphous fraction, with coercivity rising due to crystallization. 
It is anticipated that soft magnetic BMG components will be produced and commercialized once Fe-based metallic glass compositions are developed with sufficient glass forming ability to be fabricated with a fully amorphous structure using appropriate AM strategies for LPBF. \\

%corrosion resistance and biocompatibility.
\textit{Other applications ---} 
In addition to mechanical and soft magnetic applications, 
AM of MGs has emerged as a transformative approach for \textit{biomedical applications}, enabling the fabrication of complex implants and scaffolds with enhanced biocompatibility, mechanical and corrosion resistant properties~\cite{aliyu2023laser}.
Zr-based MGs and Zr-Cu-Fe-Al-Ag alloys, produced via selective laser melting, exhibit almost full amorphous structures, high strength, low Young's modulus, superior corrosion and wear resistance in simulated body fluids. Furthermore these MGs show excellent cell proliferation support and antibacterial effects~\cite{zhang20193d,larsson2022biocompatibility,Bordeenithikasem2018AM,onyeagba2025review}. Porous Zr-based BMG scaffolds can mimic bone tissue, promoting osteointegration and reducing stress shielding~\cite{zhang20193d}.
Ti-based and Mg-based MGs further expand applications, with Ti alloys offering  compatibility with magnetic resonance imaging and Mg systems providing biodegradability for temporary scaffolds in tissue engineering~\cite{li2016recent,onyeagba2025review}. Despite the difficult challenges of partial crystallization and brittleness in AM, advances in powder bed laser melting and surface modifications promise customized medical devices, bridging the gap between high-performance amorphous materials and clinical needs~\cite{liu2020crystallization}.
For \textit{catalytic applications}, 3D-printed hierarchical porous MG structures based on Zr or Fe, improve electrocatalytic performance for CO$_2$ reduction and wastewater treatment, offering high surface area, stability, and efficiency thanks to tunable amorphous phases~\cite{wu2022additive,yang20183d,zhang2019review,liang2020selective}. Recent developments in chemically complex MGs via AM further expand catalytic uses in fuel cells and hydrogen production, overcoming size limitations of traditional casting methods~\cite{zhang2022recent,zhakeyev2017additive}. Such advancements promise scalable, multifunctional components for sustainable energy and environmental catalysis.\\
%(Nuclear Applications - TO CHECK)\\ 
%\textit{Future perspectives ---} Machine learning-driven optimization of LPBF parameters shows promise for automating the dense-amorphous balance, potentially accelerating process development \cite{sohrabi2024manufacturing}. 
%These advancements will further unlock the potential of AM BMGs for industrial applications.
%Thermomechanical modeling must be integrated with feedback control to refine melt pool dynamics. This is initially done in these works.
%Latest advances~\cite{graeve2023latest}
%For details about the dynamics and temperature gradients in the molten pool the reader can refer to thermal fluid-flow simulations, presented in~\cite{wannapraphai2025quantifying}.
%Model prediction of porosity:\cite{oster2024deep}.  \\
%Designing materials by laser powder bed fusion with machine learning-driven bi-objective optimization~\cite{kononenko2024designing}.\\
%in situ monitoring is relevant: Implementing real-time thermal and compositional analysis to control crystallization and defects \cite{Lashgari2022AM_Review}.
%{Functionally graded materials}: Exploring amorphous/crystalline composites for tailored mechanical and functional properties \cite{sohrabi2021additive}. \\
%Alloy development: Designing BMGs with lower critical cooling rates to widen the processing window \cite{sohrabi2024manufacturing}.\\
%Scalability Improving build rates and reducing costs for large-scale production \cite{XP_Li2017Review}.
%Talk about composites -- High toughness metallic glass-based composites for additive manufacturing 

\section{Nanoscale modeling}

\label{sec:nano_modeling}

The purpose of modeling is to be able to understand and predict how the properties of MGs depend on their composition and processing conditions in the form of thermal and mechanical history. In this review paper, we consider modeling referring to different levels of detail, starting on the atomic scale and moving upwards to higher length scales. This section focuses on modeling at nanoscale.
A major challenge in metallic glasses, as in many areas of materials science, is to understand how bulk thermodynamic and mechanical behaviors emerge from local structure and composition. These properties are ultimately determined by atomistic interaction potentials, which describe how atoms interact with each other. At the nanoscale, computational modeling plays a crucial role by providing detailed information on these interactions, atomic positions, and chemical compositions. 

\subsection{Three categories}

Nanoscale computer simulations can be broadly categorized into three main types.

i) First-principles (\textit{ab initio}) methods: These methods calculate the interaction energy and forces between atoms directly from fundamental physics, specifically by solving equations based on the electronic structure of atoms. The most common approach is density functional theory (DFT), which approximates the quantum mechanical behavior of electrons in a material~\cite{kohn1965self}. Since \textit{ab initio} methods do not rely on fitting parameters, they are very accurate and provide deep insights into atomic interactions~\cite{car1985unified}. However, because these calculations require solving complex equations for each atom, they are computationally expensive. This limits the number of atoms that can be simulated (typically a few hundred) and the timescale of the simulations (usually only picoseconds or nanoseconds). Despite these limitations, \textit{ab initio} methods are essential for understanding fundamental properties, such as how atoms bond or diffusion processes~\cite {ma2010ab}, defect modeling~\cite{varvenne2013point,zatsepin2019local}, and how electronic structures influence material behavior~\cite{hafner1988ab,qi2019effects,marzari2021electronic}.

ii) Semi-empirical potentials: These methods aim to strike a balance between accuracy and computational efficiency. Instead of calculating interactions from first principles, semi-empirical potentials use a combination of theoretical calculations and experimental data to create effective interaction models~\cite{muser2023interatomic}. One widely used example is the embedded atom model (EAM), which considers not just direct pairwise interactions but also how an atom's local environment influences its behavior~\cite{daw1984embedded,daw1993embedded}. This makes it much more accurate than simple pair potentials (see below) while being significantly faster than \textit{ab initio} methods. Because semi-empirical potentials can simulate thousands to millions of atoms over much longer timescales (microseconds or even milliseconds), they are useful for studying bulk properties addressing thermodynamics, glass forming ability, vibrational properties, and mechanical response of metallic systems. However, their accuracy depends on how well they are parameterized, meaning they may not always generalize well to new materials without careful tuning, which opens the scope for the accurate and transferable interatomic potential models designing via machine learning approaches~\cite{hernandez2019fast,wang2024machine,jacobs2025practical}.   

iii) Simpler pair potential models: The simplest way to model atomic interactions is to assume that each atom interacts with its neighbors only through basic pairwise forces. These models ignore many complexities of real materials but can still capture essential glassy behavior. Examples include the Kob-Andersen (KA) model~\cite{kob1995testing}, which was designed to mimic the structure and dynamics of metallic glasses such as ${\rm Ni}_{80}{\rm P}_{20}$, and the Wahnström model~\cite{wahnstrom1991molecular}, which represents binary Lennard-Jones systems that can approximate alloys like Mg-Zn, Cu-Zr, and Pd-Si. Because these models use simple mathematical functions (such as the Lennard-Jones potential) to describe interactions, they require very little computational power. This allows researchers to simulate extremely large systems and long timescales~\cite{LAMMPS}, making them ideal for studying fundamental aspects of glass formation, relaxation, and mechanical responses. While these models do not capture the full complexity of real metallic glasses (i.e., role of electronic or magnetic effects), they are valuable for testing theoretical ideas and exploring general trends in glassy dynamics.

The following subsections describe in detail the three categories mentioned here, reflecting the trade-off between accurately describing metallic glasses and the computational cost of larger and longer simulations.

\subsection{Electronic Structure Insights from First-Principles Methods}

Understanding the nature of metallic glasses requires careful examination of their electronic structure, as it directly influences several key material properties, including magnetic, mechanical, and thermodynamic behavior~\cite{Ma2023, Rodriguez2022, LU2022, Evertz2020}.

The most widely used method for probing electronic structure is Density Functional Theory (DFT). DFT is a quantum-mechanical approach that allows for calculation of the ground-state properties of many-electron systems, relying on electron density rather than the many-electron wavefunction. It has become a cornerstone of modern materials science due to its balance of accuracy and computational efficiency. In the study of metallic glasses, DFT has proven to accurately capture bonding, magnetism, and thermodynamic stability, and it plays an increasingly important role in guiding and accelerating experimental research.

One of the defining features of metallic glasses is their lack of long-range atomic order, which has a significant impact on their magnetic properties. Despite the structural disorder, DFT has been successfully employed to theoretically analyze these systems. For instance, the elemental contributions to magnetism can be quantified through the electronic density of states (DOS), particularly by examining the asymmetry in spin-up and spin-down electron populations~\cite{Ma2023, Rodriguez2022, LU2022}.

Attempts have also been made to link local structure with magnetic behavior. A noteworthy example is provided in Ref.~\cite{LU2022}, which reveals a correlation between the number of perfect icosahedra and the saturation magnetization in an amorphous Fe-Co-P-C alloy.

DFT has also been instrumental in studying bonding characteristics and atomic-level stability in metallic glasses. For example, bonding strength and covalent character were analyzed in the Co-B and La-B systems using DFT~\cite{Ma2023}. Similarly, Ref.~\cite{Wang2023} provided a systematic analysis of chemical bonds within stable clusters in the Zr-Cu system. These insights are not only relevant to understanding local atomic structure, but they also aid in identifying glasses with desirable mechanical properties.
A notable study by Evertz \textit{et al.}~\cite{Evertz2020} demonstrated that DFT can be used to identify damage-tolerant and stiff metallic glasses by correlating mechanical behavior with bond energy density, providing a theoretical foundation for materials design.

To study thermodynamic properties, especially glass-forming ability (GFA), \textit{ab initio} molecular dynamics (AIMD), a DFT-based simulation technique, has been widely used~\cite{shi2023, Ma2023}. AIMD is particularly valuable because it captures many-body interactions without requiring predefined interatomic potentials~\cite{Evertz2020}. However, the high computational cost of DFT imposes limitations on system size and simulation timescales. Consequently, AIMD studies often suffer from unrealistically high cooling rates, which can limit their predictive value.

\subsection{Machine-Learning Interatomic Potentials}

While density functional theory (DFT) provides highly accurate insights into bonding, electronic structure, and thermodynamics, its high computational cost limits its applicability to small system sizes and short timescales. This becomes particularly restrictive when studying disordered systems like metallic glasses, which require large-scale simulations to capture structural heterogeneity and long-timescale dynamics.

To overcome these limitations, the materials science community has increasingly adopted machine learning methods to design and train accurate \textit{machine-learning interatomic potentials} (MLIPs)~\cite{blank_neural_1995,behler_generalized_2007,bartok_gaussian_2010,thompson_spectral_2015,shapeev_moment_2016,drautz_atomic_2019,song_generalpurpose_2024}. MLIPs are trained to datasets generated by DFT calculations and can replicate DFT to high accuracy, in the best case approaching the numerical accuracy of the DFT calculations they are trained to. Once trained, MLIPs can be used in large-scale classical molecular dynamics simulations at a fraction of the computational cost of DFT. Alternatively, MLIPs can be trained and used on-the-fly to speed up and extend the time scales of \textit{ab initio} MD simulations~\cite{jinnouchi_onthefly_2019}

The key differences between MLIPs and traditional \textit{ab initio} simulation approaches or classical interatomic potentials lie in how atomic interactions are evaluated. In DFT-based simulations, forces and energies are computed from the electronic structure obtained by self-consistently solving the Kohn-Sham equations~\cite{kohn1965self} at each time step, which is computationally expensive and scales poorly with system size. In interatomic potentials the total energy is decomposed into local atomic energies, which are computed as a function of the local bonding environment. This locality assumption with a finite interaction range leads to linear scaling with system size, making large-scale simulations possible. In classical interatomic potentials, the energy function of the local atomic environment is a simple one, depending on variables such as interatomic distance, bond angles, and coordination numbers~\cite{daw_embedded-atom_1984,tersoff_new_1988,baskes_application_1987} along with a handful parameters that are fitted to material properties from experiments or \textit{ab initio} calculations. MLIPs are also local energy models, but instead of simple analytical functions they employ advanced regression models, such as artificial neural networks, linear regression, Gaussian process regression, or other forms of nonlinear regression~\cite{behler_generalized_2007,bartok_gaussian_2010,thompson_spectral_2015,drautz_atomic_2019}. Most MLIPs rely on well-designed symmetrized \textit{descriptors} that encode the local atomic environments into vectors that form the independent variables of the regression model. MLIPs are trained by optimizing the (typically large number of) parameters of the regression model to a large database of atomic structures of the given material, with corresponding total energies, forces, and possible other quantities computed by DFT or other {\it ab initio} methods.

\begin{figure}%[Htp]
    \centering
    \includegraphics[width=0.5\textwidth]{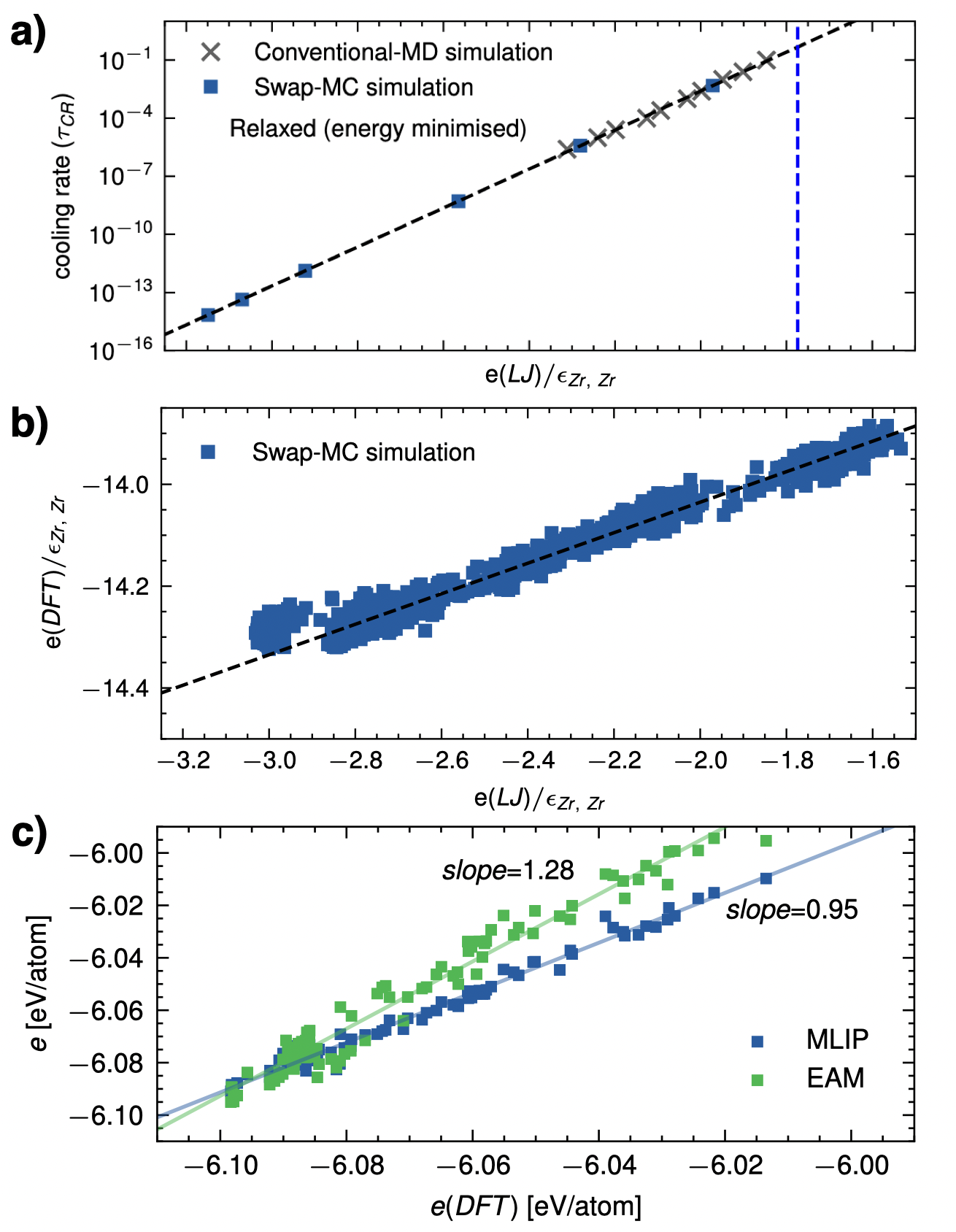}
    \caption{Supercooling extension and machine learning interatomic potential (MLIP) accuracy in CuZrAl MG~\cite{wadowski2024efficient}. 
    a) Relaxed energies per atom vs cooling rate (log scale in K/ps) for conventional MD (black crosses) and swap~MC (blue squares), spanning 14 decades. The vertical blue line marks the onset of the supercooled regime. 
    b) Swap~MC energies vs LJ surrogate parameter~$e(LJ)/\epsilon_{Zr,Zr}$. The timescales for the swap~MC are marked with the extrapolated energy-logarithmic behavior. 
    c) MLIP (blue, slope=0.95) and EAM (green, slope=1.28) energies vs DFT, showing MLIP better alignment.}
    \vspace{-0.2cm}
   \label{fig:MLIP}
\end{figure}

The development of MLIPs for metallic glasses is still emerging, but several successful examples have recently been reported. These include potentials for Pd-Cu-Ni-P~\cite{Zhao2023}, Zr--Rh~\cite{Xie2021}, and Zr--Cu~\cite{Andolina2020} systems, as well as for Zr--Cu clusters~\cite{Wang2023}.
A recent study~\cite{wadowski2024efficient} introduces a novel and efficient methodology for training MLIPs on the CuZrAl ternary system, uniquely using a Lennard-Jones surrogate model for amorphous structure, a dataset generated by swap Monte Carlo sampling (see the next subsection) across an unprecedented 14 decades of supercooling, far beyond conventional MD capabilities, followed by minimal single-point DFT corrections for accuracy.  
This surrogate-based approach significantly reduces the need for large DFT datasets, a key novelty in handling the rugged energy landscapes of disordered MGs, which require vast configurations for robust training.
%while the resulting MLIP corresponds to experimental data and EAM predictions for structural, dynamic, energetic, and mechanical properties. 
The resulting MLIP accurately reproduces experimental data (e.g., radial distribution functions, shear viscosity, elastic moduli) and improves the Embedded Atom Method (EAM) classic potential in fidelity to DFT benchmarks for dynamical (e.g., shear viscosity), energetic (e.g., supercooling energies) properties. This is illustrated in Fig.~\ref{fig:MLIP}a) that compares relaxed energies vs cooling rate (log scale in K/ps) for conventional MD (black crosses, limited to $\sim$10$^{-1}$ to 10$^{-5}$) and swap~MC (blue squares, extending to $\sim$10$^{-14}$), with a dashed line, and Fig.~\ref{fig:MLIP}b) shows the linear correlation of instantaneous liquid energies vs LJ surrogate parameter~$e(LJ)/\epsilon_{Zr,Zr}$ for efficient DFT mapping, while panel~c) plots MLIP (blue points, slope=0.95) and EAM (green points, slope=1.28) energies vs. DFT, highlighting that MLIP aligns more closely with
DFT than EAM for different supercooling.
These efforts demonstrate that MLIPs are becoming a valuable tool for bridging the gap between quantum-level accuracy and the length and time scales required to model realistic glassy systems.

\subsection{Efficient molecular simulation techniques}

This subsection mainly focuses on molecular dynamics (MD) simulations with classical potentials~\cite{frenkel2023understanding}. Although MD simulations provide valuable insight into the kinetic, thermodynamic, and mechanical behavior of materials, from an atomistic point of view, they still have limitations in terms of accessible time and length scales, compared with laboratory experiments. This is a common issue in nearly all scientific and engineering fields that use MD simulations. Yet, for metallic glasses, the timescale limitation is particularly critical, because glasses are non-equilibrium materials whose properties depend on how they are prepared, especially the cooling rate~\cite{rodney2011modeling}. Slower cooling leads to deeper annealing, resulting in more stable glasses with improved kinetic and mechanical properties.
Currently, typical molecular simulations can reach a timescale of about a microsecond, although the longest conventional MD simulations performed have been milliseconds long \cite{DAS2022,Scalliet2022}.  
In contrast, laboratory experiments can anneal glasses over much longer timescales, often around 100 seconds or more~\cite{edi96}. (Note that the conventional definition of the kinetic glass transition temperature is the temperature at which the relaxation time reaches 100 seconds.) This means that MD simulation timescales are approximately 8 orders of magnitude shorter than those in real experiments. As a result, the cooling rates used in simulations can be about $10^8$-$10^5$ times larger than in actual laboratory conditions. Hence, the kinetic, thermodynamic, and mechanical properties of glasses studied in simulations may differ significantly from those observed in real experiments~\cite{rodney2011modeling}.

To better understand this gap, we provide a pedagogical argument by translating the relaxation time into viscosity, as the two are proportional in viscoelastic materials. A material with a relaxation time of about $10^{-5}$ seconds (typical MD timescale) has a viscosity similar to soft materials, e.g., peanut butter cream~\cite{koop2011glass}. This suggests that most molecular simulations aiming to study the properties of glasses are actually examining materials with stability comparable to peanut butter cream, rather than glasses seen in various real applications.
For example, the mechanical yielding behavior of such a computational glass exhibits a ductile response, similar to soft matter systems like colloids. This contrasts sharply with real glasses (e.g., metallic glasses, silica glasses), which undergo abrupt brittle failure. 
%\sscom{However, glasses corresponding to larger relaxation times have been investigated, and indeed shown to correspond to brittle behavior.  While some of these investigations have employed brute force long MD simulations, several approaches to accelerating dynamics have been investigated, which we briefly describe here.} \MO{I agree, what I wrote is too strong. Please feel free to edit around here.}

In short, molecular dynamics simulations of glasses face a major challenge due to their limited timescale, which significantly affects the properties they can accurately capture.
The glass formation process involves extremely long timescales due to the dramatic slowing down of relaxation dynamics. This very nature of glasses makes simulations difficult and complicates direct comparisons with experiments.
To address this fundamental problem, various approaches and computational algorithms have been proposed.

One approach is to use graphics processing units (GPUs) for molecular dynamics simulations with carefully optimized parallelization. Compared to conventional CPU-based simulations, GPU-accelerated simulations are much faster. Typically, they achieve a speedup of one to two orders of magnitude compared with standard CPU simulations, allowing for correspondingly slower cooling rates or longer simulation timescales~\cite{bailey2017rumd,cos18a}.

Another approach to tackle the timescale problem involves accelerating the relaxation process in simulations. This is typically done using Monte Carlo (MC) simulations~\cite{frenkel2023understanding}, which aim to sample configurations in thermal equilibrium under a given condition. This approach takes advantage of the fact that Monte Carlo algorithms allow the use of unphysical dynamics (say, non-Newtonian dynamics), meaning that the system does not have to evolve in real physical time. Carefully designed unphysical dynamics (or Monte Carlo updates) can accelerate relaxation, leading to more annealed and aged glassy configurations.
By construction, Monte Carlo simulations are mainly aimed at preparing stable, well-annealed configurations, rather than directly studying the dynamical and mechanical properties of the system.
Once stable configurations are generated, various other simulation techniques, such as molecular dynamics (MD) with physical dynamics, are employed to investigate the system’s properties~\cite{ozawa2018random,sca22}.

A well-known example is the replica exchange method (also known as parallel tempering)~\cite{marinari1992simulated,hukushima1996exchange,sugita1999replica,yamamoto2000replica,earl2005parallel}. In this method, multiple replicas (or copies) of the system are simulated at different temperatures simultaneously. These copies are then exchanged using a Monte Carlo rule to ensure that they satisfy thermal equilibrium conditions. This effectively creates a random walk in temperature space, allowing the system to overcome large energy barriers through these temperature swaps. 
The design of this algorithm is particularly well-suited for glassy systems, where relaxation is extremely slow due to a rugged energy landscape, a feature also found in spin glasses, where this algorithm was originally developed. 
Indeed, replica exchange algorithms can accelerate simulations by reducing an effective relaxation time~\cite{jung2024normalizing}, enabling the generation of more annealed glass samples.
However, despite this improvement, a huge timescale gap between experiments and simulations still remains.

\begin{figure}[ht]
    \centering
    \includegraphics[width=0.4\textwidth]{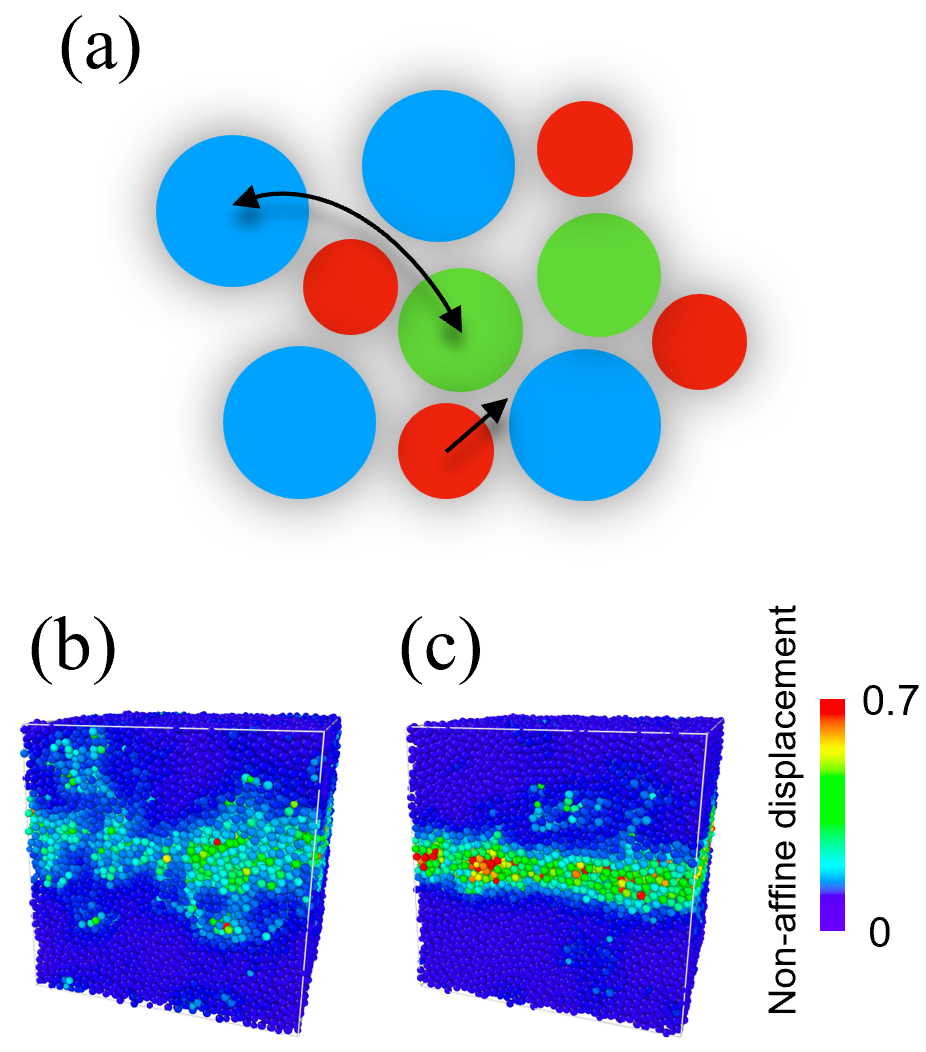}
    \caption{Swap Monte Carlo method. 
    a) Schematic illustration of the swap Monte Carlo algorithm. The straight arrow (pointing in one direction) represents a standard random Monte Carlo translational update. The curved arrows connecting two particles indicate a swap Monte Carlo update. 
    b) Non-affine displacement field  for a computational glass annealed using a standard simulation of the original Kob-Andersen mixture.  
    c) Non-affine displacement field for a glass annealed using the swap Monte Carlo algorithm applied to a Kob-Andersen mixture doped with $1\%$ medium-sized particles. (From Ref.~\cite{parmar2020ultrastable}). 
    %\CHR{Licensing info?}
    }
    \label{fig:MTGswap}
\end{figure}

Recently, particle-swap Monte Carlo simulations have gained a lot of attention as they significantly accelerate simulations for sampling, helping to bridge the gap between experiments and simulations for some model systems~\cite{ninarello2017models}.  
The swap Monte Carlo algorithm is conceptually simple. In addition to the standard random particle displacement used in conventional MC simulations, one performs an additional non-local move associated with swapping sizes of two randomly chosen particles (this may equivalent to swapping species or positions in some settings). This process is illustrated schematically in Fig.~\ref{fig:MTGswap}(a). Although the algorithm itself has a long history~\cite{tsai1978structure,gazzillo1989equation,grigera2001fast,brumer2004numerical,gutierrez2015static}, the seminal work by Ninarello \textit{et al.}~\cite{ninarello2017models} has shown that optimizing both the model and the algorithm can speed up equilibration in a class of systems by more than 10 orders of magnitude compared to conventional MC methods. This discovery has been a major breakthrough in the glass physics community~\cite{guiselin2022microscopic,scalliet2022thirty,parmar2023depleting}.  

However, this swap Monte Carlo algorithm is not generically applicable to all molecular simulation model glasses. The reason is that swapping species may lead to a large energy cost, making swap moves unlikely to be accepted (imagine swapping very big and very small particles). The first significant breakthrough was achieved for models with size polydispersity~\cite{ninarello2017models}, where particle diameters are continuously distributed. In such models, swap moves are frequently accepted. 
%Such models are particularly relevant for colloidal glasses.  

More recently, this method has been extended to discrete mixtures~\cite{parmar2020ultrastable,parmar2020stable}, which are more relevant to metallic glasses. A key innovation in this approach is the addition of medium-sized particles, which facilitates indirect swapping between larger and smaller particles. This is illustrated in Fig.~\ref{fig:MTGswap}(a). Direct swapping between larger and smaller particles is prohibited due to the very high energy cost, as mentioned above. Yet, medium-sized particles can swap with both larger and smaller particles. This enables an indirect swapping mechanism, where a sequential swap process involving medium-sized particles effectively allows the system to exchange larger and smaller particles. The efficiency of the swap Monte Carlo algorithm improves as the concentration of the medium-size particles increases~\cite{parmar2020ultrastable}.  
Moreover, the addition of medium-sized particles can be seen as a doping manipulation, similar to experimental techniques often used to tune and optimize material properties. 
For example, in the Kob-Andersen (KA) model, doping with a small amount of medium-sized particles, e.g., $1\%$ and $10\%$, resulted in simulation speedups (in terms of CPU time) of 3 and 8 orders of magnitude, respectively~\cite{parmar2020ultrastable}.
Figures~\ref{fig:MTGswap}b),c) compare the mechanical response under uniform shear deformation for computational glasses annealed using the standard Monte Carlo algorithm (which includes only translational updates) for the Kob-Andersen mixture b) and those annealed using the swap Monte Carlo algorithm for the Kob-Andersen mixture doped with $1\%$ medium-sized particles c)~\cite{parmar2020ultrastable}.  
The former exhibits gradual deformation with a diffuse shear localization, resembling ductile materials such as colloids. In contrast, the latter shows an abrupt emergence of a sharp, system-spanning shear band, similar to metallic glasses observed in laboratory experiments.

\subsection{Plasticity, shearband formation}

The mechanical properties of metallic glasses from an atomistic perspective have been extensively investigated using molecular dynamics (MD) simulations. In typical protocols, gradually cooled glass samples are subjected to external shear deformation under periodic boundary conditions, mimicking bulk samples in experiments~\cite{rodney2011modeling}. The simulated systems usually contain up to about 100,000 particles, which is still much smaller than experimental scales. Nevertheless, MD simulations provide atomistic-level insights into how metallic glasses respond to external loading and how local structures determined by interaction potentials and chemical composition influence bulk properties. In addition, several studies have employed open-boundary simulations to model axial compression~\cite{jin2023general} and nanoindentation protocols~\cite{adjaoud2021nanoindentation}, relevant to nanoscale experimental conditions.

Besides, the flexibility of molecular dynamics simulations makes it possible to investigate more complex processes, such as exploring the underlying energy landscape and its connection to elementary thermo-mechanical processes~\cite{sas98a,rodney2011modeling,fan2014thermally,fan17,zella2024ripples}. This perspective provides a unifying framework to link atomistic rearrangements with macroscopic mechanical response.

Despite their limited size and timescale, MD simulations have revealed that plastic deformation in metallic glasses arises through localized irreversible rearrangements that accommodate external stresses~\cite{ding2014soft}. These localized plastic events generate long-range elastic stress fields, which may be evaluated within the framework of linear elasticity theory~\cite{esh57}. These long-range stress fields can in turn trigger further plastic events. Such cascades of correlated rearrangements form avalanches~\cite{mal06}. Under sufficiently large loading, an avalanche can span the entire system, leading to macroscopic failure, commonly referred to as yielding~\cite{shimizu2006yield,karmakar2010statistical}. 
Yielding in metallic glasses is often accompanied by system-spanning shear localization, known as a shear band~\cite{shi05,shimizu2007theory}. Understanding how microscopic properties, including local structural features, influence not only localized plasticity but also the initiation and growth of bulk shear bands is therefore a subject of paramount importance.
Remarkably, the above mentioned phenomenologies are robust against variations in composition and interaction potentials. Specific model details primarily influence quantitative aspects, such as elastic constants and yield stress values, rather than the qualitative deformation mechanisms, which will be discussed further in Sec~\ref{sec:multiscale_modeling}.
Key aspects that will be discussed in the following include the dependence on glass preparation protocols, the identification of structurally soft regions (“soft spots”), the mechanism of the shear band formation, and the influence of loading conditions.

{\it Glass preparation protocol dependence}: A key and often undesirable feature of metallic glasses is their brittleness, i.e., the tendency of the material to fail abruptly under load~\cite{richard2021brittle}. Such behavior severely limits practical applications. Understanding the mechanisms that control the transition from brittle to ductile yielding therefore represents a major challenge for wider community~\cite{divoux2024ductile}. In particular, the influence of annealing and cooling rate on mechanical response has been a subject of central interest~\cite{utz2000atomistic,shi05,fan2017effects,ozawa2018random,ozawa2020role,BhaumikPNAS2021,YehPRL2020}. Most conventional MD simulations, however, tend to exhibit predominantly ductile yielding, largely due to the limited accessible timescales for annealing. Recent advances in simulation algorithms, such as swap Monte Carlo, now enable the preparation of glasses over a much broader range of annealing conditions, thereby covering both brittle and ductile regimes~\cite{ozawa2018random,jin2018stability,kapteijns2019fast,ozawa2020role,lamp2022brittle,kaskosz2023efficient} (see also Fig.~\ref{fig:MTGswap}). Remarkably, a brittle-to-ductile transition can be induced solely by varying the annealing protocol for a given material, while keeping composition and interaction potential fixed. This finding highlights the crucial role of local structural packing efficiency, which is directly controlled by the annealing process.
Recent studies have shown that slow oscillatory (or cyclic) shear deformation with smaller amplitudes can induce an annealing effect, even under conditions where thermal contributions are negligible, while large-amplitude oscillations lead instead to rejuvenation~\cite{leishangthem2017yielding,parmar2019strain} (see also below).
The degree of annealing across which the character of yielding changes, in the case of the cyclic shear protocol, has been argued to be associated with a change in the character of the energy landscape \cite{BhaumikPNAS2021}. This threshold energy level, marked by the vanishing of saddle points in the energy landscape, and in turn associated with the mode coupling crossover \cite{leonardoPRL,broderixPRL00}, suggests an interesting connection between yielding phenomenology and relaxation in the supercooled liquid state.

{\it Identification of soft spots}: Unlike crystalline materials, where defects can be identified unambiguously, glasses generally lack obvious structural inhomogeneities (except in partially crystallized samples). As a result, it is not straightforward to determine, from static snapshots alone, where plastic activity will occur under loading. Nevertheless, extensive studies over the past decades have demonstrated the existence of structurally “soft” spots~\cite{manning2011vibrational,din14}, where plastic rearrangements are most likely to take place.
A variety of approaches have been proposed to identify such regions \cite{Richard_PhysRevMaterials.4.113609}, including vibrational mode analysis~\cite{manning2011vibrational,jakse2012structural}, non-linear excitation protocols~\cite{ler21}, local packing \cite{TongPRX2018},
free-volume-like approaches~\cite{ding2016universal}, local yield threshold~\cite{bar18}, topological defect detection~\cite{wu2023topology}, and more recently, machine-learning-based methods~\cite{cubuk2015identifying,yang2021machine,ciarella2023finding}. 
A structural metric, termed softness, has been evaluated both using machine learning approaches \cite{cubuk2015identifying} and directly from the pair correlation function \cite{Nandi2021,Sahu2024}.
By construction, soft spots or soft regions are those most prone to undergo local deformation in the near future. Accordingly, the predictive power of different approaches for identifying soft spots has been systematically evaluated, often through correlation metrics with subsequent plastic activity~\cite{Richard_PhysRevMaterials.4.113609}. 

Closely related, but perhaps the complementary perspective, is the identification of recurrent local packing motifs, known as locally favored structures (LFS)~\cite{cos07,roy15,tan19}. These structures, such as efficiently packed icosahedral clusters, are typically more stable and less mobile under cooling or external loading~\cite{din14,pasturel2017atomic}. The prevalence and nature of LFS depend strongly on chemical composition and interaction potentials. An important direction for current research lies in connecting LFS (or conversely, locally unfavored structures) with other soft-spot identification methods~\cite{ding2014soft}, especially since LFS can now be directly probed in experiments~\cite{Yang_Nature2021}. Equally important will be efforts in relating the findings from ML and analysis of LFS to theoretically derived softness metrics \cite{Nandi2021}.

{\it Shear band formation mechanism}:
As discussed above, yielding in metallic glasses is typically accompanied by the formation of a system-spanning shear band~\cite{gre13}. At the microscopic level, amorphous materials respond to external loading through localized plastic events. These events redistribute stress, which can trigger further events in an avalanche-like manner. Shear banding can thus be viewed as an avalanche process with strong directional localization.
Molecular dynamics simulations have provided detailed insight into this process, in particular into how initial plastic activity, often concentrated in structurally soft regions (sometimes referred to as shear-band embryos), develops into a fully formed shear band~\cite{csopu2017atomic,hassani2019probing,ozawa2022rare}. The nucleation and subsequent propagation of these embryos into system-spanning shear bands can be directly visualized in simulations. 
It has been revealed that system-spanning shear bands emerge through the alignment of Eshelby-like quadrupolar stress fields~\cite{dasgupta2013yield,hieronymus2017shear}. Local plastic rearrangements generate such fields, and their repeated propagation along a preferred direction results in the accumulation of large displacements at macroscopic scales~\cite{ozawa2022rare}.

At the level of localized plastic events, both expert-designed structural descriptors and machine-learning models have been employed to predict where rearrangements will occur, as we discussed in detail above. Extending this challenge, predicting the nucleation sites and growth of shear bands has also been attempted, with some success achieved in a statistical sense~\cite{barbot2020rejuvenation,fan2022predicting}. Understanding shear-band formation from a microscopic perspective is expected to provide important clues for controlling their development in real materials and, ultimately, for tuning the ductility of bulk metallic glasses.

{\it Loading condition dependence}: The majority of MD studies investigate strain-controlled, uniformly sheared systems. Because the shear rates in MD are typically much higher than in laboratory experiments~\cite{rodney2011modeling}, athermal quasi-static (AQS) simulations~\cite{mal06} are often employed to better approximate experimentally relevant conditions. Importantly, yielding behavior is not determined solely by material properties but is also strongly influenced by the choice of loading protocol.
For instance, in strain-controlled simulations, increasing the strain rate often leads to more ductile-like mechanical responses, since rapid loading can suppress system-spanning localization, even in very well-annealed glasses~\cite{singh2020brittle,lamp2022brittle}.
In addition, stress-controlled simulations provide direct access to creep deformation~\cite{cabriolu2019precursors,dutta2023creep,chaudhuri2025athermal}, which is highly relevant for practical applications. As we have mentioned briefly, cyclic shear (as well as other related) protocols have been extensively explored~\cite{fiocco_oscillatory_2013,Regev2013,Priezjev2013,leishangthem2017yielding,kawasakiPRE16,parmar2019strain,YehPRL2020,BhaumikPNAS2021,Bhowmik_2022,
PRIEZJEV2023112230,maity2024fatigue}. As noted earlier, the degree of annealing in glasses plays a critical role in determining the nature of yielding. Beyond a certain threshold in energy, the yielding response undergoes a qualitative change, switching from brittle-like to ductile-like behavior~\cite{BhaumikPNAS2021,YehPRL2020}. Under cyclic shear, yielding is always associated with the formation of shear bands, unlike uniform shear protocols \cite{parmar2019strain,BhaumikPNAS2021,maity2024fatigue}. These deformation protocols enable the study of fatigue failure under oscillatory loading \cite{Bhowmik_2022,PRIEZJEV2023112230,maity2024fatigue}, which are directly connected to the reliability and long-term performance of metallic glass-based products. Intriguing results from recent simulations \cite{Bhowmik_2022,PRIEZJEV2023112230,maity2024fatigue} suggest the possibility of predicting failure times from monitoring dissipated work (as well as other measures of plasticity), which may be of great value in assessing performance of metallic glass-based products.

\section{Multi-scale modeling}
\label{sec:multiscale_modeling}

\subsection{Mesoscale modeling}

Mesoscale modeling of metallic glasses has relied primarily on lattice based models in which deformation is assumed to happen at evenly distributed sites based on an underlying deterministic or statistical description. A general and comprehensive review was recently published in this area \cite{Ferrero2018}, and so our goal here is to highlight subsequent significant advances. There have emerged two main research priorities driving the development and analysis of mesoscale models of glasses: better capturing the transition from quasi-elastic behavior to steady-state flow, and creating more sophisticated models that incorporate structural information and the thermal history of the material. Both of these are crucial for mesoscale models to accurately predict the mechanical behavior of the material during transient deformation, before steady state flow begins. While efforts to address these gaps typically focus on using structural information to inform predictions of materials' response at an atomistic scale, the way such information is incorporated has been largely distinct between the communities that focus on these two aspects of the problem.

Mesoscale models have provided insights toward understanding and predicting the transition point between ductility and brittle catastrophic failure in MGs. Atomistic simulations have also extensively been deployed to study the effect of strain rate and glass preparation on these behaviors \cite{shi2005strain,shi2006atomic,singh2020brittle}. Well-annealed materials often undergo abrupt failure with a pronounced stress drop accompanying shear banding, while poorly annealed materials undergo a smooth crossover. This transition in response is strain rate dependent. From analyses of mesoscale elastoplastic models, it has been hypothesized that the transition in yielding behavior is governed by an underlying critical point~\cite{ozawa2018random,rossi2022finite}. Their work suggests that the critical point is related to the universality class of a random field Ising model, which suggests that the difference between the two yielding behaviors constitutes a true phase transition. 
Both atomistic simulations and mesoscale elastoplastic models, ranging from analytically solvable mean-field theories to finite-dimensional implementations~\cite{ozawa2018random,ozawa2020role,rossi2022finite,rossi2022emergence,rossi2023far,parley2024ductile,mutneja2025finite}, have been employed to support this conclusion, although the very existence of such a critical point is still debated ~\cite{Barlow2020,richard2021finite,pollard2022yielding}.

Barlow \textit{et al.}~\cite{Barlow2020} describe the yielding behavior of amorphous materials subjected to shear forces as a function of increasing levels of initial sample annealing prior to shearing. In the systems subjected to annealing at elevated temperature for extended periods, they see a progression from smooth ductile yielding to abrupt brittle yielding which is strongly shear banded. This work appears to predict smooth transitions from brittle to ductile behaviors in contrast to mean field calculations \cite{ozawa2018random, Popovic2018,rossi2022emergence,parley2024ductile} that suggest that a random critical point and a sharp transition. This difference might arise from finite size effects of the simulations, or fundamental differences between the models pointing toward the importance of bridging the gaps between model scales and better understanding the limits of mean-field models~\cite{richard2021finite,rossi2022finite,pollard2022yielding}.

In the case of cyclic shear deformation, a qualitative change in yielding behavior has been observed in MD simulations \cite{BhaumikPNAS2021,YehPRL2020}, as a function of the degree of annealing. However, a key difference is that under cyclic shear, shear band formation always accompanies yielding. Although the shear band width is small for poorly annealed glasses, it remains finite \cite{BhaumikPNAS2021} even when the system size increases, scaling with the system size \cite{BhaumikShearband}. Several aspects of the cyclic shear yielding behavior are captured by theoretical models including computationally investigated elasto-plastic models \cite{sastry_models_2021,mungan_metastability_2021,parley_mean-field_2022,debargha_meso2,liu_fate_2022,Kumar2022,PushkarEPM}. Nevertheless, statements regarding the finiteness of the shear band width, and its significance, are not fully clear. Whereas analytical results ({\it e.g.}, \cite{parley_mean-field_2022}) seem to suggest that the shear band width goes to zero at the threshold degree of annealing, computational results \cite{liu_fate_2022,PushkarEPM} indicate otherwise.  

Transition graph theory is deployed using elastoplastic modeling in Ref.~\cite{Kumar2022} to obtain additional insights regarding yielding under cyclic loading, by determining the number of cycles that separate distinct structural states.
The authors characterize plastic strain as a function of quench to determine the conditions upon which cyclic shear can revert the system to a prior state. Once this is no longer the case the system is said to have undergone an irreversibility transition. This transition depends on system size and the prior aging of glass.

Elasto-plastic models also show promise for describing the behavior of MGs in the transient region before the onset of steady-state flow. 
Liu \textit{et al.}~\cite{Liu2018} deploy mesoscale elasto-plastic models to describe the transient dynamics of amorphous materials in athermal regimes. This study provides insight into the nature of the slowing (creep) and speeding up (fluidization) of the strain rate. 
%Fluidization rate is observed to increase as a power-law of the externally applied stress. 
Within the framework of the models two time scales arise from different underlying physical processes: the stress distribution around the marginal stability threshold, the nature of the subsequent plastic activations, and the resulting spatial cooperativity of the plastic events.
Castellanos \textit{et al.} \cite{Castellanos2022} extend elasto-plastic models of glasses beyond capturing the physics of stationary flow states to describe the mechanical response in the transient regime for a wide range of initial system and degrees of stability. The model is calibrated from local atomistic data extracted from glasses produced at various quench rates via simulation. The evolution of slip thresholds are based on a statistical model of local plastic strain increments.

\begin{figure}[h]
	\centering
	\includegraphics[width=0.5\textwidth]{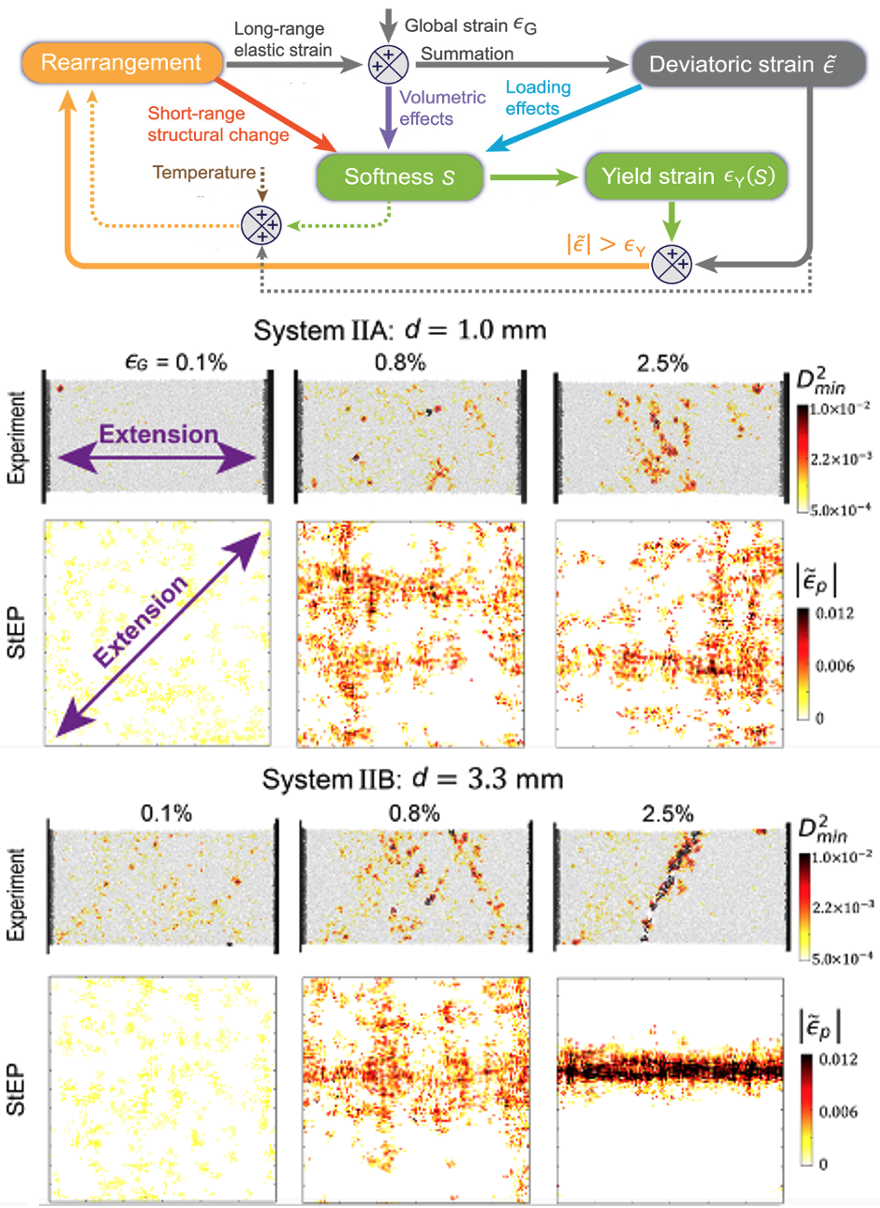}
 \caption{At top, a schematic illustration of the Structuro-elasto-plasticity (StEP) model \cite{Zhang2022}, adapted from Ref.~\cite{xiao2023identifying}. Below, the particle evolution of an experimental granular raft, consisting of a 2D monolayer of polydisperse Styrofoam spheres floating at an air-oil interface under a quasistatic tensile test, was investigated. Here, $\epsilon_G$ is the driving global strain, $\tilde{\epsilon}_P$ is the plastic strain, and $D^2_\text{min}$ is the particle non-affine displacement. System IIA represents a more ductile system and System IIB a more brittle one. Both cases were compared with the corresponding StEP model simulations. In both cases, the simulations showed statistical agreement with the experimental results. 
 %\CHR{Licensing info?}
 \label{figure1}
 }
\end{figure}

Establishing a clear link between the many parameters in constitutive models to the underlying physics is a grand challenge in the field of glass mechanics. Xiao \textit{et al.} \cite{Zhang2022} have deployed the structural quantity of softness, which is machine-learned and correlates with imminent plastic rearrangement, as a framework with which to quantify glass structure. Softness has been shown to correlate with local yield strain (see Fig.~\ref{figure1}). Within this framework a structuro-elasto-plastic model was developed that reproduces particle simulation results well. The connection to underlying atomistic structural information of the material helps bridge the gap between atomistic modeling and mesoscale modeling.

Recent work by Xu \textit{et al.}~\cite{Xu} presents a direct-from-data modeling approach for elasto-plastic models. Instead of fitting parameters to atomistic data to create a meso-scale model, this approach harvests a large statistical database of elastoplastic events from glasses with different thermal histories. This database is utilized to construct a stochastic representation in two state variables: the shear stress and the intrinsic potential energy that encodes the stored energy of cold work. This meso-scale model can then be interpreted as a dynamical map that shows how an MG with a given thermal history and will traverse the state space. This map provides a window into the dynamical state of the glass through the deformation process, from the initial elastic or anelastic behaviors, to yielding, and finally to steady-state flow. This novel approach to meso-scale modeling opens a new pathway to generating descriptive models of materials by taking advantage of the large amount of data that can be generated with simulation and high-performance computing.

%There has been substantial additional work in the area of elastoplastic modeling that is relevant, but which we cannot review in detail here \cite{Ferrero2019, Tyukodi2019, Tyukodi2023,Castellanos2021, Castellanos2022, LeGoff2021, Pollard2022, Aguirre2018, Tanguy2021, Loock2021}. \CHR{I suggest revising this.  Any author of these works would likely see their work viewed as an afterthought. I suggest either expanding this out, moving the citations elsewhere, or deleting entirely.}

\subsection{Continuum modeling}

Continuum models of MG have primarily been built on two different plasticity theories, the free volume theory and the shear transformation zone (STZ) theory. 
Free volume purports to quantify the excess space among atoms that allows atomic movement in a material and is closely related to density. Spaepen~\cite{SPAEPEN1977407} introduced the concept of free volume as a fundamental aspect of flow defects, emphasizing its significance in enabling atomic diffusion and plastic shear flow. Cohen and Turnbull~\cite{Turnbull1961,Turnbull1970} assumes that free volume is a natural descriptor of the structure of metallic glasses, and use it to develop a flow defect theory. Johnson, Lu, and Demetriou~\cite{JOHNSON20021039} extended the free volume-based theory to describe deformation and flow at high homologous temperatures. Free-volume based theories have turned out to be useful despite the challenge of rigorously defining free volume.

Building on the theories, Anand and Su~\cite{ANAND20051362,ANAND20073735} formulated a free volume based constitutive model for BMGs based on the Mohr-Coulomb theory. The model takes into account both shearing and dilatation mechanism for plastic flow. The model has been successful in accurately capturing the deformation of BMGs in different loading scenarios, such as indentation and strip bending. The original work uses explicit time stepping scheme, where the time step size is restrictively small. Tandaiya \textit{et al.}~\cite{2011Tandaiya} reformulated and implemented their model with an implicit backward Euler numerical scheme, enabling efficient quasi-static simulations over a long duration. More recently, Kamble \textit{et al.}~\cite{KAMBLE2024103915} extended Anand and Su's work~\cite{ANAND20051362} to model the split Hopkinson pressure bar (SHPB) setup, capable of simulating the deformation of BMGs at high strain rates and elevated temperatures. 

In contrast to free volume theory, STZ theory assumes that a population of bistable defects called STZs exist in an otherwise elastic material, which are localized regions susceptible to configurational changes. Argon~\cite{ARGON197947} proposed the existence of shear transformations in amorphous solids, drawing an analogy to dislocation glide in crystalline materials. Falk and Langer~\cite{falk1998dynamics,Falk2011Chi} introduced the concept that plastic flow is related to an underlying and evolving defect population susceptible to such transformations. The STZ population is deployed to describe the structural state of the system, incorporating the orientation and number density of defects as key parameters. Further developments by Bouchbinder and Langer~\cite{Bouchbinder2009} and Kamrin and Bouchbinder~\cite{Kamrin2014thermo} have placed the STZ theory on a thermodynamic foundation, allowing a formal definition of an effective temperature of the structural degrees of freedom. This thermodynamic approach allows for a more comprehensive modeling of the atomic kinetics and plastic deformation in BMGs, linking microscopic mechanisms to macroscopic observable. 

This framework has been implemented numerically by Rycroft \textit{et al.}~\cite{Rycroft2015BMG2D} who developed a general framework for simulating elastoplastic materials in quasi-static loading scenarios over long duration. Within this framework the effective temperature STZ model can be efficiently solved using a numerical projection method. Employing the effective temperature STZ theory as the plasticity model, the model successfully captures the shear banding behaviors of MGs. It has been further used to systematically study fracture toughness of glasses~\cite{Vasoya2016}, and has been extended to a 3D implementation to study shear band features uniquely in 3D~\cite{BOFFI2020_3D}.

%\subsection{Recent trends} 

%\subsubsection{Traditional continuum modeling}

By incorporating more thermodynamic degrees of freedom in modeling plastic deformation Rao \textit{et al.}~\cite{RAO2022103309} developed a finite deformation constitutive model within the new framework of irreversible non-equilibrium thermodynamics.  Based on the two-temperature continuum thermodynamics theory developed by Kamrin and Bouchbinder~\cite{Kamrin2014thermo}, they consider an energy exchange between the configurational and kinetic subsystems, which drives the subsystems towards equilibrium. Furthermore, by introducing a local transformation energy, their work considers not only shear transformation, but also reverse shear transformation, caused by the relaxation of a strain energy field accumulated due to shear transformation. 

Advances to this modeling schema has also been sought by incorporating additional thermally activated mechanisms into the continuum theory to account for material evolution and interactions between plastic events.
Zhu \textit{et al.}~\cite{ZHU2021104216} noted that the established continuum models have inadequately addressed time-dependent plastic deformation, and cannot accurately model the creep or relaxation behaviors of BMGs. They developed a new free-volume based chemo-mechanical continuum model based on the laws of thermodynamics. This model considers atomic motion, and directly relates plastic deformation to atomic kinetics by linking atomic concentration (free volume) and stress state to the local chemical potential. The gradient of the local chemical potential causes atomic flux, which induces plastic deformation.  The gradient of the local chemical potential is used directly in the calculation of the plastic flow magnitude of the material. The model is further verified with numerical simulations, including the creep of a BMG. The model is able to capture stress dependence of creep behavior, and a critical stress transitioning from stable to unstable creep. 

Mo \textit{et al.}~\cite{MO2023103673} focused on the non-local interactions among neighboring plastic events. The medium-long-range (MLR) interaction of structures due to plastic events is crucial for understanding avalanches and failures in BMGs. A plastic event at a site can create a surrounding stress-impacted zone on a micron scale. However, the MLR correlation among plastic events have not been directly formulated in established continuum mean field theories, including the free volume theory and the STZ theory. Specifically, the evolution of local state variables is not directly related to neighboring elements. Mo \textit{et al.} introduced a general MLR correlation mechanism into the free volume theory. At a site, the free volume increment includes not only the structural deformation by thermal mechanical coupling, but also the diffusion to its neighboring regions that have lower free volume values, and the equal amount of annihilation of free volume itself. This introduces a direct non-local effect between plastic events. Compared to the traditional free volume simulations, the new model can capture the self-organized criticality of BMGs and self-adaptive evolution of shear bands.

\subsection{Linking microscopic/mesoscopic simulations and macroscopic continuum models}
There have recently emerged significant advancements in multi-scale modeling, particularly in linking microscopic/mesoscopic simulations and macroscopic continuum models. The main effort lies towards using microscopic and mesoscopic simulations to inform and improve plasticity modeling in macroscopic continuum models. The ultimate objective is to combine physical accuracy of microscopic models with the scalability of continuum models, enabling the simulation of larger systems over longer durations. 

In multi-scale modeling of BMGs, a notable challenge arises in accurately aligning atomic configurations obtained from molecular dynamics (MD) simulations with macroscopic descriptors used in continuum simulations. To address this challenge, Hinkle \textit{et al.}~\cite{Hinkle2017} developed a coarse-graining methodology to translate MD simulation data into alignment with continuum simulation data. Specifically, Gaussian coarse-graining techniques are applied to atomic strain fields and atomic potential energy. Analyzing the coarse-grained fields allows for the extraction of the shear-banding region and the background region in the simulation, as well as the appropriate length scales to use for coarse-graining. They further developed an equation to map the coarse-grained atomic potential energy in MD to the effective temperature field in the STZ theory for continuum simulations. By employing such techniques, researchers can better integrate information from different simulation scales. For example, Hinkle \textit{et al.} employed the coarse-graining techniques to a MD simulation, extracting an initial condition for the effective temperature field suitable for continuum simulations. They conducted two-dimensional, quasi-static numerical simulations based on the STZ theory using the initial condition. Their findings demonstrated that, at certain coarse-graining length scales, the continuum simulation reached good agreement with the MD simulation. The work paved the way for future research in using the coarse-grained atomistic data to parametrize and validate continuum plasticity models. 

Another challenge in multi-scale modeling arises from the disparity in boundary conditions between microscopic MD simulations and continuum simulations. The boundary conditions employed in MD simulations, such as the widely used Lees-Edwards boundary condition, differ from those typically utilized in continuum simulations. Boffi and Rycroft~\cite{boffi20} proposed a coordinate transformation methodology that enables the implementation of consistent boundary conditions for continuum simulations that mirror the Lees-Edwards boundary condition in MD simulations. The work has laid the groundwork for future research in the direct comparison and integration between the two simulation scales.

\subsection{Data-driven approaches}
%Furthermore, we have observed increasing efforts in utilizing data-driven approaches in modeling, particularly machine learning (ML) techniques. 
Data-driven methods have the potential to bypass several limitations present in traditional continuum modeling techniques, leading to more accurate modeling of the plasticity. For example, BMGs exhibit complex plastic deformation behavior. Their plasticity is also influenced by different factors, such as microstructure, composition, and processing history. Traditional continuum models often struggle to capture these complexities accurately. Data-driven methods, however, can learn this complexity from large datasets, and capture the non-linear and non-local effects in plastic deformation.

In one direction, these data-driven methods can serve as powerful tools to inform existing plasticity models. For example, parameters in theoretical models can be learned and fitted based on computational or experimental data, enabling more accurate and realistic predictions of material behavior. This is a broader trend in the materials science field. Acar~\cite{Pinar2020} used an artificial neural network to calibrate a crystal plasticity model to simulate the mechanical behavior of a titanium–aluminum alloy under larger deformations. Baltic \textit{et al.}~\cite{BALTIC2021109604} proposed a machine learning supported calibration of a ductile fracture locus model for ductile materials. Yang \textit{et al.}~\cite{met13010166} used a Bayesian neural network-based surrogate-assisted genetic algorithm optimization method to calibrate constitutive parameters for crystal plasticity model that best reproduce the mechanical response in experiments and simulations. 

For BMGs, building upon the groundwork laid by the coarse-graining and coordinate transformation methodologies, Kontolati \textit{et al.}~\cite{KONTOLATI2021117008} applied machine learning (ML) techniques to further investigate multi-scale modeling of BMGs. In particular, they introduced Grassmannian efficient global optimization (Grassmannian EGO), a general ML framework, to probabilistically determine optimal parameters for the STZ plasticity model based on MD atomistic simulation data. Employing the coordinate-transformation methodology, they ensured exact matching of boundary conditions between continuum and MD simulations. Additionally, they utilized the coarse-graining methodology to directly compare fields from MD simulations with those from continuum simulations. Continuum simulations were initialized with the effective temperature field extracted from MD simulations. Through the Grassmannian EGO framework, trained on MD simulation data, they identified optimal parameters for the coarse-graining length scale and the STZ model. These optimal parameters resulted in good agreement between MD and continuum simulations, regarding the development and location of the shear bands.

ML techniques can also be used to create entirely new, data-driven plasticity models.
%In another direction, other than using ML techniques to inform existing plasticity models of BMGs, they can also be used to create entirely new, data-driven plasticity models.
This is an important current trend in materials science research. Deep learning models have been used to predict plasticity in materials other than BMGs. For instance, Mozaffar \textit{et al.} \cite{Mozaffar2019} developed a deep learning algorithm that was capable of predicting plasticity in simulated representative volume elements of aluminum alloys with elliptical rubber fillers. Huang \textit{et al.}\cite{Huang2020} developed a feedforward neural network that is capable of recreating the Von Mises yield surface for 3D materials, based solely on data collected from multi-axial loading tests. Nascimento \textit{et al.}~\cite{NASCIMENTO2023103507} developed a deep-neural-network-based surrogate model for predicting the anisotropic yield surfaces of polycrystalline materials suited for sheet metal forming.

For BMGs, recently, Wen and Wei~\cite{WEN2024105629} developed ML models for plastic flow based on experimental data across a wide range of temperatures and strain rates~\cite{LU20033429}. Using the gradient boosting regression method~\cite{Natekin2013,Friedman2000} within the tree-based algorithm, their plasticity model learns complex mappings between input parameters (stress, plastic deformation history, temperature, and strain rate) and the output variable, equivalent plastic strain rate. 
The ML plasticity model is integrated into continuum constitutive models. By refining the simulation mesh, they showed that the model has the capability to resolve nanoscale-sized shear bands.
While the above approaches focus on data-driven constitutive learning, Mäkinen \textit{et al.}~\cite{makinen2025growth} proposed a physics-informed framework that predicts plastic deformation and yield in metallic glasses from early plastic strain accumulation. Their Bayesian approach enables early and interpretable prediction of bulk plastic response from small-strain data. These complementary approaches underscore the emerging role of hybrid, interpretable frameworks for predictive modeling of plasticity and failure in metallic glasses.

\section{Machine Learning Prediction of Glass-Forming Ability}
\label{sec:ML_GFA}
%Machine learning (ML) is playing an increasingly important role in materials science, with applications ranging from phase prediction to the inverse design of complex materials. In the context of MGs, ML methods have been adapted to address both general challenges in glass science and specific issues related to multi-element metallic systems, often involving a mix of metallic and non-metallic elements (e.g., phosphorus).One of the central questions in metallic glass research is the prediction of glass-forming ability (GFA)—the tendency of a composition to form an amorphous rather than crystalline structure upon cooling. Understanding GFA is crucial for designing bulk metallic glasses (BMGs), where large sample sizes without crystallization are desired. In addition to conventional MGs, in high-entropy metallic alloys, where many components are mixed, phase prediction becomes particularly challenging due to competition between crystalline and amorphous phases. 

One of the central challenges in metallic glass research is determining the glass-forming ability (GFA), i.e., the propensity of a given alloy composition to avoid crystallization and form an amorphous structure upon cooling. There is no single universal parameter for evaluating GFA. Rather, as highlighted above, many empirical and theoretical criteria have been proposed. Understanding GFA is crucial for designing bulk metallic glasses, where large sample sizes without crystallization are desired. A reliable prediction of GFA is also essential for discovering novel BMGs, as experimental exploration alone is impractical due to the huge compositional possibilities. Schroers \textit{et al.} estimated that around 3~million binary, ternary, quaternary, and quinary BMG alloys potentially exist, based on empirical rules, such as atomic size ratio, heat of mixing, and liquidus temperature applied to 32 metallic elements~\cite{li2017many}. Consequently, in recent years, GFA prediction has become a canonical problem for machine learning (ML) and data-driven approaches in materials science. These methods have been shown to effectively address complex problems arising from the intricate interplay of thermodynamic, kinetic, and structural factors, enabling extrapolation beyond the training data~\cite{butler2018machine,schmidt2019recent}.
This section reviews the key efforts that apply ML to the prediction of GFA. We divide the scope into three interconnected parts: data generation, model selection, and validation technique, reflecting the typical workflow of ML studies and highlighting how progress in each area has shaped current capabilities. \\

\textit{Data generation}: This is the cornerstone of any ML approach for predicting GFA. Four main sources of GFA data can be identified. \textit{Literature data} coming from experiments and reference handbooks, provide one source. The most widely used is the dataset assembled by Ward \textit{et al.}~\cite{ward2018machine}, which presents measured critical casting diameters in many alloy systems. Other important sources are handbooks collecting phase diagrams and thermophysical properties~\cite{okamoto2000phase,kawazoe1997nonequilibrium,kawazoe2022phase}. Although this is the most followed strategy to obtain data for ML, yet it remains a bottleneck.  
Experimental GFA data may be imbalanced, e.g., biased toward high-GFA alloy systems and underrepresenting failures or marginal cases, leading to overfitting and poor generalization~\cite{zhou2021rational}. Data augmentation techniques mitigate imbalance~\cite{yao2022balancing}: oversampling (duplicating minorities) or nonlinear transformations. Overall, if we include all available data, the GFA dataset could reach $\sim 8000$ entries~\cite{zhou2022critical}.
\textit{Thermodynamic data from CALPHAD calculations.} Computational thermodynamics allows extraction of parameters such as the liquidus temperature $T_l$ from assessed phase diagrams. From those, one can extract GFA from calculating the eutectic depth, as done by Dasgupta \textit{et al.}~\cite{dasgupta2019probabilistic}, or a liquidus depression parameter~$\Psi$, as defined by Greer's group~\cite{houghton2024calphad}. CALPHAD's limitations, such as high cost and incomplete databases for rare elements, however, persist.
\textit{Synthetic data from MD simulations} generate thousands of virtual alloys by modeling atomic trajectories during cooling, yielding features that relate to GFA such as icosahedral fraction or bond order parameters~\cite{afflerbach2021molecular,hu2023data}. 
However, a limitation may be the simulation timescale which might not be able to tackle other important thermodynamical parameters in the GFA prediction.
\textit{High-throughput synthesis techniques.} Advanced combinatorial fabrication methods, such as magnetron co-sputtering from multiple targets, can produce thin films with continuous composition gradients covering hundreds to thousands of alloy compositions in a single deposition run~\cite{ren2018accelerated,xie2023application}. These composition-spread libraries can then be screened for GFA indicators (e.g., via XRD). However, as commented above, it is difficult to relate amorphous thin film GFA with bulk~\cite{houghton2024calphad} and also, the field is still in its infancy.  

 \begin{figure}[!htbp]
	\centering
	\includegraphics[width=0.5\textwidth]{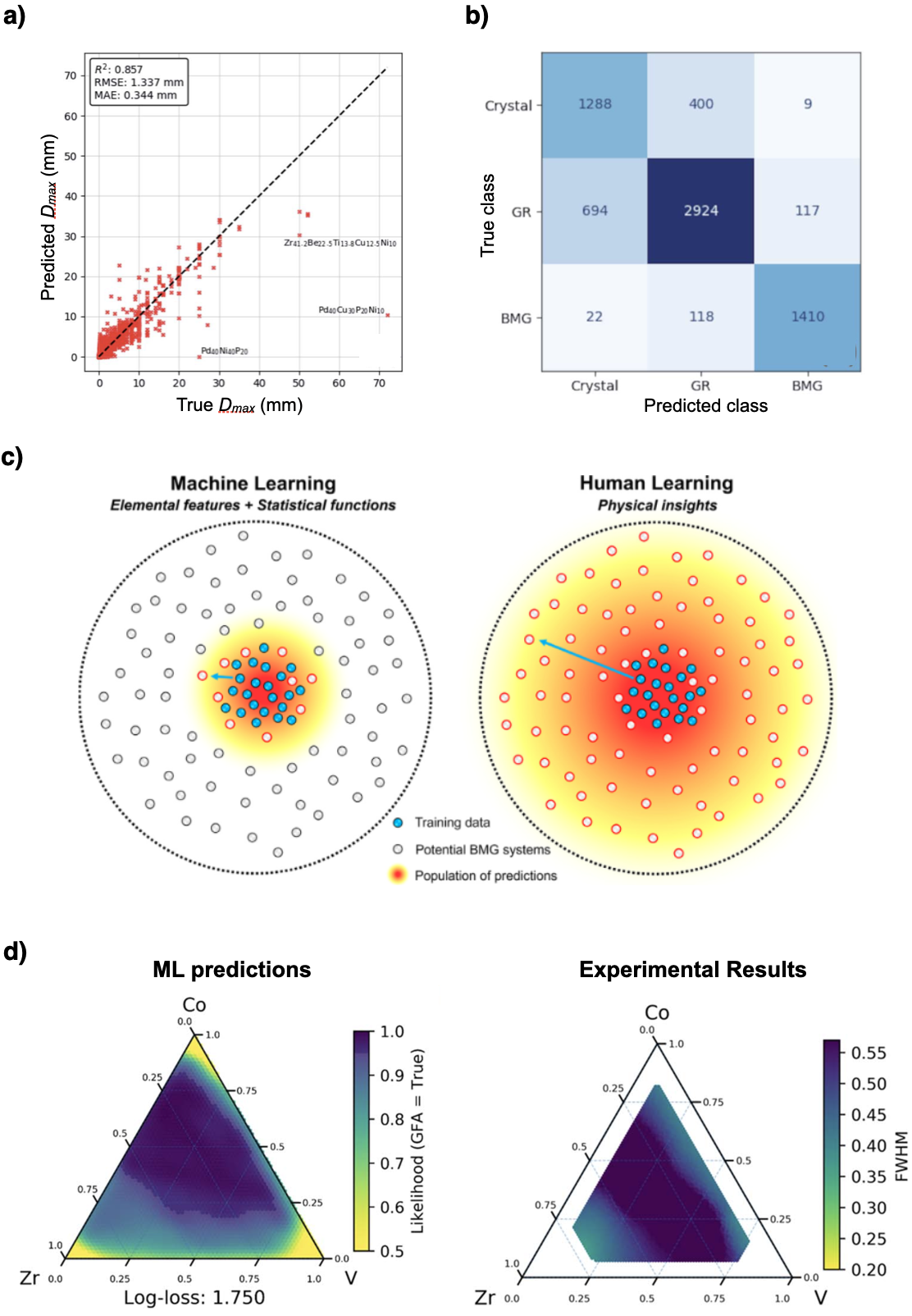}
	\caption{
    %Example of data generation for ML prediction of GFA using: 
    %(a)~Literature available data for GFA, $D_{max}$ and supercooled liquid range~$\Delta T_x$~\cite{ward2018machine}, 
    %(b) CALPHAD to compute the relative liquidus depression~$\Psi$ for AuCuSi liquidus surface. Reprinted from~\cite{houghton2024calphad} under CC~BY~4.0 license.
    Example of 
    a)~regression and b)~classification methods within a single output neural network model. Figures adapted from~\cite{forrest2022machine} under CC~BY~3.0. 
    Regression is used to predict the maximum diameter of a glassy rod~$D_{max}$ compared to true values. Classification confusion matrix is used to predict the GFA category of an alloy composition through a confusion matrix: crystalline (non glass forming), glassy ribbon (GR), or bulk metallic glass (BMG), showing counts of true labels (rows) versus predicted labels (columns).
    c) Role of (Left) Machine learning features, (elemental and statistical function) and (right) human features in predicting GFA. Figure reprinted from~\cite{liu2023machine}, under CC~BY~NC~ND licence. 
    d) (Left) ML prediction of glass forming likelihood in the Co-V-Zr ternary system, optimized for sputter deposition with principal component analysis (PCA) parameters included (yellow: low likelihood, dark blue: high). (Right)~Experimental validation mapping the full width at half maximum (FWHM) of the first sharp diffraction peak via high-throughput sputtering and XRD (yellow: low, indicating crystalline. Dark blue: high, indicating amorphous). Adapted from~\cite{ren2018accelerated}, under CC~BY~4.0 license.}
	\label{ML_GFA}
\end{figure}

\textit{Model selection}: Machine learning models for GFA prediction fall in two categories, \textit{regression} or \textit{classification}, sharing core steps like descriptor construction and hyperparameter optimization. The choice between regression and classification depends on the type of target data available, continuous values such as cooling rate~$R_c$, reduced glass transition temperature~$T_{rg}$ or critical casting diameter~$D_{max}$ (see Fig.~\ref{ML_GFA}a)),
as opposed to discrete labels like crystalline, ribbon or bulk (see Fig.~\ref{ML_GFA}b)).
A variety of algorithms have been explored for GFA regression. Support vector regression (SVR) to predict $D_{max}$ from thermodynamic descriptors, demonstrating good generalization in small datasets~\cite{Xiong_Zhang_Shi_2019}.
Tree-based ensembles, including random forest regression and gradient boosting, to capture correlation of key features with $D_{max}$ prediction~\cite{VERMA2024122710}.
Artificial neural networks (ANNs), from shallow feed-forward architectures to deeper models, capable of learning complex feature hierarchies, but more prone to overfitting in the limited-data regime.
Gaussian process regression (GPR) offers probabilistic predictions and uncertainty estimates, used by Zhou et al.
to guide active-learning cycles~\cite{zhou2021rational}. 

In the classification setting, the task is to assign each composition to discrete categories, most often ``glass'' or ``non-glass” (binary classification)~\cite{ward2016general,yao2022balancing}. Multi-class schemes also exist~\cite{ward2018machine} based on threshold values of $D_{max}$ or $R_c$. Classification is also used as an initial large-scale screening, where the objective is to discard clear non-glass formers early, reducing the candidate pool before more computationally intensive modeling~\cite{zhou2021rational}. Promising examples include pioneering works by Wolverton \textit{et al.}~\cite{ward2016general,ward2018machine} that employ random forest classifiers with input literature datasets, as well as hybrid classification-regression models for broader compositional exploration~\cite{zhou2021rational}. 
Building on these models, ML enables efficient exploration of compositional spaces for discovering new MGs.
This is achieved, for example, by predicting the GFA across all possible ternary alloys derived from combinations of elements in the training dataset~\cite{ward2016general}, or by deep searches in higher-order (quaternary to senary) alloy spaces~\cite{zhou2021rational}.

A more recent study~\cite{LIU2023118497} compares the role of different features in the prediction of GFA.  
In this work, the authors compare four distinct feature sets in random forest models for GFA prediction, using a database of 6816 unique alloy compositions (spanning 55 elements, binary to octonary) labeled as BMG ($R_c < 10^3$ K/s), ribbon ($R_c < 10^6$ K/s), or non-ribbon ($R_c > 10^6$ K/s) formers. They reconstruct Wolverton's model with 201 alloy features from six statistical functions (e.g., mean, range, mode) on 31 elemental properties. Surprisingly, this model achieves ~89\% accuracy, but its performance is indistinguishable from two baseline ``unphysical'' models: one built on randomly generated (meaningless, non-physical) values for five arbitrary elemental properties (then converted to alloy features via the same six statistical functions), and another using no features at all, relying solely on compositional information (i.e., the atomic percentages of each element in the alloy as direct inputs). Yet, in predicting a vast ternary space ($\sim$2.6 million alloys from 24 elements), the general-material model favors BMGs near training data, lacking extrapolative power. Conversely, a human-learning model with three physical features: i)~liquidus reduction $\Delta T$ (from binary pairs), ii)~atomic size difference $\delta$, and iii)~maximum mixing heat $\Delta H_{\max}$ (Miedema model), excels by $>$20 times in recovering withheld BMGs at high probability ($p>$0.95). This is illustrated in Fig.~\ref{ML_GFA}c), which contrasts the two approaches: the left panel shows predictions clustered tightly around training data (blue dots) in a sparse space, indicating limited exploration. The right panel depicts a broader, gradient-filled distribution of predictions (yellow-to-red) that extend far beyond the training data, enabling discovery in the vast composition space.
The study highlights the limitations of feature-agnostic ML for GFA, where huge spaces require physical insights (thermodynamics, topology) to avoid overfitting and enable extrapolation beyond proximity predictions. The integration of human rules for an accurate representation of alloys is emphasized, as elemental statistics miss key mechanisms such as eutectic suppression or crystallization resistance.

Recent developments have also explored \textit{inverse design} using deep learning. In particular, \textit{Generative adversarial networks (GANs)} have been employed to generate novel, compositionally complex BMG compositions, like high-entropy types~\cite{zhou2023generative}. Trained on curated datasets and physical descriptors, GANs can suggest new compositions likely to exhibit high GFA. The generated candidates are evaluated by comparing them to known alloys using techniques such as \textit{principal component analysis (PCA)}, offering a visual and quantitative assessment of similarity and novelty.

\textit{Validation}:
%Once a model demonstrates reliable predictive performance, the next step involves \textit{``human-centered ML''}—interpreting which descriptors are most relevant and why~\cite{VERMA2024122710,D2DD00026A}. Techniques like \textit{feature ablation} (removing or adding descriptors and measuring changes in accuracy) help rank feature importance. 
%While this approach helps identify key descriptors, it does not always clarify more complex effects such as \textit{cocktail effects}, where interactions between multiple elements jointly influence GFA in nontrivial ways.
The reliability of ML models for GFA prediction is evaluated through several methods, ranging from pure computational techniques to rigorous experimental confirmation. Here we distinguish these methods in two main categories: computational self-assessment, which focuses on statistical and internal model checks, and experimental validation, which involves the synthesis and testing of predicted alloys to practically validate the results.

\textit{Computational Self-Assessment.} Many studies rely on internal ML validations to evaluate performance without physical synthesis. Common techniques include k-fold cross-validation (CV) for overfitting mitigation, as in Wolverton's random forest classifiers~\cite{ward2016general,ward2018machine}, achieving $\sim$89\% accuracy via 10-fold CV. 
For Gaussian process regression (GPR)~\cite{zhou2021rational}, the uncertainty quantification provides probabilistic outputs, to guide active learning cycles and tag unreliable predictions. Once a model demonstrates reliable predictive performance, the next step involves \textit{"human-centered ML"}-interpreting which descriptors are most relevant and why~\cite{VERMA2024122710,D2DD00026A}. Techniques such as \textit{feature ablation} (removing or adding descriptors and measuring changes in accuracy) help rank feature importance, alongside permutation methods. Liu \textit{et al.}~\cite{LIU2023118497} apply these to compare physical vs. unphysical features, pointing out the shortcomings of capture mechanisms such as eutectic suppression. Although efficient, these methods miss the synthesis challenges, kinetic factors, or data biases inherent in glass formation. The glass forming ability is also highly sensitive to the processing method, e.g., vapor deposition can produce ultra-stable thin film glasses with enhanced GFA in compositions that might not form glasses in bulk via melt quenching and direct extrapolation between the two is challenging~\cite{ding2014combinatorial,kube2022compositional}.

\textit{Experimental Validation.} To address these gaps, some works proceed to experimental verification of the ML results, synthesizing novel predicted alloys via melt spinning, suction casting, and characterizing them via XRD for amorphicity, differential scanning calorimetry (DSC) for thermal properties, or mechanical testing. For instance, the new Zr-based BMGs predicted by the hybrid model of Zhou \textit{et al.}~\cite{zhou2021rational} were cast into amorphous rods up to 5~mm in diameter and validated via XRD. The GAN framework~\cite{zhou2023generative} generated high-entropy BMGs, with candidates cast into 2~mm rods and confirmed amorphous. Other examples include inverse design via variational autoencoders (VAE), generating MGs with targeted ductility in Cu-Zr systems and verifying properties experimentally~\cite{li2024inverse}. In biocompatible Ti-Zr-Cu-Pd alloys, ML guided low-Cu compositions, were assessed for GFA via casting showing improved biocompatibility~\cite{douest2024machine}. For Fe-based BMGs, models predicted $D_{max}$ using thermophysical features, validated through arc-melting and suction casting of new amorphous ribbons~\cite{jeon2021inverse}. 

More importantly, some approaches use experiments to iteratively refine the ML model, creating active learning loops where new data from synthesis feeds back to improve predictions. A seminal example is Ren \textit{et al.}'s iterative ML high-throughput experimentation in Co-V-Zr ternaries, discovering new MGs by refining models with experimental outcomes~\cite{ren2018accelerated}.
This process is shown in Fig.~\ref{ML_GFA}d), where the left panel shows the first generation ML prediction of glass-forming likelihood in the Co-V-Zr ternary composition, optimized for sputter deposition with PCA parameters.  The right panel depicts the experimental validation via high-throughput sputtering and XRD, mapping the full width at half maximum (FWHM) of the first sharp diffraction peak (FSDP), confirming numerical predictions.
Experimental validation confirms the predictions, but also reveals some limitations, such as the overestimation of $D_{max}$ due to unmodeled impurities. 

%Despite these advances, fundamental challenges of data-driven science remain. For instance:\textit{Data quality and coverage}: Experimental GFA data may be sparse, uncertain, or biased toward specific alloy systems. \textit{Extrapolation limitations}: ML models trained on biased data may struggle to generalize to unseen compositions.
%To address these issues, two strategies are emerging:\textit{Artificial surrogate data generation} and \textit{advanced sampling techniques}~\cite{YAO2022114366}, which improve data diversity and reduce overfitting risks. \textit{Simulation-based datasets}, where GFA is estimated from computational models, offering a controlled alternative to experimental data.

\section{Future outlooks and challenges}
\label{sec:future}

In this review, we have summarized recent progress in the field of metallic glasses, covering advances in experimental characterization, multiscale modeling approaches, and the growing role of machine learning in understanding and designing these complex materials. Despite these significant achievements, metallic glasses still present numerous scientific and technological challenges that must be addressed to fully realize their potential in practical applications. Below, we outline key directions and open questions that will likely define the next phase of research in this field.

\subsection{Mechanical behavior}

We have made great advances in understanding how the mechanical behavior of metallic glasses is controlled by their processing history, thermodynamic energy state, glassy structure, and loading conditions. However, our ability to precisely control the structure and properties of metallic glasses remains poor and further research must be conducted to develop detailed process-structure-property relationships that are comparable to our knowledge for crystalline materials. This will require research on many fronts, including improvements in process control, structure characterization, and understanding of the mechanisms controlling mechanical properties.  

\begin{itemize}
    \item {\it Process control}: Although the influence of factors such as cooling rate, structural relaxation, and rejuvenation on the mechanical properties of metallic glasses is qualitatively understood, controlling these effects in practical component manufacturing remains difficult. Casting metallic glasses directly from the melt is inherently unstable, often producing uncontrolled gradients in cooling rate and glass structure. Achieving reliable control over the properties of complex-shaped castings therefore requires not only advanced process modeling tools but also new fundamental insights into how structure and properties evolve from the supercooled liquid, as dictated by the local thermal history and flow behavior. Besides, while additive manufacturing provides a high degree of process control through the various process parameters (e.g., laser power, laser scanning speed, etc.), the layer-by-layer nature produces strong gradients and mesoscale structures related to the melt-pools and their effect on the mechanical properties is still relatively unknown. Thermoplastic forming can offer relatively good control over the structure and properties of small dimensioned metallic glass parts \cite{Kumar:2010fo}, but its scalability to larger-scale, engineering components remains limited. Future applications of metallic glasses will likely involve components processed by all of these routes depending on the dimensions of the part and the application, and research should focus on using modern developments in modeling and machine learning to achieve new levels of control in the structure and properties that arise from any of the common fabrication routes. 
    \item {\it Structure characterization}: This review has highlighted some key advances in characterizing the atomic-level structure of metallic glasses \cite{Yang_Nature2021, Yuan_NaterMat2022, Nomoto2025APT} and making correlations to mechanical properties \cite{Nomoto_MaterTod2021, Li_JALCOM2025, Li_MSEA2022, Nomoto2025APT}. However, such studies are too often carried out with a single characterization method in isolation, and/or new characterization methods are not readily available to a broad range of researchers. Solving the structure of metallic glasses is a multiscale problem that requires several characterization tools (e.g., TEM, synchrotron X-rays, neutrons, atom probe, etc.) working together to reveal the full glassy structure. Research is needed to make more characterization tools routine and accessible to glass researchers, for example through user facilities and open source software for data analysis. Moreover, creating a composite picture of a metallic glass structure from multiple characterization tools is not a straightforward task, and this will benefit greatly from new research into modeling and machine learning tools that can help interpret the data and create a three-dimensional structural picture. Similarly, there is a large scope for new modeling and interpretable machine learning approaches to help reveal new structure property relationships, as is currently done for crystalline metals \cite{Liu_ADDMAN_2024}.  
    \item {\it Mechanisms of deformation and fracture}: At the nanoscale, deformation carriers in metallic glasses are generally described in terms of free volume and shear transformation zones (STZs). However, their relative importance under different loading and temperature conditions remains poorly understood, as does the manner in which they interact and eventually coalesce to form shear bands. The latest experimental developments in studying shear band behavior \textit{in situ} will help advance our knowledge \cite{Glushko_NatComm2024}, but more work is needed to understand the multi-scale nature of the deformation and failure processes in metallic glasses. This will require further experimental advances for the \textit{in situ} study of nanoscale deformation mechanisms, as well as new modeling and machine learning tools to link the observations across multiple length scales. At macroscopic length scales, our ability to understand and predict the fracture of metallic glasses is limited by a lack of adequate analysis tools. Existing analytical constructs for elastic-plastic fracture mechanics assume the strain hardening material behavior of crystalline metals, while metallic glasses are instead strain softening materials. This creates well-known problems in measuring and predicting the fracture behavior of metallic glasses \cite{Gludovatz:2014cp}, and new developments are required in the field of strain softening, elastic-plastic fracture mechanics to accurately predict their behavior.  
\end{itemize}

\subsection{Additive Manufacturing}
Future perspectives on AM of MGs include the integration of advanced computational and monitoring techniques to overcome persistent challenges (crystallization, defect minimization, and scalability). 
Machine learning driven optimization of LPBF parameters are promising to automatize the balance between density and amorphicity~\cite{graeve2023latest,kononenko2024designing}. These methods could result in process acceleration and adjustments, predicting optimal conditions without performing lengthy trial-and-error experiments. Porosity prediction models are useful for defect predictions~\cite{vastola2018predictive}.
To refine the molten pool dynamics, thermomechanical modeling should be integrated with feedback control, to possibly enable precise management of temperature gradients~\cite{scime2019using,oster2024deep,wannapraphai2025quantifying}. All these advances pave the way for producing more reliable MG components with tailored properties, expanding their adoption in several high-performance applications. 

\subsection{Efficient simulations and Machine learning}

Developing efficient simulation techniques is a vital issue, especially for nanoscale modeling, since key features of metallic glasses involve long-timescale behavior. The recent surge of machine learning methods is also making a strong impact on nanoscale modeling, including the development of machine-learning interatomic potentials and the prediction of glassy dynamics.

\begin{itemize}
    \item {\it Machine-Learning Interatomic Potentials}: The current state of the art of machine-learning interatomic potentials (MLIPs) can be roughly divided into two groups; (1) targeted and accurate material-specific MLIPs, and (2) large general-purpose MLIPs for wide applicability to many materials and properties. The first group represents the conventional strategy of developing interatomic potentials, where accurate simulations for a specific material and properties are desired. In very recent years, the MLIP community has increasingly focused on the second group by developing \textit{foundation potentials}, i.e., universal interatomic potentials for the entire periodic table~\cite{chen_universal_2022,deng_chgnet_2023,batatia_foundation_2023,yang_mattersim_2024,fu_learning_2025,grasselli2025uncertainty}. These advances have been driven by improved machine-learning regression architectures, particularly equivariant graph neural networks, together with the development and public availability of large, consistently calculated materials databases~\cite{deng_chgnet_2023}. 
    Nevertheless, a central challenge for foundation-model approaches is to obtain sufficiently diverse and high-quality reference datasets, especially for multicomponent bulk metallic glasses, where experimental atomic structures are scarce and quantum-level calculations remain expensive. Despite their broader training scope, current foundation-model MLIPs still exhibit limited transferability across composition, temperature, and pressure, often performing reliably only near the states represented in the training data. 
    Moreover, fine-tuning or the development of smaller, system-specific MLIPs is still required to reach the accuracy needed for quantitative insight into the structure-property relations of metallic glasses.

    \item {\it Efficient molecular simulations}: At the level of classical molecular simulations, advanced algorithms such as the swap Monte Carlo method have been established, but their applicability remains largely restricted to specific model systems. Extending such algorithms to more general metallic glass formers with realistic interaction potentials is therefore an important direction for future research. 
    For example, a very recent study has demonstrated the effectiveness of swap Monte Carlo in more realistic systems, such as metallic alloys modeled with EAM potentials~\cite{kaskosz2023efficient}. 
    Another promising avenue is to combine different efficient sampling algorithms in order to exploit possible synergies among them~\cite{jung2025numerical}. Moreover, there are also attempts to develop machine-learning–assisted sampling algorithms aimed at accelerating the equilibration process~\cite{jung2024normalizing,galliano2024policy}.

    \item {\it Machine learning prediction of dynamics}: An exciting line of current research is the prediction of glassy dynamics-whether thermally activated or driven by external loading-directly from static structural snapshots using machine learning~\cite{jung2025roadmap}. A wide spectrum of machine learning models has been explored, ranging from simple linear regression with carefully designed descriptors to state-of-the-art deep learning architectures~\cite{cubuk2015identifying,boattini2021averaging,bapst2020unveiling,pezzicoli2024rotation}. These approaches have already achieved high levels of predictive accuracy. However, machine learning models, in particular, deep learning models, are often regarded as “black boxes”, making it difficult to connect predictions with underlying physical mechanisms. For this reason, interpretability and explainability are now viewed as essential next steps, as the community increasingly seeks to extract genuine physical insight from machine-learning-based predictions~\cite{wang2020predicting,liu2023concurrent,swain2024machine,sharma2024selecting,liu2024classification,sharma2025interpretability,yoshikawa2025graph}.
\end{itemize}

% HERE CITE: @article{makinen2025growth,
%\

\subsection{Multiscale modeling and data-driven approach}

Integrating data-driven techniques with multi-scale modeling offers great promise for advancing our understanding of the complex, non-linear plastic deformation of bulk metallic glasses (BMGs). 
Data-driven methods trained on atomistic simulations or experiments can both refine existing plasticity models and guide the development of new ones. 
Embedding these mesoscopic models into continuum-scale simulations enables physically grounded predictions over larger spatial and temporal scales, opening opportunities for improved design and engineering of BMG-based components. 
Key challenges that remain are outlined below.

%A few works already mentioned have shown progress in this direction. Zhang \textit{et al.}~\cite{Zhang2022} uses structural information to train a machine-learned parameter, softness, that correlates with a given region's susceptibility to plastic deformation. This parameter allows their model to span the ranges from atomistic data all the way to the mesoscale modeling of plasticity. 
%Xu \textit{et al.}~\cite{Xu} harness the power of modern computational technology to harvest a large dataset of MGs with various thermal and loading histories, which can then be utilized to generate a dynamical map of MGs as they are loaded with stress. This stochastic evolution elasto-plastic model (SEEM) gives deep insight into the stochastic evolution of MGs from the initial atomic configuration, to yielding, and then to steady-state flow. 
%The work by Kontolati \textit{et al.}~\cite{KONTOLATI2021117008} used ML techniques to learn from MD data to refine existing STZ plasticity model for continuum simulation. Wen and Wei~\cite{WEN2024105629} employed ML techniques on experimental data to create a new, data-driven plasticity model for continuum simulation.

\begin{itemize}
    \item {\it Sample size and boundary conditions}: Determining the optimal sample size for conducting MD simulations or experiments to gather data relevant to plastic deformation can be challenging. When the sample is too small, excessive stochastic noise may obscure meaningful patterns in the data, rendering them less informative. Conversely, if the sample is excessively large, isolating data of individual plastic events becomes difficult and the resulting dataset may be overly averaged. A strategy for optimizing the sample size is needed which is better than trial-and-error. Besides, matching boundary conditions exactly between simulations at different scale can be tricky, but necessary, in order to compare the simulations directly. Boffi and Rycroft~\cite{boffi20} have contributed in this regard by devising a coordinate transformation methodology for continuum simulations in rectangular Eulerian grid to match the Lees-Edwards boundary condition in MD simulations. Further developments are needed to handle more complex geometries and other boundary condition cases. 

    \item {\it State variables}: Mapping state variables between models at different scales is another challenge, but essential if we want to integrate the models. As mentioned, Hinkle et. al.~\cite{Hinkle2017} contributed by proposing a mapping from atomistic potential energy in MD to effective temperature in STZ theory for continuum simulations. However, the map was done by choosing an equation, which in itself is a limitation as we cannot know what the true mapping function is. In the future, data-driven methods may be employed, like in Xu \textit{et al.} \cite{Xu}, to find optimal equation-free data-driven mappings between the variables. 

    %item Simulation evolution variables: Aligning evolution variables between simulations at different scales presents another challenge. Continuum simulations typically employ time-stepping methods, whereas MD simulations and experiments often control strain rates. Thus, integrating a plasticity model derived from data gathered in MD simulations or experiments into a continuum simulation requires the development of consistent and accurate strategies to compute the correct strain rate with each timestep in the continuum model.  

    \item {\it Heterogeneity of material properties}: In continuum models, material properties like the shear modulus are typically assumed to be uniformly distributed across the entire domain. However, data-driven mesoscopic plasticity models may exhibit sensitivity to local variations in material properties, determined by the sample size. Therefore, the need arises to introduce heterogeneous material property fields into continuum simulations to achieve greater physical accuracy in plasticity simulation. 
    
    \item {\it Plasticity lengthscales}: Determining the appropriate lengthscales for the mesoscopic plasticity model within the continuum framework presents a challenge. This entails defining the size of the local domain within the continuum code to collect input for the plasticity model. Moreover, when plastic deformation occurs as predicted by the plasticity model, we need to determine the spatial scope to which the plasticity affects the continuum domain. Additionally, careful consideration need to be given to the computation of input from the continuum model for the plasticity model, as well as the application of output from the plasticity model into the continuum model. Common approaches include mean averaging or Gaussian smoothing, which require testing to ensure physically accurate results.
    
    %\item Adaptivity in continuum model: We may need to resolve the lower-scale plastic events with adaptive spatial discretization in the continuum code, to capture their local features. Additionally, to effectively capture discrete plastic events given by the plasticity model -- events that may lack differentiability over time -- adaptive time stepping methods may be employed. Careful considerations should be placed in dynamically adjusting the resolution of the continuum model, both temporally and spatially, to accurately capture the evolving plastic deformations. 
\end{itemize}

%\section{Applications}
%For a more comprehensive review of the manufacturing processes, look Ref.~\cite{sohrabi2024manufacturing}. For applications, see Ref.~\cite{gao2022recent}.

\section*{Author Contributions}
JJK, IG, and RB contributed equally to the writing of the first draft of the experimental section. JE contributed to the writing, reviewing, and editing. GM contributed to the writing, reviewing and editing of the draft. JL and SF contributed to the first draft and figure for the mesoscale and continuum modeling sections, and these were further edited and contributed to by MLF and CHR. JCD contributed to the reviewing and editing. JB contributed to the Nanoscale modeling section.
MO, ADSP, and SS contributed to the Nanoscale modeling and multi-scale modeling sections, and to overall editing of the manuscript. SB organized the entire review paper, integrating all contributions, and led the writing, reviewing, and editing of the manuscript.

\section*{Acknowledgments}
This paper originates from discussions at the ``First International Workshop on Complex Glasses'', held in October 2023 in Warsaw, organized by SB and NOMATEN CoE (at the National Center for Nuclear Research, NCBJ). We would like to thank all the participants.
In particular, we thank Łukasz Żrodowski and Bartosz Morończyk from AMAZEMET  for valuable discussions on additive manufacturing from an industrial perspective.
We also thank Walter Kob for discussions and helpful suggestions on the draft.
The discussions in the above Workshop continued at the MRS Meeting in Boston in December 2024 and culminated in a Symposium on bulk metallic glasses held at Warsaw University of Technology in January 2025.

This work has been supported by the European Union Horizon 2020 research and innovation program under grant agreement no.~857470 and from the European Regional Development Fund via the Foundation for Polish Science International Research Agenda PLUS program grant No.~MAB PLUS/2018/8.

JJK would like to acknowledge financial support from Australian Research Council grant DP240101127 and from the Alexander von Humboldt Foundation Friedrich Wilhelm Bessel Award. 

IG and RB acknowledge the financial support of the European Innovation Council through the HORIZON-EIC-2021-PATHFINDEROPEN-01 grant (101046870) and the Deutsche Forschungsgemeinschaft
 (DFG). IG and RB also acknowledge DESY (Hamburg, Germany) and ESRF (Grenoble, France) for the provision of experimental facilities.

JE thanks additional support provided through the European Research Council under the Advanced Grant “INTELHYB – Next Generation of Complex
Metallic Materials in Intelligent Hybrid Structures” (Grant No: ERC-2013-ADG-340025,) the Austrian Science Fund (FWF) and Land Steiermark, Project No. P31544-NBL.

JCD acknowledges support from the VILLUM Foundation's \textit{Matter} grant VIL16515.
 
SB thanks support from SONATA BIS number DEC-2023/50/E/ST3/00569 by National Science Centre Poland (NCN) and FIRST TEAM FENG.02.02-IP.05-0177/23 by Foundation for Polish Science (FNP).

MLF, CHR, SF and JL acknowledge support by the National Science Foundation under Grant Nos. 2323718/2323719/2323720 through the Designing Materials to Revolutionize and Engineer our Future (DMREF) program. CHR was supported in part by the U.S.\@ Department of Energy, Office of Science, Office of Advanced Scientific Computing Research's Applied Mathematics Competitive Portfolios program under Contract No.~AC02-05CH11231.

JB acknowledges funding from the Research council of Finland through the OCRAMLIP project, grant number 354234.

G.M. acknowledges support by the project GLAXES ERC-2021-ADG (Grant Agreement No. 101053167) funded by the European Union. 

M.O. thanks the support by MIAI@Grenoble Alpes and the Agence Nationale de la Recherche under France 2030 with the reference ANR-23-IACL-0006).
 
S.S. acknowledges SERB(ANRF) (India) for support through the JC Bose Fellowship (JBR/2020/000015) SERB(ANRF), DST (India) and a grant under SUPRA (SPR/2021/000382). 

\bibliography{biblio_arxiv}
\bibliographystyle{unsrt}
\end{document}